\documentclass[twocolumn,superscriptaddress,showkeys,floatfix,jcp]{revtex4-1}
\usepackage{amssymb,amsfonts,amsmath}
\usepackage{graphicx}
\usepackage{color}

\usepackage{hyperref}

\begin{document}

\title{The structure of the genotype-phenotype map strongly constrains the evolution of non-coding RNA}

\author{Kamaludin Dingle }
\affiliation{Rudolf Peierls Centre for Theoretical Physics, University of Oxford, Oxford, OX1 3NP, United Kingdom}
\affiliation{Systems Biology DTC, University of Oxford. UK}
\affiliation{Department  of Mathematics and Natural Sciences, Gulf University for Science and Technology, Block 5, West Mishref, Kuwait.}
\author{Steffen Schaper}
\affiliation{Rudolf Peierls Centre for Theoretical Physics, University of Oxford,  Oxford, OX1 3NP, United Kingdom}
\affiliation{Bayer Technology Services GmbH, BTS-TD-ET-CSM, Building B106, 51368 Leverkusen, Germany}
\author{Ard A. Louis}
\email{ard.louis@physics.ox.ac.uk}
\affiliation{Rudolf Peierls Centre for Theoretical Physics, University of Oxford, Oxford, OX1 3NP, United Kingdom}

\begin{abstract}
The prevalence of neutral mutations implies that biological systems typically have many more genotypes than phenotypes.
 But can the way that genotypes are distributed over phenotypes determine evolutionary outcomes?  Answering such questions is difficult because the number of genotypes can be hyper-astronomically large.   By solving the genotype-phenotype (GP) map for RNA secondary structure for systems up to length $L=126$ nucleotides (where the set of all possible RNA strands would weigh more than the mass of the visible universe) we show that the GP map strongly constrains the evolution of non-coding RNA (ncRNA).  Simple random sampling over genotypes  predicts the distribution of properties such as the mutational robustness or the number of stems per secondary structure found in naturally occurring ncRNA with surprising accuracy.
Since we ignore natural selection, this strikingly close correspondence with the mapping suggests that  structures allowing for functionality are easily discovered, despite the enormous size of the genetic spaces.  The mapping is extremely biased: the majority of genotypes map to an exponentially small portion of the morphospace of all  biophysically possible structures. Such strong constraints provide a non-adaptive explanation for the convergent evolution of structures such as the hammerhead ribozyme.   These results presents a particularly clear example of bias in the arrival of variation strongly shaping evolutionary outcomes and  may be relevant to Mayr's distinction between proximate and ultimate causes in evolutionary biology.
\end{abstract}

\date{\today}
\maketitle

\noindent {\bf \large 1. Introduction} 

 Many  questions about the limits of evolution hinge not only what has happened in natural history, but also on {\em counterfactuals}:  what did not happen, but perhaps could have.  Re-run the tape of life~\cite{gould1989wonderful,morris2003life} and what parts of  phenotype space  -- 
 the set of all possible phenotypes~\cite{mcghee2006geometry} --  would be occupied?  
   Typically, only a minescule fraction of the phenotype space has been explored throughout natural history. The reasons given for this phenomenon usually combine adaptive arguments:  some parts of phenotype space yield higher fitness than others, with  arguments based on contingency: nature hasn't had time to explore all of phenotype space.  However, evolutionary search does not occur by uniform sampling over phenotypes, but rather by random mutations in the space of genotypes.    Does this basic fact alter the way that phenotype space is explored and occupied?  To answer such whole genotype-phenotype (GP) map questions is difficult, in part because the number of possible genotypes typically grows exponentially with length, and so rapidly becomes unimaginably vast~\cite{salisbury1969natural,smith1970natural,wagner2005robustness}, and in part because biological systems with sufficiently tractable GP maps are rare.

\begin{figure}
\includegraphics[height=5.5cm,width=7cm]{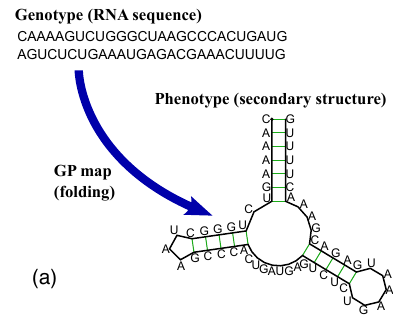} \hspace{0.5cm}
\caption{{\bf  Schematic of the mapping from an RNA sequence to a SS.} Here for an $L=55$ type III hammerhead ribozyme with three stems. Any sequence that folds to the same SS topology is considered to map to the same SS phenotype.  \label{fig:schematic}  } 
\end{figure}

One system where progress can be made is  the mapping from sequences to the structure of RNA.   Although simpler than many biological phenotypes, RNA is interesting and important because it can fulfil multiple roles both as an information carrier (e.g.\ messenger RNA and viral RNA), or as functional non-coding RNA (ncRNA)~\cite{mattick2006non}, including  chemically active catalysts (e.g.\ ribozymes) and key structural components of larger self-assembled structures (e.g.\ ribosomal RNA). One reason RNA is so versatile is that it can fold into complex three-dimensional (3D) structures.   The bonding pattern of these structures is called the secondary structure (SS), which is an important determinant of the 3D structure and biological activity of ncRNA molecules, and so can be treated as a (simplified) phenotype in its own right~\cite{carothers2004informational,bailor2011topological} (Fig.~\ref{fig:schematic}).

Fast algorithms exist to predict the free energy minimum SS for a given sequence, and these are thought to be fairly accurate, especially for shorter strands~\cite{hofacker1994fast,zuker1999algorithms}.   Moreover, extensive databases exist for functional ncRNA~\cite{mattick2006non}.  For these reasons, this computationally tractable yet biologically relevant sequence to structure mapping  is  uniquely suited for 
investigating `whole genotype-phenotype map'  properties that may point towards general principles  relevant for a wider class of systems, and has been extensively studied over the last few decades~\cite{schuster1994sequences,fontana1993statistics,schultes2005compact, wagner2005robustness,schuster2006prediction,smit2006natural,cowperthwaite2008ascent,jorg2008neutral,aguirre2011topological,schaper2014arrival}.

Nevertheless, the RNA sequence to SS GP map also suffers from the exponential growth of the size of genetic space, which  has limited prior comprehensive GP studies to fairly small lengths, making direct comparison to evolutionary outcomes difficult.    Here we show that many detailed properties of this GP map can, in fact, be calculated, even for lengths as long as $L=126$, where the set of all possible strands would weigh more than the mass of the visible universe (see Methods).  
   One reason this is possible is because the map is highly biased towards a small fraction of phenotypes that take up the majority of genotypes.  This bias means that  sampling over genotypes (which we will call G-sampling)  generates significantly different outcomes than sampling over the morphospace~\cite{mcghee2006geometry} of all shapes (phenotypes) (which we will call P-sampling).  The existence of such highly peaked distributions in the mapping from genotypes to phenotypes,  also explains, in a way that is reminiscent of statistical mechanics, why sampling a relatively small number of genotypes is enough to determine certain key properties of RNA SS.  
   
   Perhaps most strikingly, we find that the distributions of several properties of natural ncRNA taken from the function RNA database (fRNAdb)~\cite{kin2007frnadb},  including the number of stems, the mutational robustness and the number of genotypes per SS phenotype, are very similar to what we obtain from random sampling over genotypes, and significantly different from uniform sampling over phenotypes. This result does not mean that natural selection can be ignored, but rather, as we will argue below,  that the mapping strongly prescribes which parts of morphospace are presented to natural selection as potential variation. Variation can only be selected if it arrives~\cite{devries1904mutation,wagner2014arrival}.

\medskip

  \begin{figure}
\includegraphics[height=5.6cm,width=8cm]{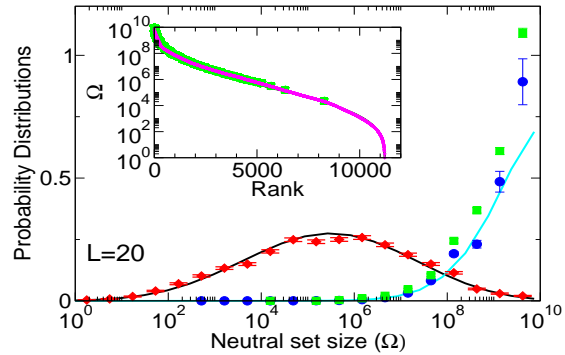}
\caption{{\bf Comparison of P-sampled and G-sampled distributions to natural data for $\bf L=20$ RNA}.    The P-sampled $ \displaystyle P_P(\Omega)$ (red diamonds) measures the probability  distribution for a phenotype  to have a given NS size $\Omega$. It differs markedly from  G-sampled $ P_G(\Omega)$  (blue circles), generated by random sampling over genotypes.  Error bars arise from binning data. The black and cyan lines are theoretical approximations to $P_P(\Omega)$ and $P_G(\Omega)$ respectively (see Methods).  The probability distribution of $\Omega$ for the SS  all $7327$ (non-trivial) $L=20$ sequences for \textit{Drosophila melanogaster} from the  fRNAdb database~\cite{kin2007frnadb}  (green squares) is much closer to the G-sampled $P_G(\Omega)$ than to the P-sampled $P_P(\Omega)$.  {\bf Inset:}  All $11,218$  SS phenotypes (purple triangles) ranked by NS size $\Omega$. There is strong bias, just $5 \%$ of phenotypes take up $58\%$ of all genotypes.   The $7327$ natural data points (green squares) are clustered at lower rank (larger $\Omega$). \label{figL20} } 
\end{figure}

  \begin{figure*}[htp]
\includegraphics[height=3cm,width=4.3cm]{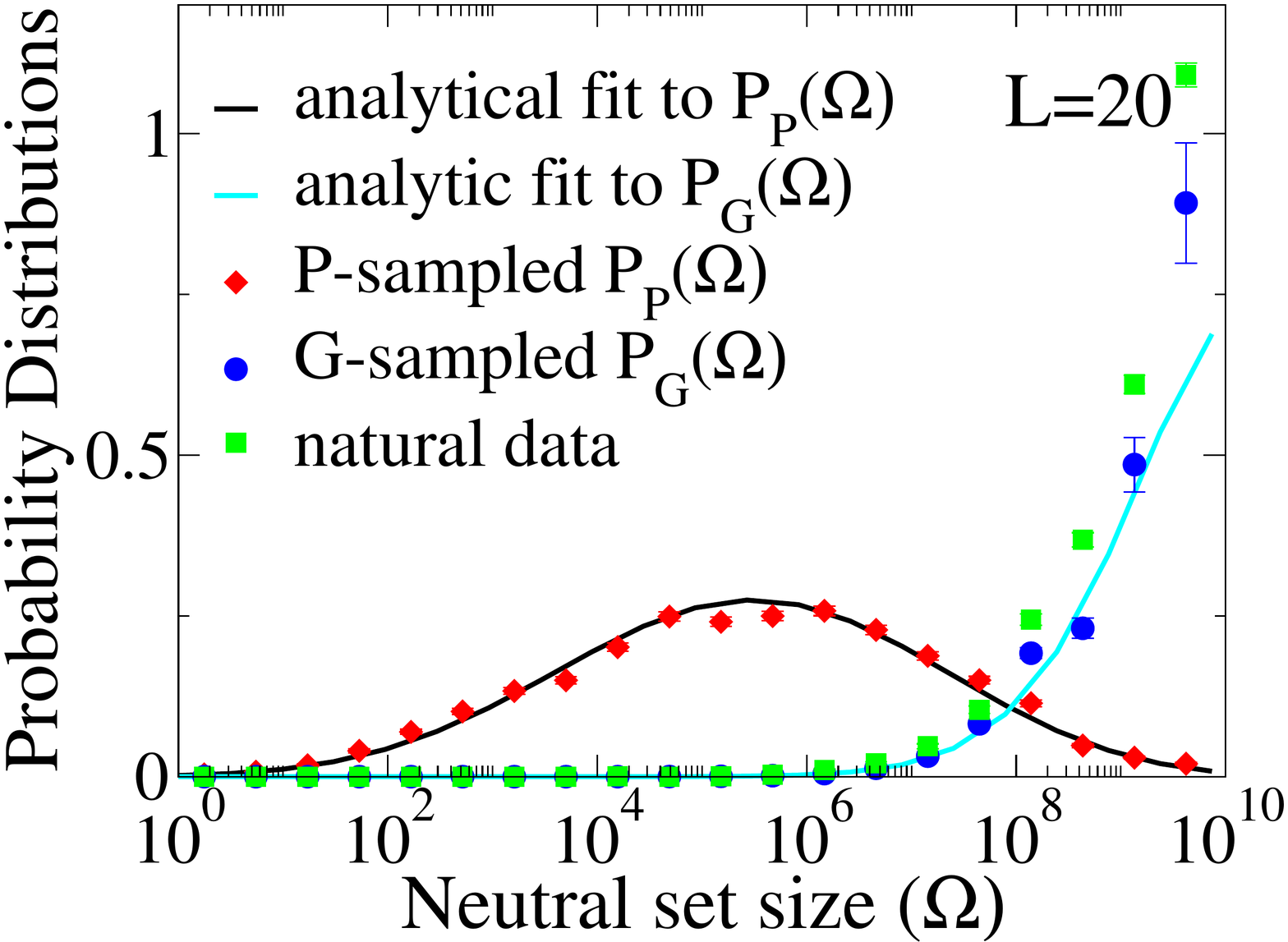}
\includegraphics[height=3cm,width=4.3cm]{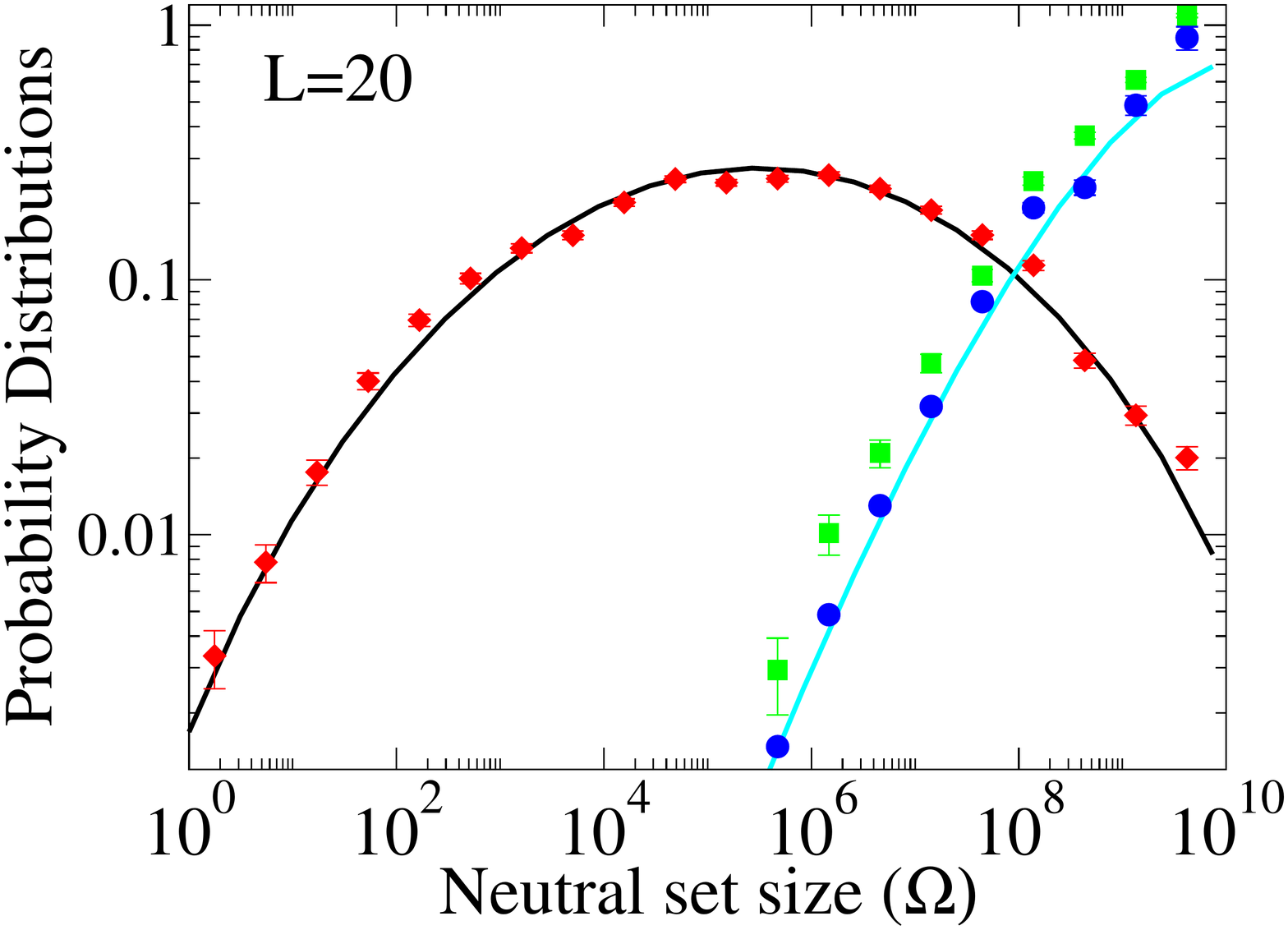}
\includegraphics[height=3cm,width=4.3cm]{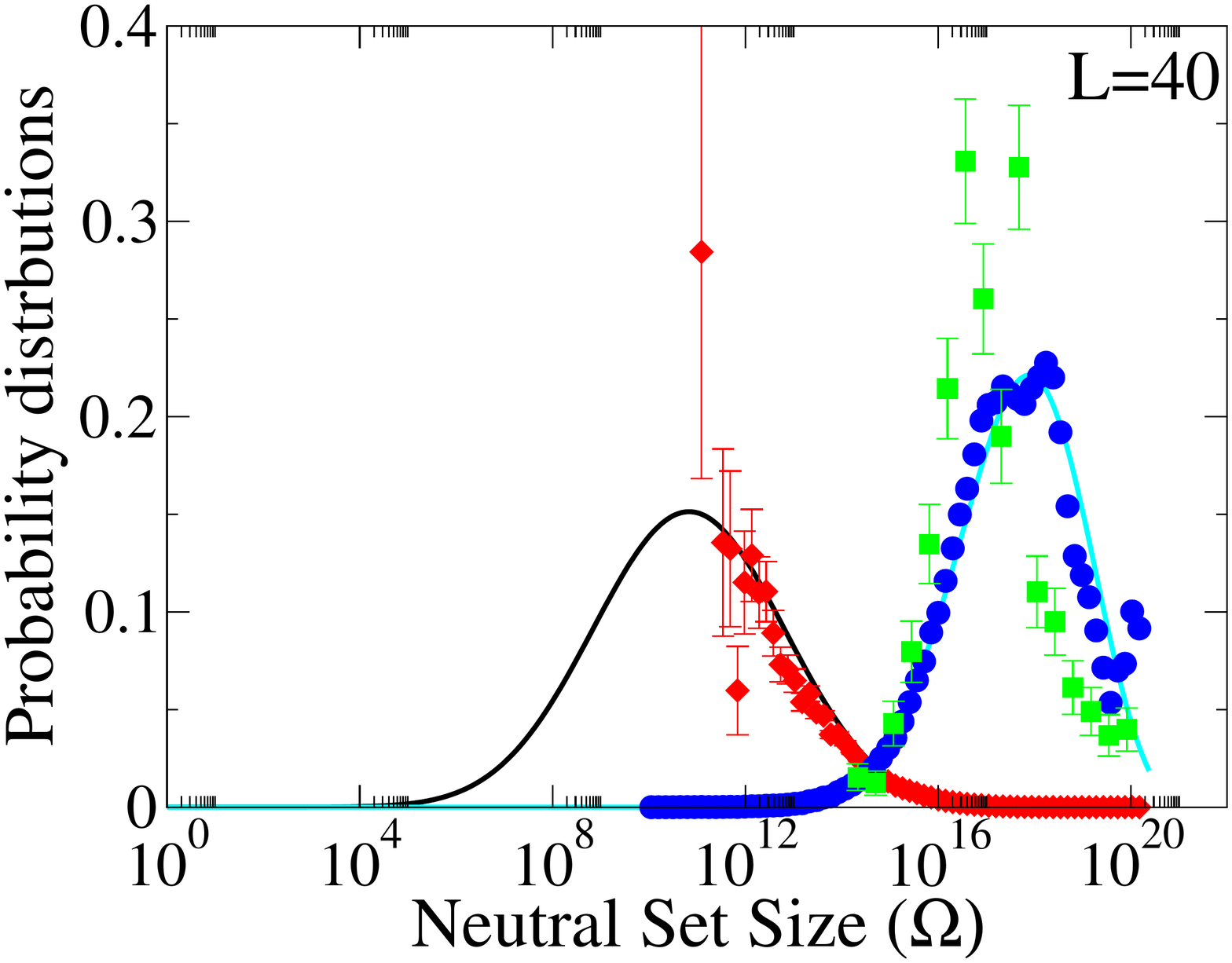}
\includegraphics[height=3cm,width=4.3cm]{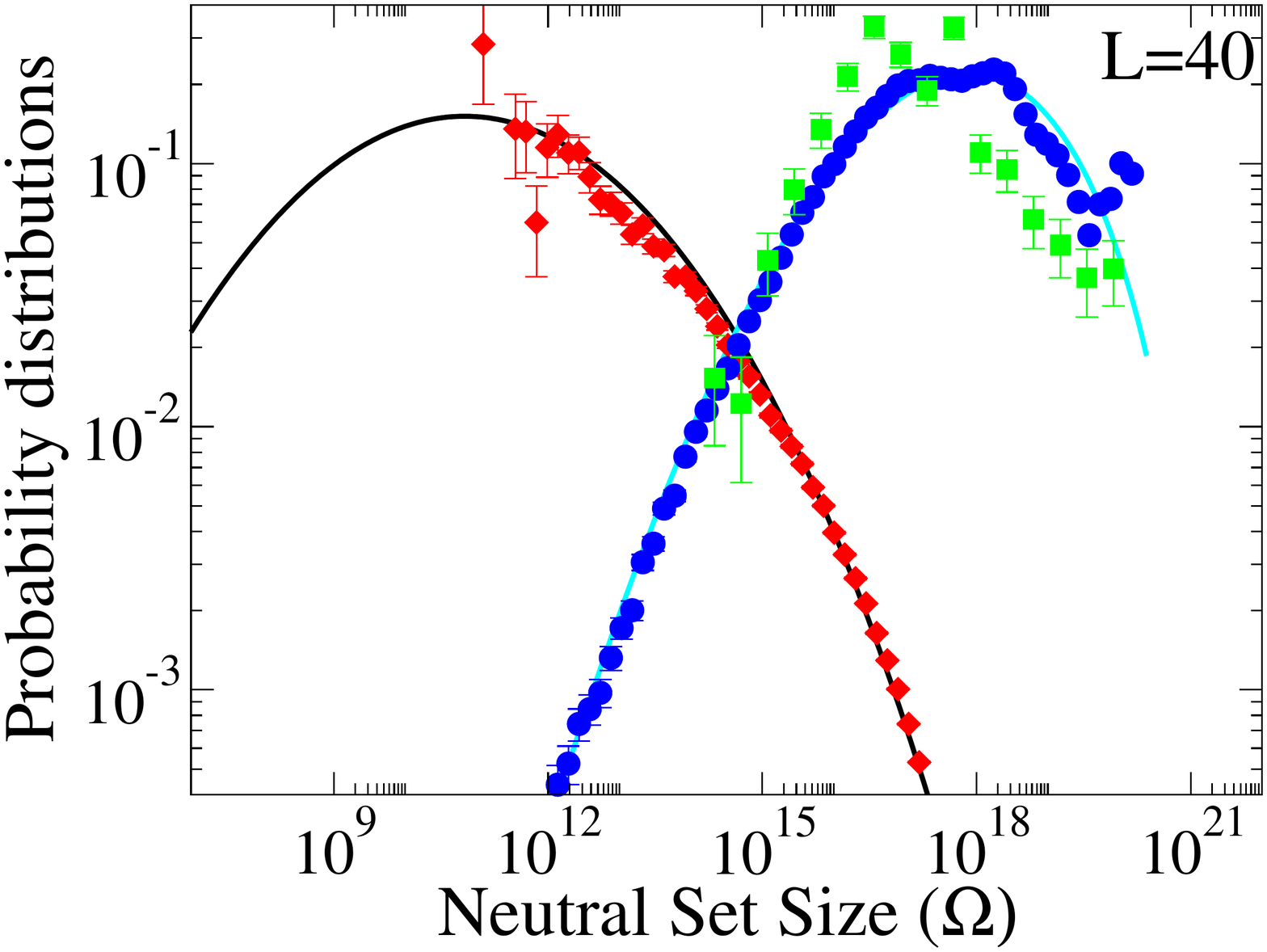}
\includegraphics[height=3cm,width=4.3cm]{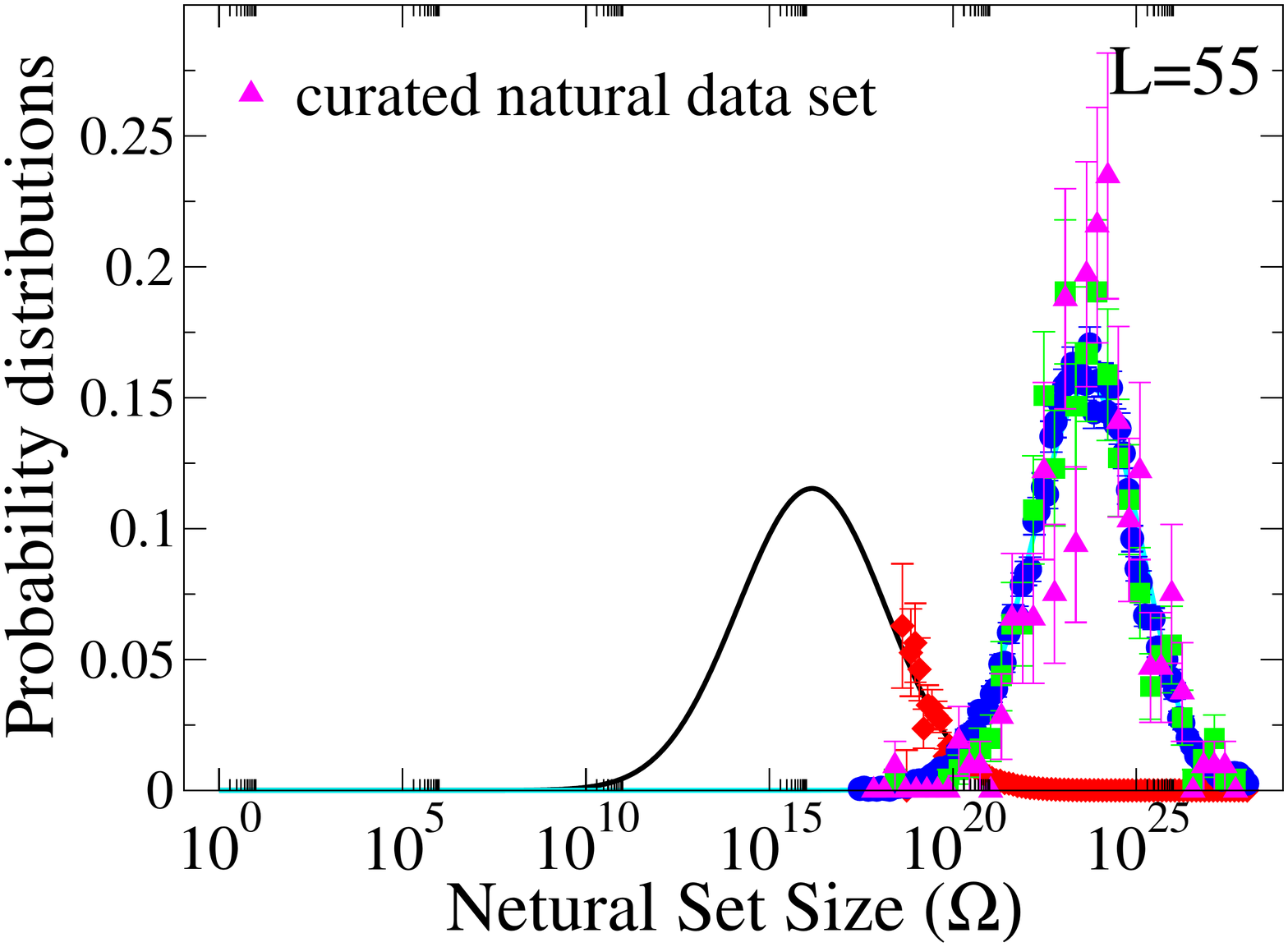}
\includegraphics[height=3cm,width=4.3cm]{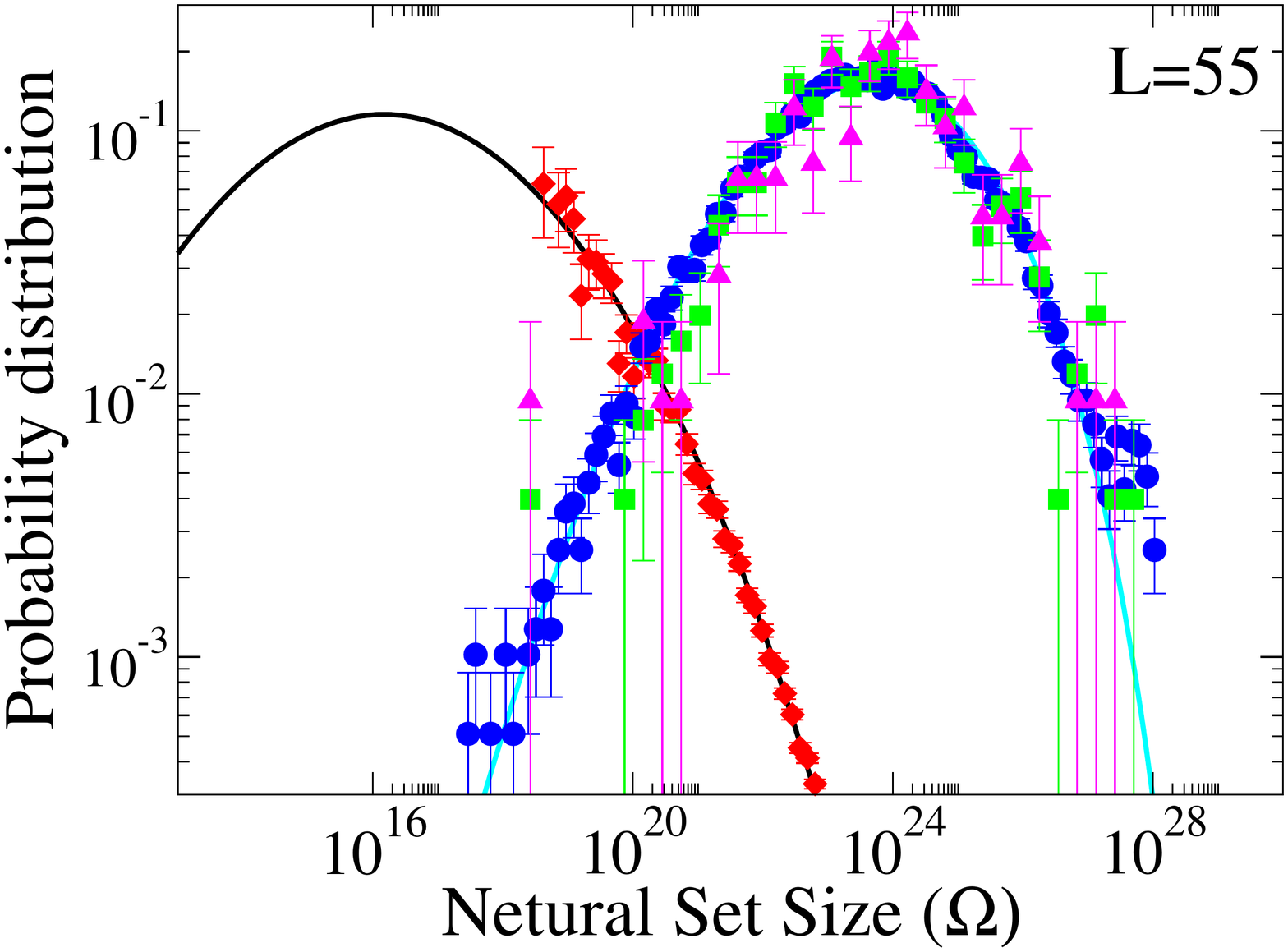}
\includegraphics[height=3cm,width=4.3cm]{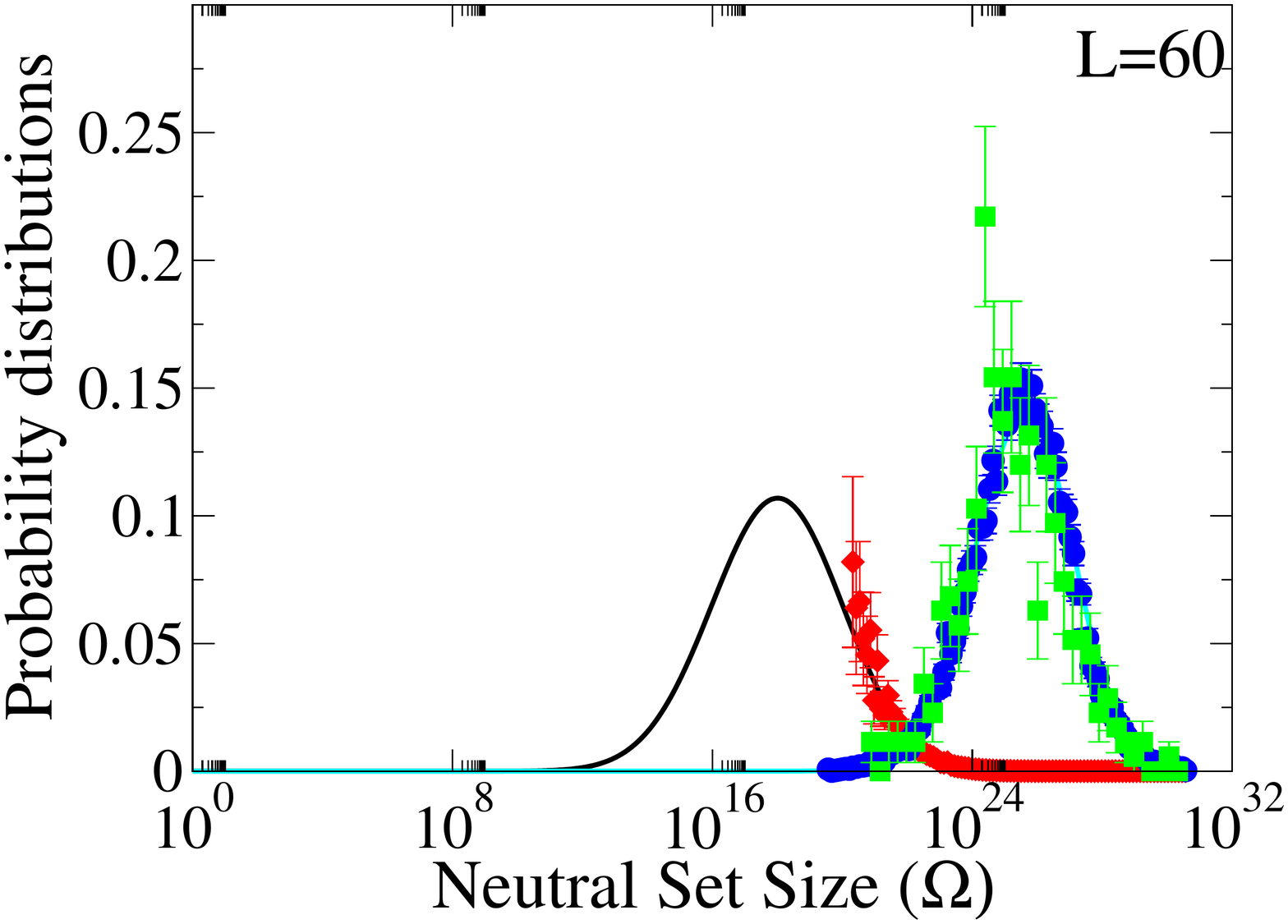}
\includegraphics[height=3cm,width=4.3cm]{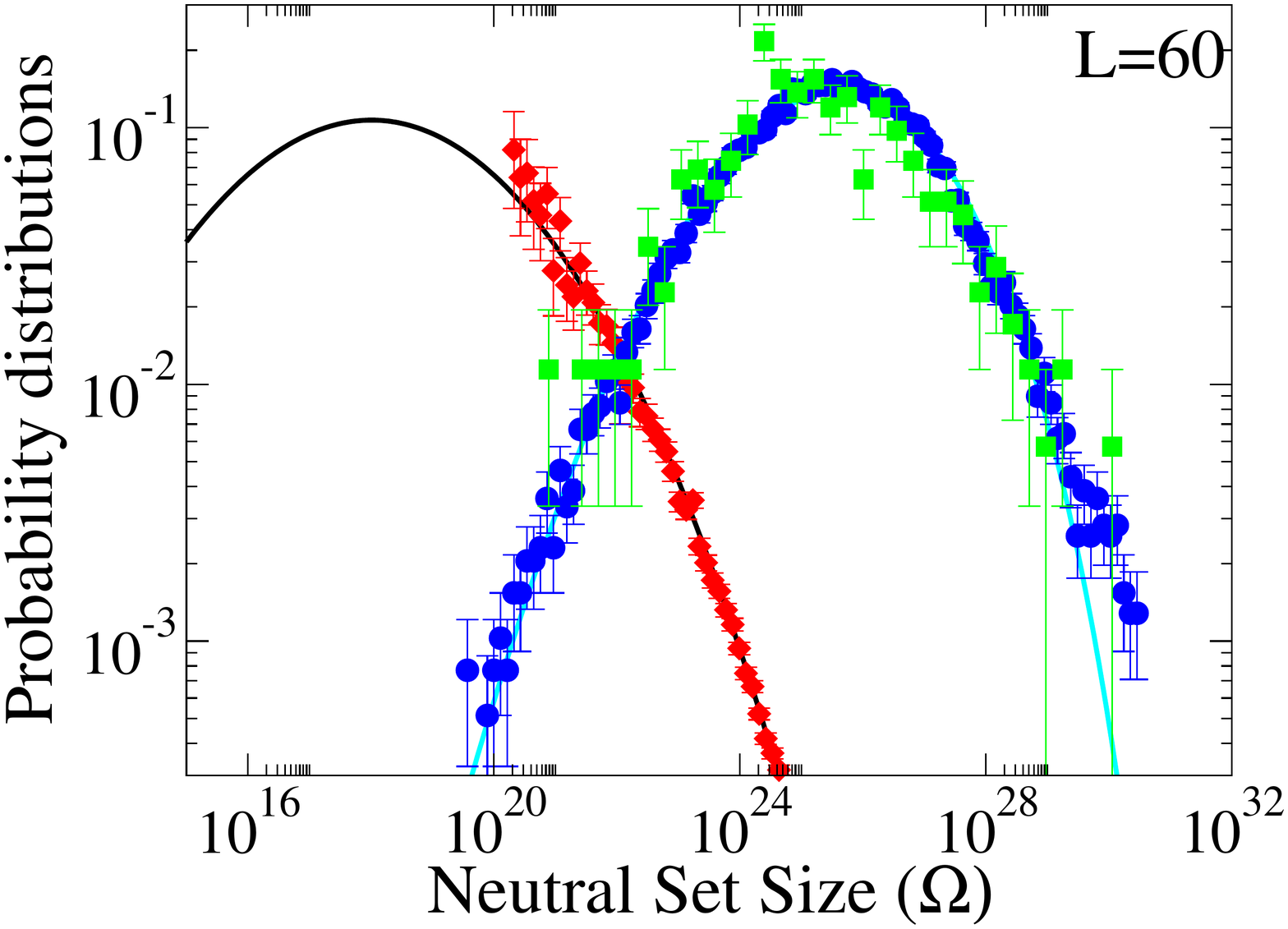}
\includegraphics[height=3cm,width=4.3cm]{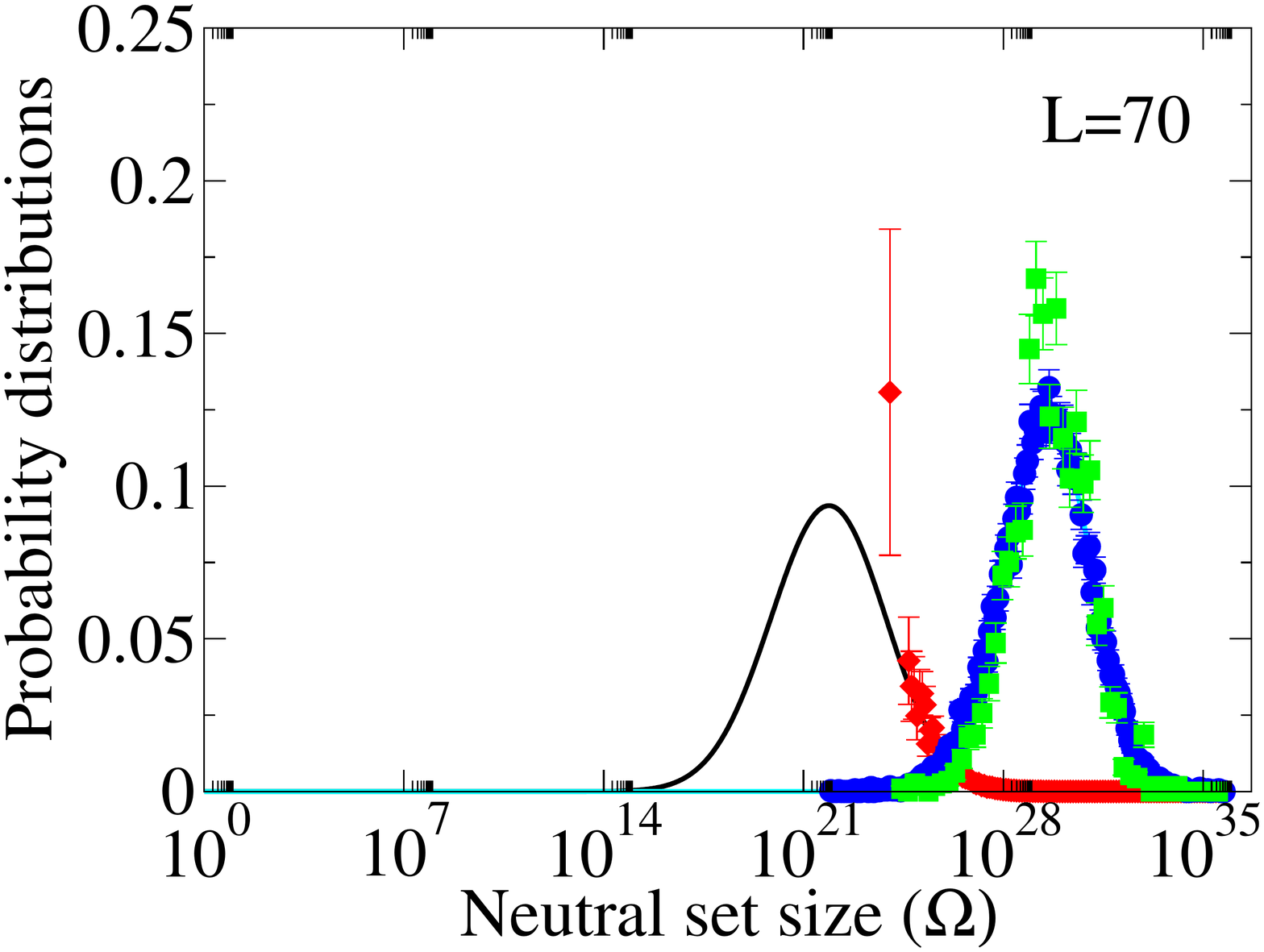}
\includegraphics[height=3cm,width=4.3cm]{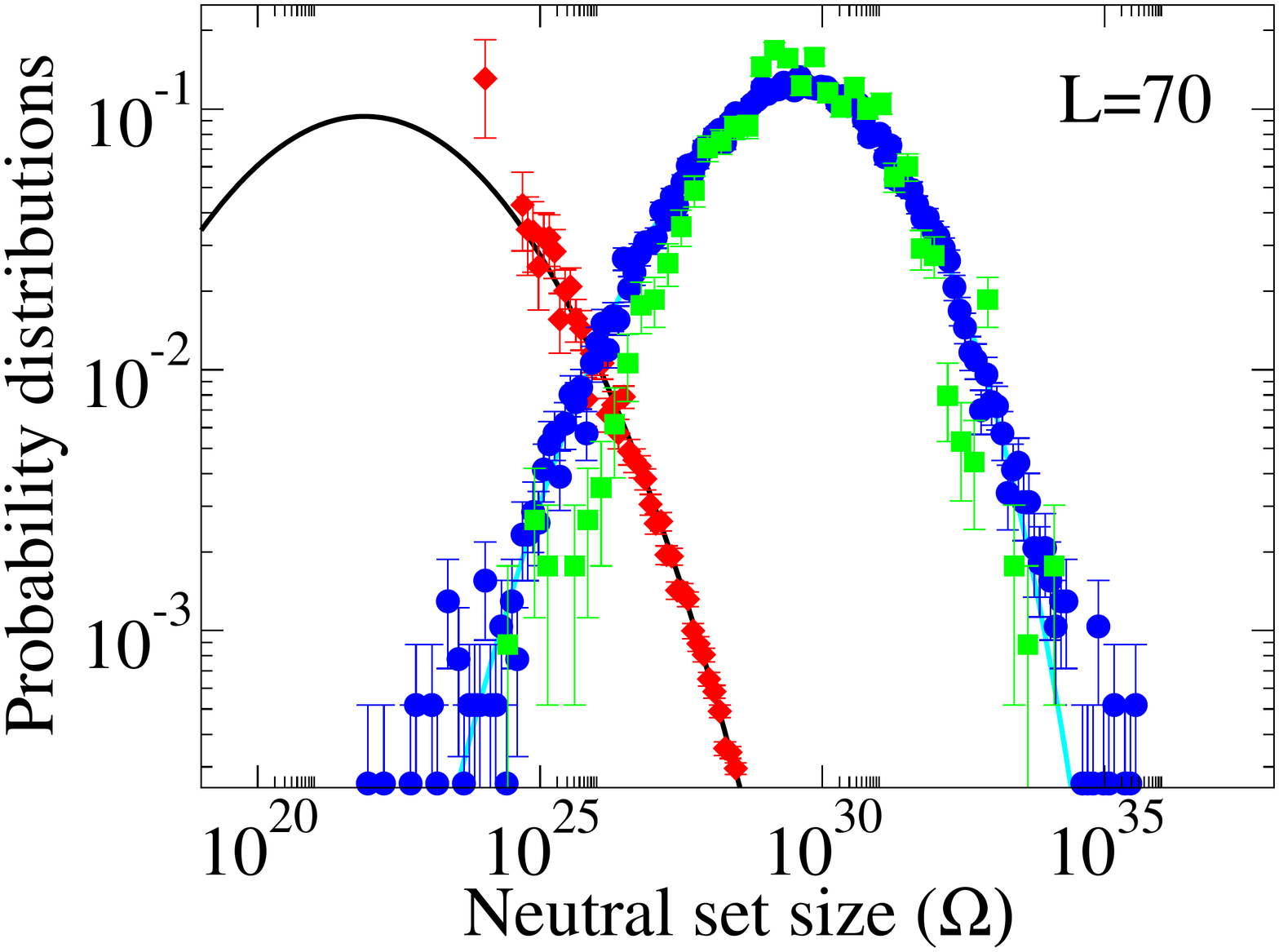}
\includegraphics[height=3cm,width=4.3cm]{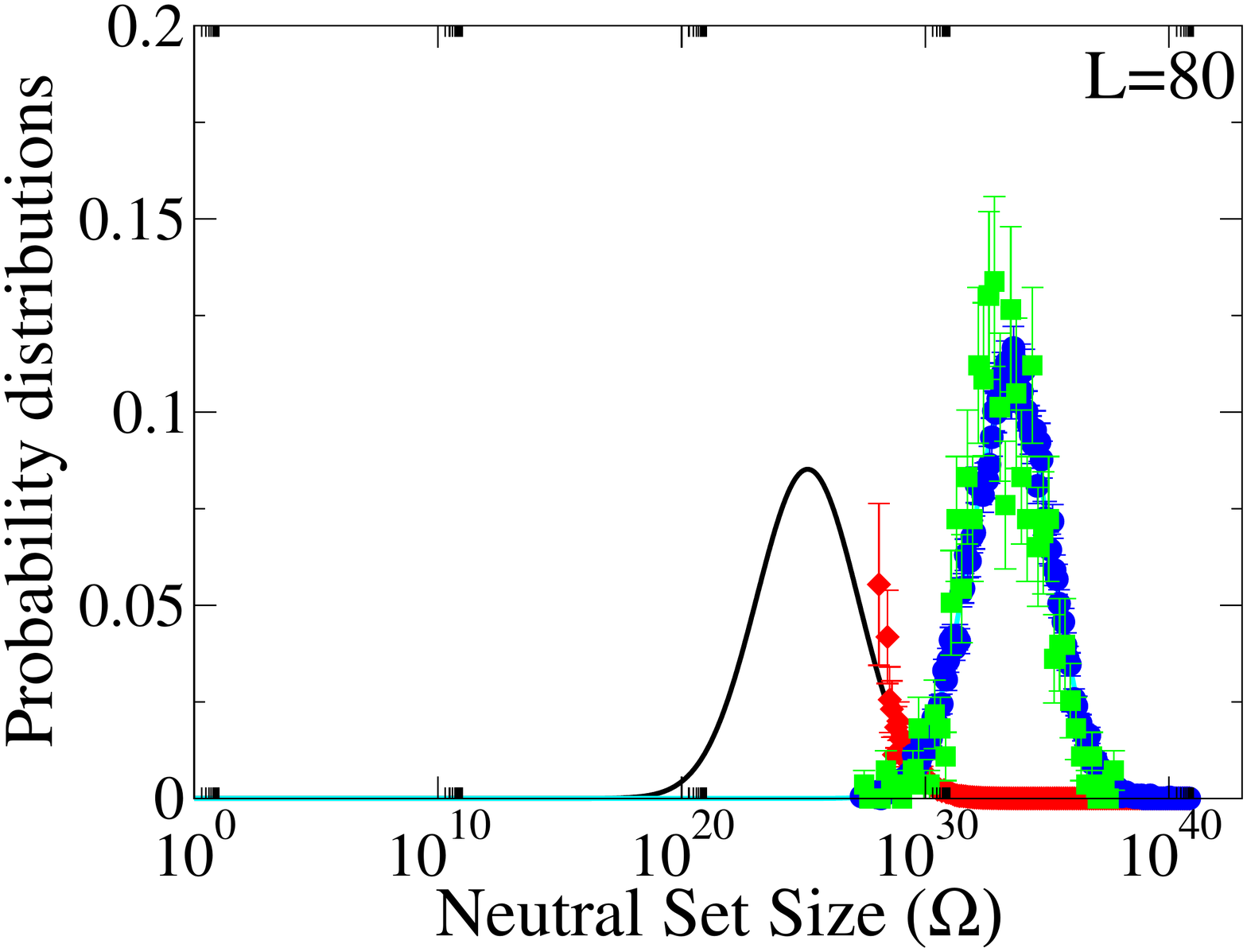}
\includegraphics[height=3cm,width=4.3cm]{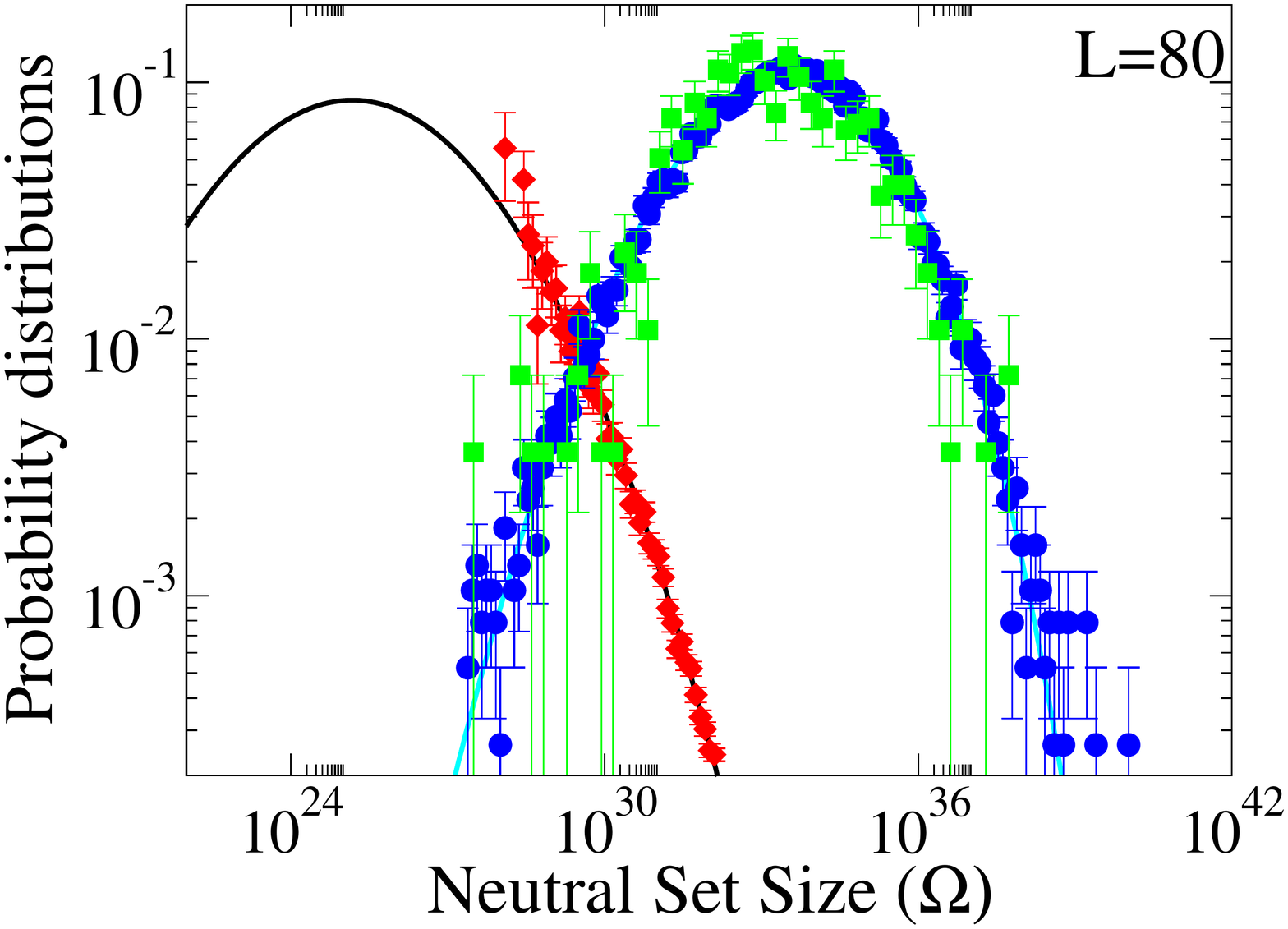}
\includegraphics[height=3cm,width=4.3cm]{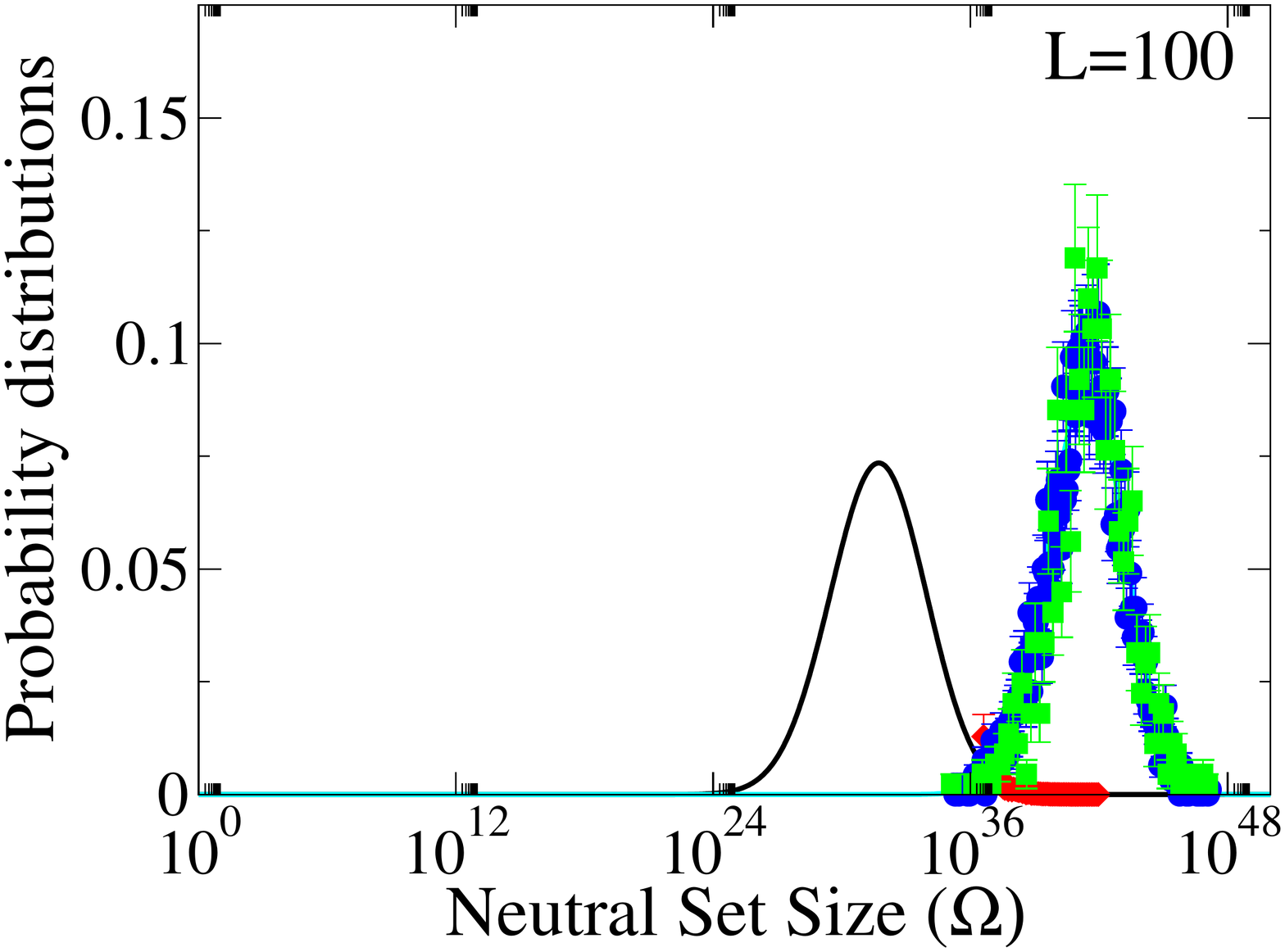}
\includegraphics[height=3cm,width=4.3cm]{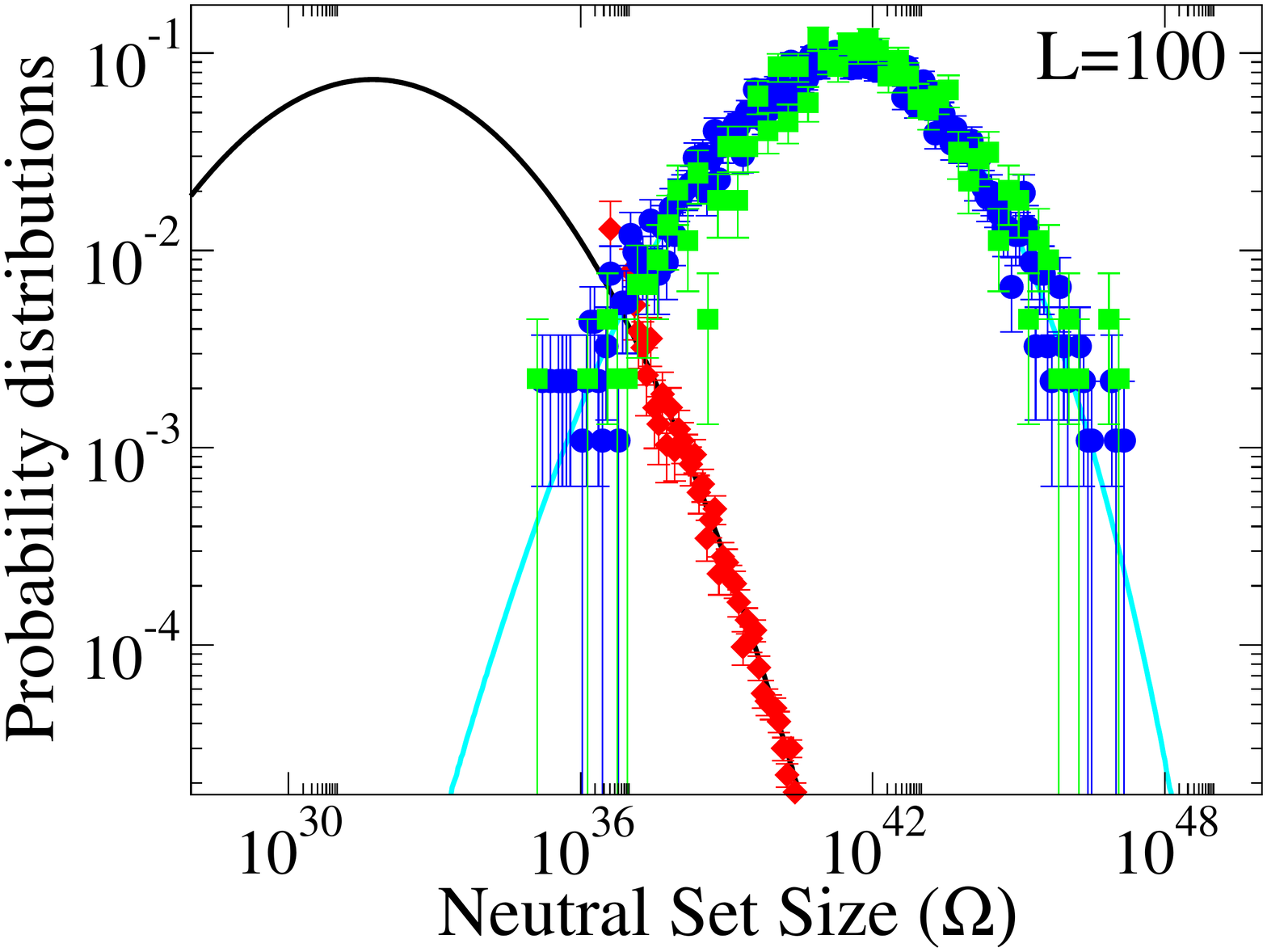}
\includegraphics[height=3cm,width=4.3cm]{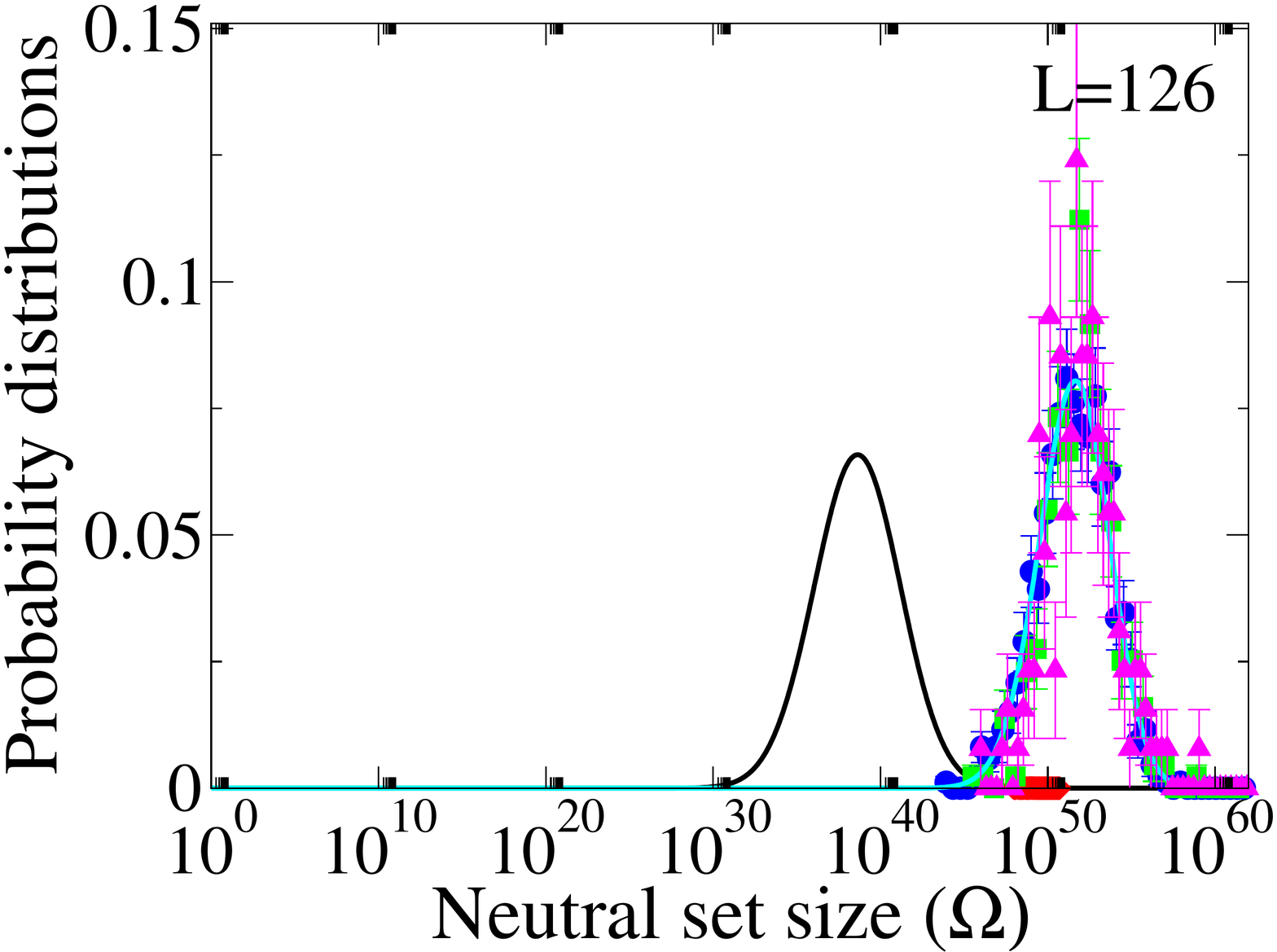}
\includegraphics[height=3cm,width=4.3cm]{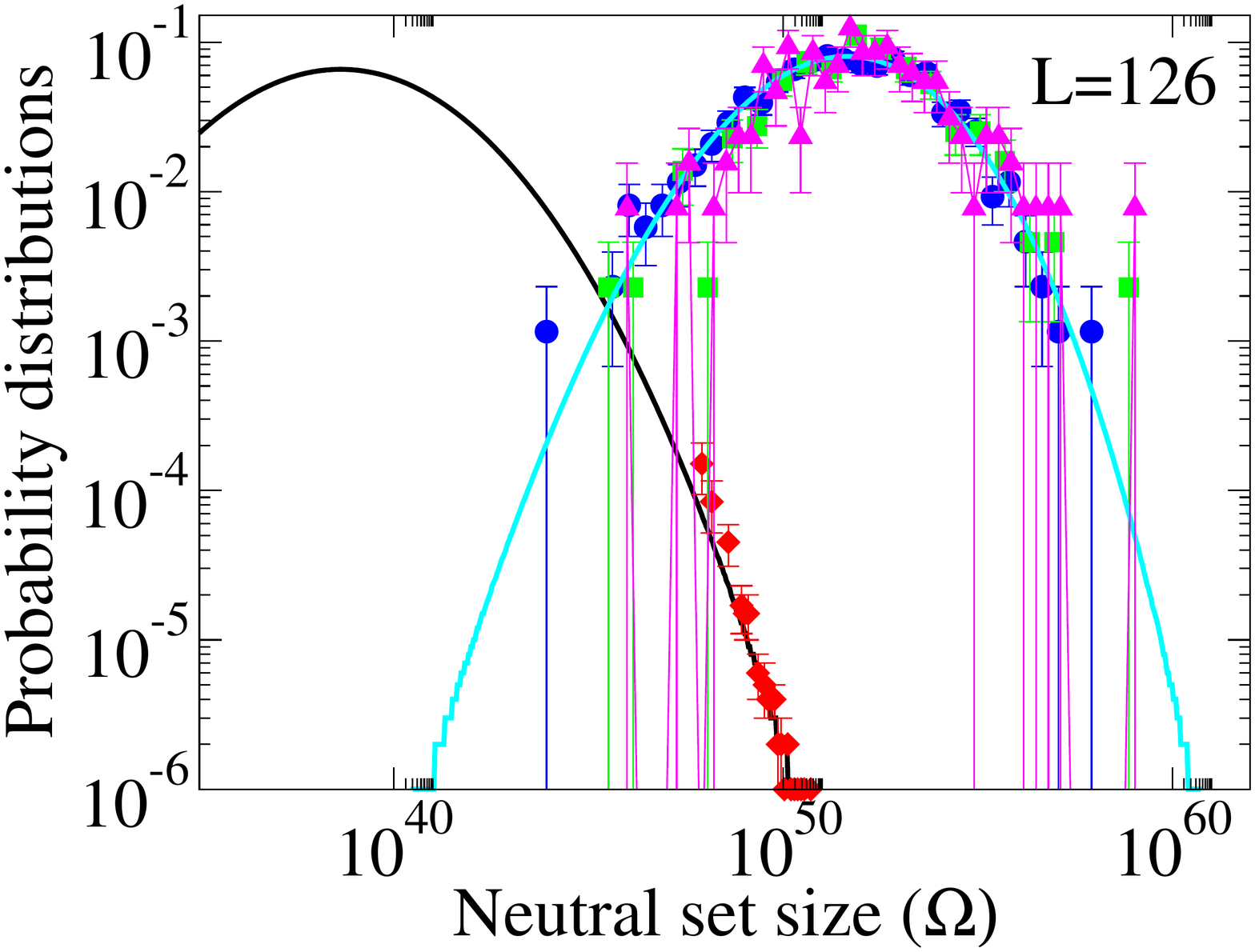}
\caption{{\bf Neutral set size distributions illustrate how bias constrains natural RNA secondary structures}.  
 Sampled distributions $P_P(\Omega)$ (red diamonds) and $P_G(\Omega)$ (blue circles) and analytic approximations to  $P_P(\Omega)$ (black lines) and $P_G(\Omega)$ (cyan lines)  are compared to  ncRNA from the  fRNAdb  database~\cite{kin2007frnadb} (green squares).   For each length (denoted in the top corners)  the data is shown both on a lin-log and log-log graph to highlight different parts of the distributions.  The natural data is remarkably clsoe to the   G-sampled $P_G(\Omega)$, but quite far from the P-sampled $P_P(\Omega)$. The number of natural structures plotted are:  $7327$ for $L=20$ (which is repeated from Fig 2. for clarity of comparison);\,\, $658$ for $L=40$;\,\, 
 $472$ for $L=50$;\,\,  $350$ for $ L=60$;\,\,  $2263$ for $L=70$;\,\,  $553$ for $L=80$;\,\,  $731$ for $L=90$;\,\,  $891$ for $L=100$;\,\, and $291$ for $L=126$. 
  In each case all structures in the  fRNAdb~\cite{kin2007frnadb} database are used, except for $L=20$ where only structures for {\it Drosophila melanogaster} are used, and for $L=55$ and $L=126$ where we also plot a smaller curated data set  of 213 and 172 structures respectively  (magenta triangles) where the SS is known to be important (see Supplementary Information).  Smaller numbers of data lead to larger binning error bars, but  the curated sets are clearly very similar to the full set of fRNAdb structures. \label{fig2}    }
\label{fig:X1}
\end{figure*}

\noindent {\bf  \large 2. Results} 

\noindent {\bf 2.1~P-sampling over phenotypes differs significantly from G-sampling over genotypes.} 

\noindent We first analyse an exhaustive enumeration of all sequences for $L=20$ RNA, the largest system for which this has so far been accomplished.   The $4^{20}\,$$\approx$$\, 10^{12}$ sequences were folded with the Vienna Package~\cite{hofacker1994fast}, and map to $N_P$$=$$11,218$ unique bonded SS and one trivial structure with no bonds, as their free-energy minima~\cite{schaper2014arrival}.  The set of all sequences that map onto a given SS is called the neutral set (NS), and we use  $\Omega$ to denote the NS size (the number of sequences in the NS). 
 This mapping exhibits strong phenotype bias~\cite{schuster1994sequences,fontana1993statistics,wagner2005robustness,cowperthwaite2008ascent,jorg2008neutral,aguirre2011topological,schaper2014arrival}. For example,  $\Omega$  varies by over ten orders of magnitude for $L=20$  (Fig.~\ref{figL20}).   
  
Next, we introduce the concept of P-sampling, that is, uniformly sampling over the set of possible phenotypes (the morphospace), and  define  $P_P(\Omega)$ as the probability distribution that a randomly chosen phenotype has NS size $\Omega$. We calculate distributions for fixed $L$ and bin data uniformly in $S=\log_{10}(\Omega)$, but write $P_P(\Omega)$  and $P_G(\Omega)$ for simplicity.    $P_P(\Omega)$  has a maximum when $\Omega$  is about half of the maximum value of the exponent $\log(\Omega) \equiv U \approx 10$ (Fig.~\ref{figL20}).   
 We  ignore the trivial structure with no bonds for which $\Omega \equiv 10^{T}$. For $L=20$, 
 $T\approx 11.56$; for  larger $L$,  $T/U \rightarrow 1$ and  the probability of finding the trivial structure tends to zero (see Methods). 
 
  Instead of P-sampling, one could also sample uniformly over sequences (genotypes), which we refer to as G-sampling, giving  $P_G(\Omega) \propto \Omega P_P(\Omega)$ which 
highlights structures from the large $\Omega$ tail of $P_P(\Omega)$, as can be seen in Fig.~\ref{figL20}.

  Novel variation does not arise by uniform random sampling in the morphospace of all physically permissible phenotypes (P-sampling),  but instead by processes such as mutation that change  genotypes.  While evolutionary dynamics do not proceed by simple uniform G-sampling either,      recent detailed population genetics calculations for RNA~\cite{schaper2014arrival}  have shown that the rate at which novel variation (a particular SS) arises in an evolving population is, to first order, directly proportional to the NS size of the SS phenotype, and so also to $P_G(\Omega)$.    This proportionality holds for a wide range of mutation rates and population sizes.  
  It was also demonstrated explicitly  that strong phenotype bias can overcome fitness differences in an RNA system~\cite{schaper2014arrival}.   Bias should therefore affect outcomes for multiple evolutionary scenarios. Indeed, important prior studies have suggested that natural RNA could have larger than average $\Omega$~\cite{jorg2008neutral,cowperthwaite2008ascent}.  We took  every $L=20$ sequence in the fRNAdb database~\cite{kin2007frnadb}  for \textit{Drosophila melanogaster} (see Methods and  Supplementary Information)   and calculated the NS size $\Omega$ of its associated SS using the neutral set size estimator (NSSE) from ref.~\cite{jorg2008neutral}. Not only are SS with larger $\Omega$ overrepresented, but 
      the entire natural distribution is remarkably close to the genotype sampled $P_G(\Omega)$ (Fig.~\ref{figL20}). 

\smallskip

\noindent {\bf 2.2~Sampling for lengths up to $\bf L=126$.} 

\noindent While these results for $L=20$ are suggestive,  distributions for larger  length RNA are needed to ensure we are not just observing database biases or artefacts of the short length.    Since the number of sequences grows exponentially with increasing $L$,  exhaustive enumeration is not an option for lengths much larger than $L=20$. 
Instead, we  estimate the NS size distributions by randomly sampling genotypes, folding them into a SS, and then measuring  their NS size with the NSSE (Methods).  This process naturally generates $P_G(\Omega)$; $P_P(\Omega)$ can be backed out by dividing by $\Omega$ and  normalising.   However, it is hard to sample SS with small $\Omega$ so $P_P(\Omega)$ is only partially determined.    To make progress we use a simple analytical ansatz based on a log-binomial approximation to the distributions (see Methods), which works  well both $P_G(\Omega)$ and $P_P(\Omega)$ for $L=20$ (Fig.~\ref{figL20}).
We compare this approximation to sampled data for $L=20$ up to  $L=126$ (Fig.~\ref{fig2} and Supplementary Figs 1,2).  In each case the analytic fit to $P_G(\Omega)$ is excellent, and the fit to $P_P(\Omega)$ works well, giving  confidence that this form provides a reasonable approximation to the full $P_P(\Omega)$.

 \begin{figure}[htp]
\includegraphics[height=5.6cm,width=8cm]{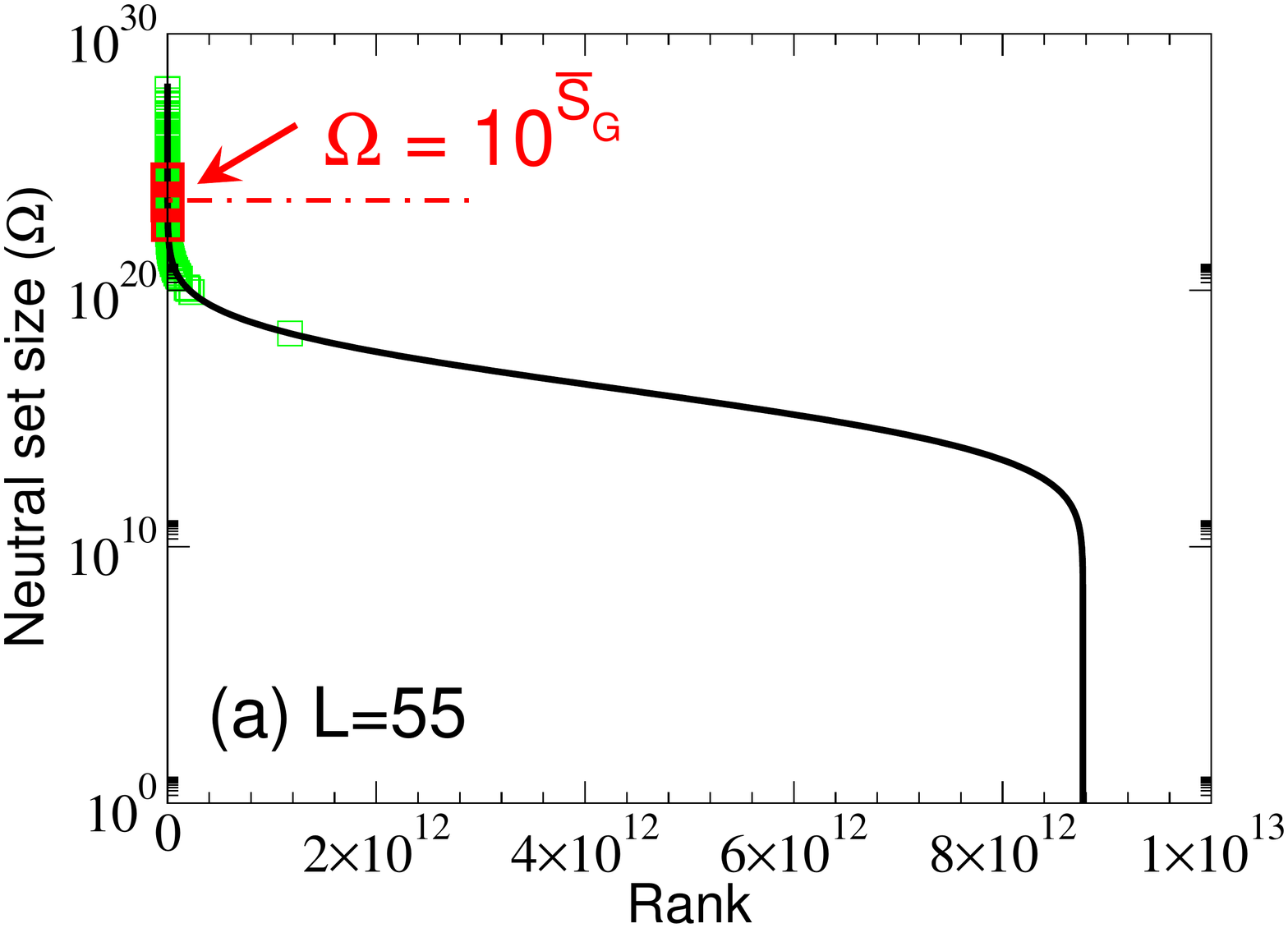}
\includegraphics[height=5.6cm,width=8cm]{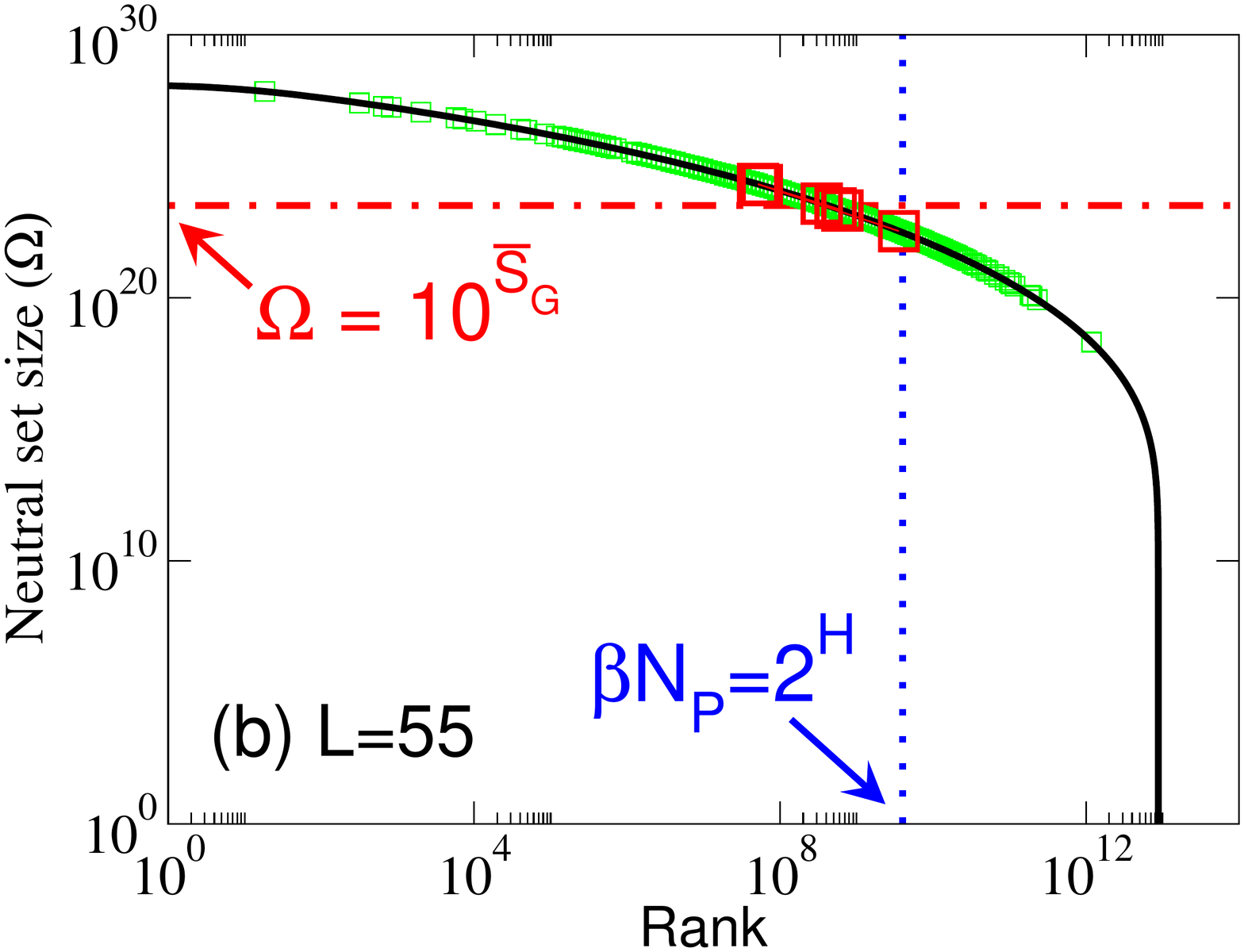}
\caption{{\bf Rank plots for $L=55$ RNA}.  
 $\Omega$ v.s.\ rank plot for all $N_P \approx 4 \times 10^{12}$ possible SS structures for $L=55$ shown as as (a) log-lin and (b) log-log plots. The black line is from the analytical approximation and the green squares denote the natural data.   The horizontal red dot-dashed line is $ \Omega =10^{\bar{S}_G} \approx 3 \times 10^{23}$, near the peak of the $P_G(\Omega)$ distribution  for L=55 in Fig 3.  The 13  $L=55$ hammerhead ribozyme structures from the fRNA database (red squares)  are clustered near this peak.  The vertical blue dotted line in (b)  denotes  $\beta N_P = 2^H\approx 4 \times 10^9$, the  ``effective'' number of SS. This set of $\beta \approx 0.1\%$  of all structures captures the majority ($\approx 75\%$) of natural structures. 
\label{fig:rank}}
\end{figure}

 \begin{figure}
\includegraphics[height=5.6cm,width=8cm]{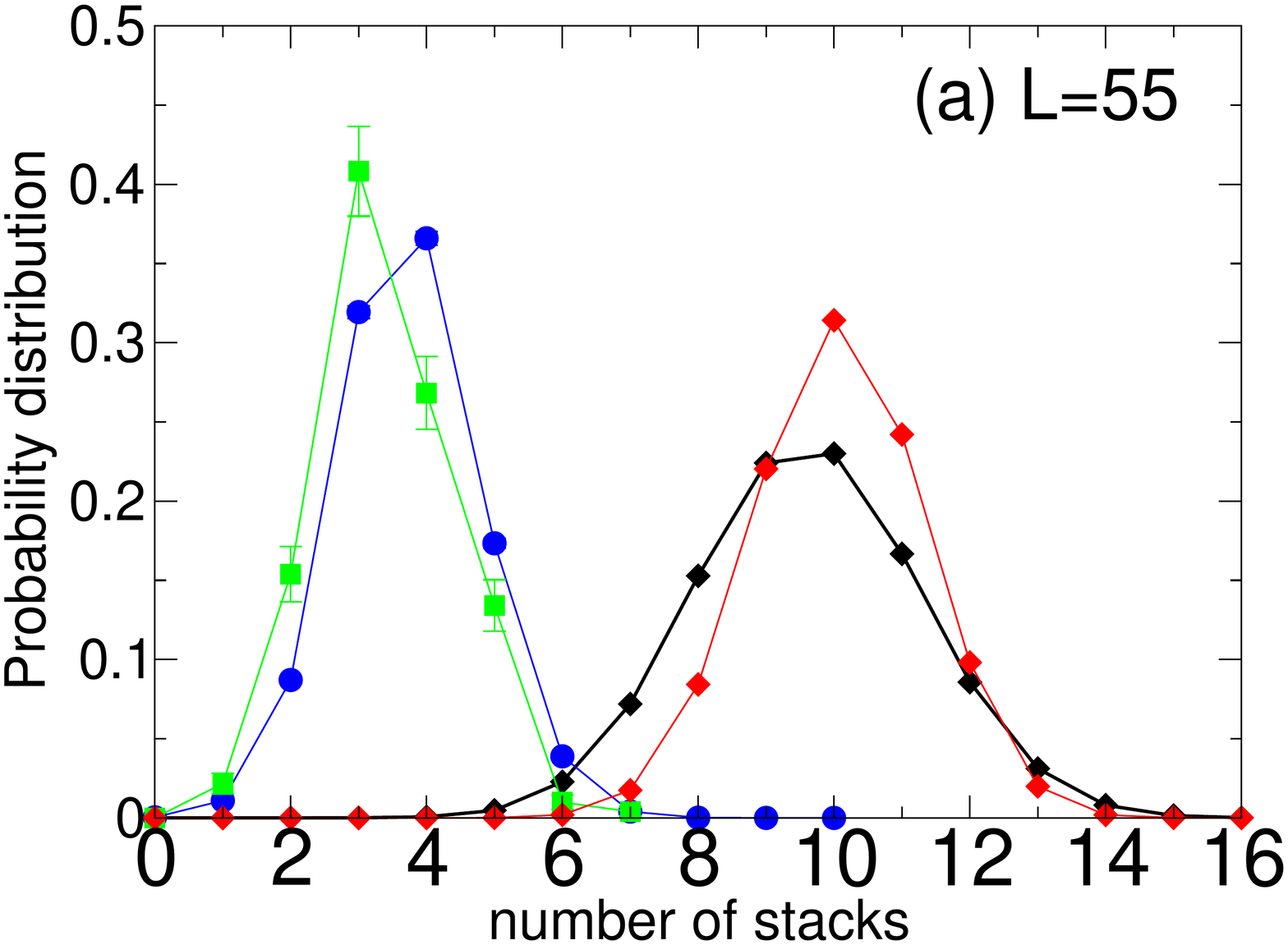}
\includegraphics[height=5.6cm,width=8cm]{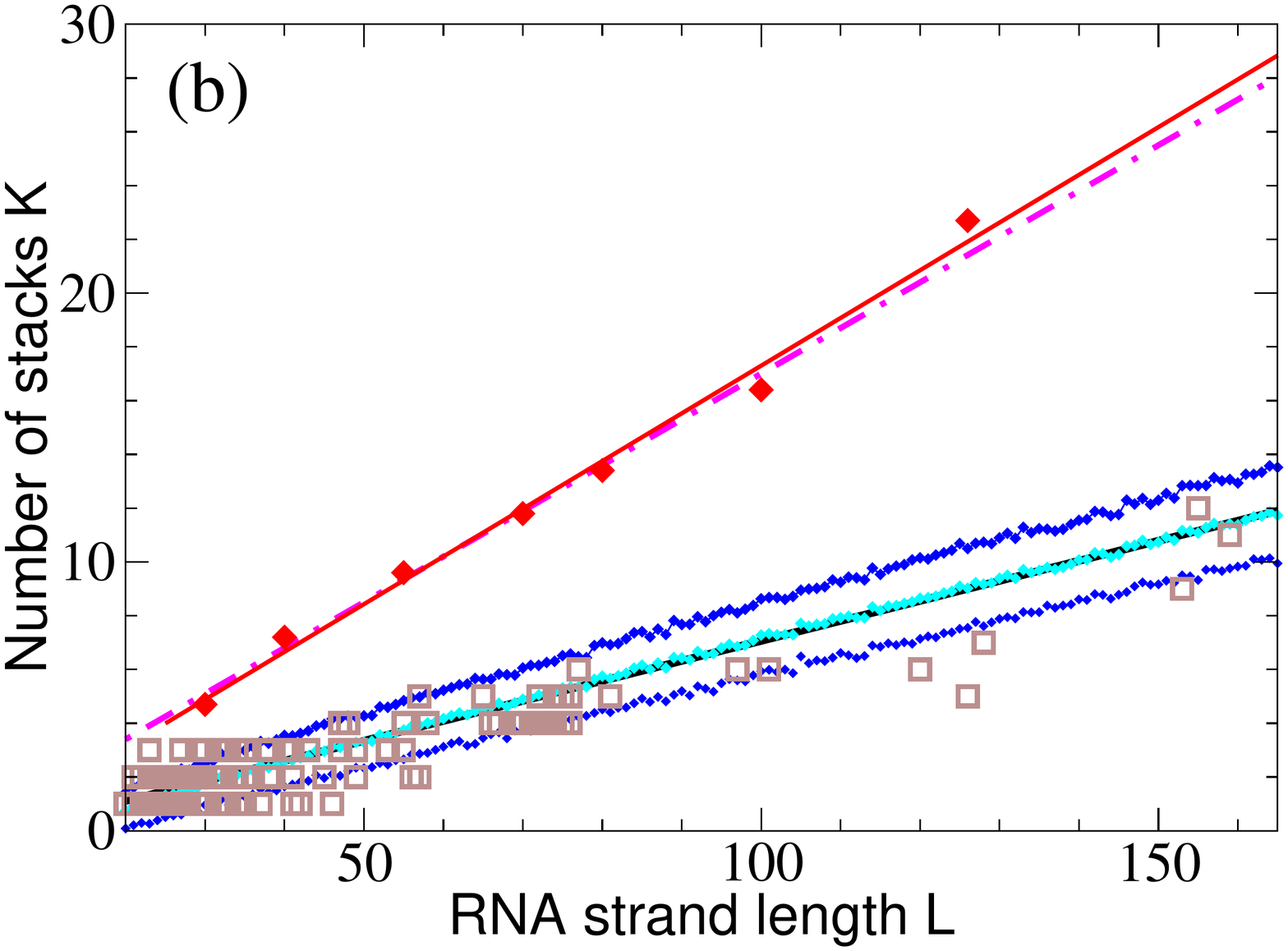}
\caption{{ \bf Natural stack distributions for ncRNA correlate with G-sampling but not with P-sampling.} {\bf (a)} Distribution of the number of stacks  for $L=55$ via G-sampling (blue circles) and P-sampling (black diamonds).  Natural data (green squares) are remarkably close to the G-sampled distribution, and  are drawn from a tiny fraction of the full morphospace of structures (the P-sampled distribution) with small numbers of stacks.  The red diamonds are from a combinatorial estimate of stack number~\cite{hofacker1998combinatorics} that helps corroborate our  P-sampling result. {\bf (b)}   Comparison to experimentally measured structures: The brown squares denote the number of stacks experimentally determined for  $214$ non-pseudoknotted structures with $L> 20$ which were taken from RNA STRAND database~\cite{andronescu2008rna}.   The cyan diamonds show $\bar{K}_G$, the mean number of stacks calculated by G-sampling with the Vienna package~\cite{hofacker1994fast}, which can be accurately fit with    $\bar{K}_G = 0.074 L - 0.377$.   The blue diamonds show the G-sampled number of  stacks one standard deviation above or below from the mean.   The natural data (brown squares) from the RNA STRAND database are consistent with the G-sampled theoretical data. By contrast these natural data are far away from the expected number of stacks from P-sampling, shown here  for an estimate that uses the linear relationship between the P-sampled distribution of $ \log(\Omega)$ and $K$ (Supplementary Fig. 6)  to infer $\bar{K}_P$ (red diamonds), which are well described by a linear fit $\bar{K}_P =   0.177 L -0.443 $  (solid red line).  Independent estimates of $\bar{K}_P$ come from reference~\cite{hofacker1998combinatorics},  and include an asymptotic measure $\bar{K}_P \approx 0.1717 L$ (dash-dotted line) (see Supplementary Information). 
The close agreement between the two independent methods for estimating the mean number of stacks from P-sampling gives us further confidence in our fits to the full P-sampled distribution. 
\label{fig:3}}
\end{figure}

 \begin{figure}
\includegraphics[height=5.6cm,width=8cm]{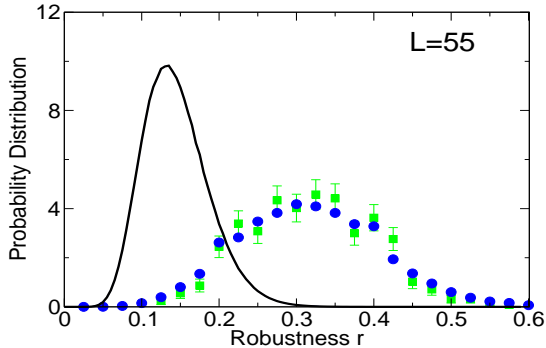}
\caption{{ \bf Natural robustness distributions for ncRNA correlate with G-sampling but not with P-sampling.} 
  The distribution of robustness, defined as the fraction of mutations $r$  that retain the same SS phenotype is given for $L=55$ via P-sampling (black line) and  G-sampling  (blue circles).  Natural data (green squares) are remarkably close to the G-sampled distribution, and are considerably more robust than the average of structures in the full morphospace. We also note that most phenotypes have a robustness that is above the threshold ($ r> \delta \approx 0.0061$) needed for the formation of connected neutral networks~\cite{greenbury2015genetic}.
\label{fig:robustness}}
\end{figure}

 \smallskip
 
\noindent{\bf 2.3~Entropy and the bias ratio.}  

\noindent The  $P_G(\Omega)$  distributions can be further quantified by the Shannon entropy   $H=-\sum^{N_P}_{k=1} P(p_k) \log_2 \left(P(p_k)\right)$,
where  $P(p_k)$ is the probability of choosing one of the $N_P$ possible phenotypes $p_k$ by G-sampling.  
 The exponential of the entropy, $2^H$, is 
used in statistical physics and information theory~\cite{mezard2009information} as a  measure of the effective number of states. Since  $P(p_k)=\Omega(p_k)/4^L$, it follows straightforwardly that 
\begin{equation}
2^H = 4^L/10^{\bar{S}_G}, \label{eq:2H}
\end{equation}  where $\bar{S}_G$ is a G-sampled average of  $S = \log(\Omega)$.  Thus to measure the `effective number' of phenotypes that take up the majority of genotypes, one only needs to determine $\bar{S}_G$.  This can be done rapidly because the G-sampled distributions are  sharply peaked in a manner reminiscent of statistical mechanics~\cite{mezard2009information}.  

As an example of how the highly peaked nature of these distributions facilitates the calculation of averages, consider the case of  $L = 126$.  We find that $\bar{S}_G \approx  51.5$ with standard deviation $\bar{\sigma}_G \approx 2.1$ so that $95\%$ of SS structures found by G-sampling  have $S$ in range $51.5 \pm 4.2$ which is very narrow compared to the full range $[0,64.5]$, and comparatively far (about  $6$ standard deviations) from the phenotype averaged  $\bar{S}_P \approx 38.6$.  Rather strikingly, this implies that even though the set of all possible  $L=126$ nucleotide sequences would weigh more than the mass of the observable universe (see Methods), sampling the NS size for  just $10$  randomly chosen $L=126$ structures is enough to fix the exponent  $\bar{S}_G$ to within about $1.3\%$ relative accuracy, which allows us to determine, for example the number of relevant states $2^H$ via Eq~\ref{eq:2H}.

To further quantify the bias, we introduce the 
\emph{bias ratio} $\beta\in(0,1]$ as  
\begin{equation}
\beta = 2^H/N_P \approx 0.25 \times 0.91^L
\end{equation}
which can be interpreted as the ratio of the effective number of phenotypes that  take up most of the sequences to the total number of phenotypes $N_P$.  For example, if all $\Omega$ are equal then $2^H$=$2^{\log_2(N_P)}$=$N_P$,  $\beta = 1$, and bias is weak.  On the other hand, if just one phenotype accounts for  nearly all the genotypes then $2^H$$\approx$$2^0$=1, $\beta \rightarrow 0$ and bias is very strong. 

 The number of phenotypes $N_P$ can be estimated from $P_P(\Omega)$; $N_P \approx 0.13 \times 1.76^L$ fits the data well (see Methods).  From this we find
  $\beta\approx$0.25$\times0.91^L$ which shrinks with increasing $L$.    Typically we find that about $75\%$ of genotypes map to the `effective number' $\beta N_P = 2^H$  of structures.
 Nevertheless, $\beta N_P = 2^H \approx0.033 \times 1.60^L$, and continues to grow exponentially.  For $L=55$, $\beta \approx 0.001$  and $\beta N_P\approx4 \times 10^9$ of the $N_P\approx4 \times 10^{12}$ phenotypes take up about $75\%$ of the genotypes. For $L=126$, $\beta\approx2 \times 10^{-6}$ and $\beta N_P\approx1 \times 10^{24}$.  Even though bias greatly reduces the effective number of  phenotypes accessible to mutations, their number continues to grow exponentially, and so can still be extremely large for this system.

\medskip

 \noindent {\bf 2.4~Comparison to fRNAdb database of functional non-coding RNA.} 
 
 \noindent To explore how this bias plays out in nature, we took, for lengths ranging from $L=20$ to $L=126$,  every sequence in the fRNAdb database~\cite{kin2007frnadb} and calculated the NS size $\Omega$ of its associated SS.  The correspondence between $P_G(\Omega)$  from random sampling and the distribution of $\Omega$ from the database is striking (Fig.~\ref{fig2} and  Supplementary Figs.~1,2).   Not only are  these natural RNAs mainly drawn from the minuscule fraction $\beta$ of SS with large $\Omega$, but their distribution is remarkably close to what one would obtain from  randomly sampling genotypes.   Rank plots for $L=55$ (Fig.\ref{fig:rank}) further illustrate how natural ncRNA are constrained mainly to the set of $\beta N_P = 2^H$ `relevant structures'.  
 
These results are, at first glance, surprising because many ncRNAs have well-defined functional roles and so  will have been subject to natural selection for which there exists extensive evidence~\cite{mattick2006non}.   Some insight may be gleaned from an exchange that occurred not long after the start of the molecular revolution in biology.  Frank Salisbury~\cite{salisbury1969natural} expressed doubt that  evolutionary search could work, since genetic spaces are typically exponentially large, and, he claimed, are probably sparsely populated with functional phenotypes.
    In an famous response~\cite{smith1970natural}, John Maynard Smith pointed out that evolution is aided by neutral mutations because these imply many fewer phenotypes than genotypes and also connect genotypes into neutral networks that can be explored by evolving populations. Phenotype bias may further facilitate evolutionary search by limiting potential outcomes.  But taken together, this still does not counter the heart of Salisbury's objection, namely that functional phenotypes are extremely rare and so hard to find. 
    
Here we instead suggest that the reason the `null model' $P_G(\Omega)$ 
 predicts  what is found in nature so accurately is that `good enough' SS  are relatively easy to find.    Of course the full functional ncRNA phenotype typically has a smaller NS size than the SS does  because it requires further constraints on the sequence to achieve functionality~\cite{carothers2004informational}.  We postulate that while natural selection creates function by acting on such sequence constraints, it also automatically draws from a palette of easily accessible SS variation that is strongly pre-sculpted by the mapping (see also ref.~\cite{schultes2005compact} for a similar suggestion).
    Another possibility might be that the correspondence with $P_G(\Omega)$ simply means that the SS has no adaptive value.  However, this scenario is unlikely, since it is well established that the SS is an important determinant of the 3D structure~\cite{wagner2005robustness,carothers2004informational,bailor2011topological} which, in turn, helps determine function.  Moreover, the correlation with $P_G(\Omega)$ remains strong when we curate the database for structures where SS is known to be important. 
 Altogether this suggests that RNA SS that facilitate biological function are -- {\it contra} Salisbury -- not rare, at least  not within the set of 
  `relevant structures' most easily accessible to random mutations.

Salisbury's arguments are also undermined by {\em in-vitro} evolution experiments that selected random RNA strands for  self-cleaving catalytic activity and  found that the hammerhead ribozyme repeated emerged, suggesting convergent evolution~\cite{salehi2001vitro}.  The hammerhead ribozyme appears so frequently in all three kingdoms of life that it has recently been termed ubiquitous~\cite{hammann2012ubiquitous}.  We examined the 13 hammerhead ribozymes from the natural data set~\cite{kin2007frnadb} for $L=55$, finding that $S=\log(\Omega) \approx 23.87 \pm 0.64$~\ref{fig:rank}, very close to  the peak of the G-sampled distribution at $\bar{S}_G \approx 23.4$ (see L=55 panel in Fig.~\ref{fig2} and Fig.~4).  We postulate that evolutionary convergence is observed in these experiments and in nature not so much because the hammerhead ribozyme is fitter than  other possible self-cleaving enzymes, but rather because it is particularly easy to find.  
There may even be many other self-cleaving ribozymes among the  $99.9 \%$ of  $L=55$ structures that evolution is unlikely to search through~\cite{salehi2001vitro}. 
It would in fact be interesting to devise artificial methods to search for such undiscovered ribozymes~\cite{Gan03exploringthe,athavale2014experimental}.

\smallskip

 \noindent {\bf 2.5~Distribution of stacks.} 
 
 \noindent Are the $2^H$ relevant structures different from the whole set of $N_P$ possible structures?  Our arguments above suggest that this should be the case for properties that correlate with neutral set size.  Previous studies (typically on much smaller data sets) have found that structural features (e.g.\ distributions of stack and loop sizes) of the natural and random SS are  quite similar~\cite{fontana1993statistics}, and that  natural and random rRNA share strong similarities in the sequence nucleotide composition of SS motifs such as stems, loops, and bulges~\cite{smit2006natural}, although natural RNA are more stable than random RNA~\cite{schultes2005compact}. Indeed we find that the natural RNA have slightly more bonds than in G-sampled structures (see Supplementary Information Fig. S3 (c),(d)).  
 
 We find that $\Omega$ correlates negatively with the number of stacks (i.e. sets of contiguous base-pairs) $K$
 (see Supplementary Information Fig. S6).  The natural distribution of stacks closely follows the G-sampled distribution, but differs markedly from the P-sampled distribution (Fig 4a).  For example,  the hammerhead ribozyme has $3$ stems, close to the most likely number by random G-sampling for $L=55$, but much less than  the P-sampled average of $\approx 10$. Bias means that it will be difficult for evolution to find $L=55$ structures with a large number of stacks, again raising the question  of what kind of functionality is possible in principle that cannot be reached by  evolution because of such phenotype bias constraints.

As an independent check on our stack predictions, we also obtained  214 natural experimentally determined SS from the STRAND RNA database~\cite{andronescu2008rna} and plot the experimentally determined number of stacks versus length $L$  in Fig.~\ref{fig:3}b.
The majority of experimentally determined  structures have numbers of stacks that are within one standard deviation of the mean calculated from G-sampling, as one would expect if our theoretical predictions were accurate.   By contrast, the number of experimentally determined stacks differs significantly from P-sampling estimates.

\smallskip

\noindent {\bf 2.6~Distribution of mutational robustness.} 

\noindent
 Interestingly, the bias towards larger $\Omega$ also leads to structures with larger mutational robustness~\cite{schuster1994sequences,wagner2005robustness}, and again natural data closely follow G-sampled distributions (Fig.~\ref{fig:robustness}).  Larger robustness is considered to be advantageous~\cite{wagner2005robustness} so that in this important way phenotype bias facilitates evolution.    It is also interesting to note just how large the mutational robustness is for this data.
 For $L=55$ there are on the order of $N_P \approx 8 \times 10^{12}$ phenotypes, so that the mean probability of finding a phenotype by randomly picking a genotype is on the order of $1 \times 10^{-13}$.   Instead, for both the G-sampled and the P-sampled structures, the probability of a nearest neighbour generating the same phenotypes is on the order of $10^{12}$ times higher than random chance. As recently emphasised in ref~\onlinecite{greenbury2015genetic}, this large difference arises from genetic correlations, and typically lifts the robustness well over the minimal threshold $\delta = 1/3L$~\cite{smith1970natural} ($\delta \approx 0.0061$ for $L=55$) needed to generate large connected neutral networks.
 
\medskip
\noindent{\bf \large 3. Discussion} 

By solving properties of the GP map from sequences to RNA secondary structures for strands up to $L=126$ nucleotides in length, we show explicitly that the vast majority of sequences map to an exponentially small fraction of all possible phenotypes (a summary of scaling forms for key properties of RNA can be found in Table~\ref{table:X1}).   One consequence of this  strong bias is that only an exponentially small proportion of the morphospace of possible structures can ever be presented to natural selection.  Even if one could re-run the tape of life over again multiple times, many structures that are  physically feasible and probably biochemically functional are extremely unlikely to appear simply  because they are inaccessible to evolutionary search.   

While the existence of bias in the RNA GP map has been known for quite some time~\cite{schuster1994sequences,fontana1993statistics,schultes2005compact, wagner2005robustness,schuster2006prediction,smit2006natural,cowperthwaite2008ascent,jorg2008neutral,aguirre2011topological,schaper2014arrival}, and studies using smaller amounts of  natural data have suggested that aspects of the mapping are reflected in nature~\cite{fontana1993statistics,smit2006natural,schultes2005compact,jorg2008neutral,cowperthwaite2008ascent}, our ability to calculate GP map properties for much larger $L$ allows us to make a comprehensive comparison to the fRNAdb database for functional ncRNA~\cite{kin2007frnadb}.   While one might expect that, due to bias, large $\Omega$ structures are relatively more plentiful in nature~\cite{cowperthwaite2008ascent,jorg2008neutral},  perhaps the most surprising result  of this study is just how closely the natural data follows the prediction of uniform G-sampling over genotypes.
   We find that the distribution of neutral set size $\Omega$, the number of stacks $S$ and the mutational robustness $r$ of naturally occurring ncRNA all closely follow the G-sampled distributions, and deviate significantly from the P-sampled distributions.   The distribution of bonds also deviates strongly from the P-sampled distribution, but in contrast to the other distributions above, it also exhibits a small deviation from the G-sampled distribution, with natural RNA having slightly more bonds~\cite{schultes2005compact} (see Supplementary Information Fig S3 (c),(d)).
   
     Why does it appear as if we can virtually ignore natural selection with G-sampling?  We postulate that even though the number of relevant structures remains extremely large, secondary structures that are good enough for function are nonetheless abundant, most likely because small differences between them don't matter that much. Instead, some broader coarse-grained structural features are likely to be sufficient for many functional roles.  We suggest that natural selection works on this pre-sculpted variation mainly by further refining parts of the sequence.  Some evidence for this picture can be gleaned by the fact that  natural structures have slightly more bonds than G-sampled RNA structures do, suggesting selection for greater thermal stability~\cite{schultes2005compact}.  

Interestingly, the fact that many properties of ncRNA SS so closely follow the G-sampled distribution is direct evidence against claims that the hyper-astronomically large size of genotype space makes  functional structures virtually impossible to find.  Such assertions were perhaps most famously made by Frank Salisbury~\cite{salisbury1969natural} over 45 years ago. Despite an illuminating response by John Maynard Smith~\cite{smith1970natural},  and much experimental evidence to the contrary, such ``arguments from large numbers'' remain a popular trope in anti-evolutionary polemics today. 

The ease with which we can calculate distributions of RNA properties  suggests an analogy to the concept of ergodicity in  statistical physics, which means that ensemble averages are equivalent to time-averages.  When ergodicity (approximately) holds, then it is often true in practice that sampling (via computer simulation for example) a relatively small number of "typical" states (small compared to the total number off possible microstates) is sufficient to accurately calculate ensemble averages of key properties.  Something akin to ergodicity may be operating in our case case because statistical averaging via G-sampling is close to the snapshot of many trajectories over time that we observe in the database, the latter being analogous to  a time-average.  On the other hand, evolution is not an equilibrium process.  Indeed, the exponential bias in the GP map suggests that evolutionary waiting times for more rare phenotypes to appear also grows exponentially~\cite{schaper2014arrival}.  Thus the number of novel phenotypic possibilities will continue to  (slowly) increase with time. With large waiting times contingency is more likely play an important role. So if a certain evolutionary change is predicated on one of these rare phenotypes fixing, then the process of waiting for these innovations may be non-ergodic. 

The suggestion that biases in development or other internal processes could strongly affect evolutionary outcomes has traditionally been highly contentious, see e.g.~refs.~\onlinecite{yampolsky2001bias,amundson2005changing} for an overview.  
For example, in a recent exchange, entitled: ``Does evolutionary theory need a rethink?'', Laland and colleagues~\cite{laland2014does} argue in favour of the thesis by, amongst other things, advocating for the importance of developmental bias.  In their rejoinder, Wray, Hoekstra and colleagues, write~\cite{wray2014does}: ``Lack of evidence also makes it difficult to evaluate the role that developmental bias may have in the evolution (or lack of evolution) of adaptive traits'', and call for new evidence:  ``The best way to elevate the prominence of genuinely interesting phenomena such as \ldots developmental bias \ldots is to strengthen the evidence for their importance.''  While the RNA system we study here is much simpler than a typical developmental system, that vice is also a virtue because it allows us to make detailed calculations of the whole morphospace which can then be closely compared to natural data.  Thus RNA secondary structure provides perhaps the clearest and most unambiguous evidence for the importance of bias in shaping evolutionary outcomes. 

Bias in the GP map constrains outcomes and so naturally suggest one mechanism for homoplasy (where similar biological forms  evolve independently)~\cite{losos2011convergence}.   The causes of homoplasy are sometimes elaborated in the context of the difference between parallel evolution, where homoplasy is thought to occur because two organisms share a common genetic heritage, and convergence proper, where the same solution is found by different genetic means, and where the primary causal force is usually attributed to selection.  While this binary distinction may be too simplistic (see e.g.~refs~\onlinecite{losos2011convergence,scotland2011parallelism,pearce2011convergence,powell2012convergent} for some recent discussion), very roughly, parallel evolution is thought to be considerably weaker evidence than true convergent evolution is for the idea that re-running the tape of life would generate similar outcomes.   Interestingly, the bias in the GP map  discussed here does not fit into this two-fold demarcation at all.  Re-run the tape of life, and as long as RNA graces the replay, so will a very similar suite of molecular shapes.  But the reason  for this repetition is not a contingent common genetic history, nor the  Allmacht  of selection~\cite{weissmann1893all}, but rather a different kind of `deep structure in biology'~\cite{morris2008deep}.

Our ability to make detailed predictions about evolutionary outcomes as well as counterfactuals for RNA may also shed light on Mayr's famous distinction between proximate  and ultimate causes in biology~\cite{mayr1961cause}, which has been the subject of  much recent debate in the literature~\cite{laland2011cause}.  This distinction has  historically been used to argue against the role of developmental bias in determining evolutionary outcomes (see e.g.~\onlinecite{amundson2005changing,schollproximate}).  While, as also mentioned above, the RNA GP map is much simpler than a typical developmental system, it is instructive to consider how the bias described in the current paper plays into Mayr's ultimate-proximate distinction: 
For example, what is the cause of  convergent evolution of the hammerhead ribozyme~\cite{salehi2001vitro}? The ultimate cause that this ribozyme emerges in  populations or in in-vitro experiments is surely natural selection for self-cleaving catalytic activity.  But {\em why} is a 3-stack structure repeatedly found, and not say a 10-stack structure, even though the latter are much more common in the morphospace?   The cause here is not natural selection {\em per se} since it is unlikely that  an efficient 10-stack ribozyme is biophysically or biochemically impossible to make.  Instead the explanatory force~\cite{schollproximate,schwenk2004relativism} for the ``why" question is mainly carried by a proximate  GP mapping constraint, namely that  the frequency with which 10-stack structures arise as potential variation is many orders of magnitude lower than the frequency with which  3-stack structures do.  Since the fittest can only be selected and survive if they arrive in the first place~\cite{devries1904mutation,wagner2014arrival},
 the evolutionary mechanism that leads to convergence  here might be better termed  {\em the arrival of the frequent}~\cite{schaper2014arrival}.

We also note that the mapping constraint described here differs from classical physical constraints, which would act on the whole morphospace, and from phyletic constraints, which are contingent on evolutionary histories~\cite{gould1979spandrels}.  This mapping constraint has some resemblance to classical morphogenetic constraints which also bias the arrival of variation~\cite{gilbert2000developmental}. But it also differs because the the latter are conceptualised at the level of phenotypes and developmental processes, and may have been shaped by prior selection, while the former constraint is a fundamental property of the mapping from genotypes to phenotypes and was not selected for (except perhaps at the origin of life itself).

Finally, strong  phenotype bias is also found in model GP maps for protein tertiary~\cite{li1996emergence,england2003structural} and quaternary structure~\cite{greenbury2014tractable}, gene regulatory networks~\cite{nochomovitz2006highly,raman2011evolvabilitysig}, and development~\cite{borenstein2008end}, suggesting that the some of the results discussed in this paper for RNA may hold more widely  in biology.  While all these GP maps  
only capture a tiny fraction of the full biological complexity of an organism,  if the bias is also (exponentially) strong, then  its effects on the rate with which novel variation appears are likely to persist even when further biological details are included.  Although more evidence needs to be gathered before making firm pronouncements, it may well be that we need to let go of the commonly held expectation that variation is isotropic in phenotype space or morphospace~\cite{gerber2014not}. Or to put it more provocatively, perhaps our null models  should  by default assume  that variation is  highly anisotropic and biased towards certain outcomes over others, unless there is direct evidence to the contrary.

\bigskip
\bigskip

 \noindent{\bf \large 4. Methods}

\noindent {\bf Folding RNA structures} To map a sequence to a SS, we use the Vienna package~\cite{hofacker1994fast} with  all parameters set to their default values (e.g.\, the temperature  $T=37^\circ$C). This software employs dynamic programming techniques to efficiently fold sequences based on thermodynamic rules.   Methods of this type are widely used, have been extensively tested, and are thought to be relatively accurate.  For example, a related method~\cite{mathews2004incorporating} was recently shown to  correctly predict  $73 \pm 9\%$  of canonical base pairs for a database of known RNA structures up to lengths of 700 nucleotides.   Generally these methods are expected to work better for shorter strands~\cite{reuter2010rnastructure}, and should work well for the lengths we explore. However, they typically cannot correctly predict pseudoknots. 
   Knowledge based methods that also take input from known structures and other information may be more accurate for predicting the structures of individual sequences~\cite{gardner2004comprehensive}, but such methods  could introduce biases for genotype-phenotype maps because they take input from natural structures.  Thermodynamically based methods such as the Vienna package are therefore probably better suited for working out  global properties of the entire genotype-phenotype map, including structures that have not (yet) been found in nature.  For these reasons, this package has been most frequently used in studies of full genotype-phenotype maps~\cite{schuster1994sequences,fontana1993statistics,schultes2005compact, wagner2005robustness,smit2006natural,schuster2006prediction,cowperthwaite2008ascent,jorg2008neutral,aguirre2011topological,schaper2014arrival}.      To check our Vienna package results, we compared RNAstructure package~\cite{mathews2004incorporating} calculations for the G-distributions of stacks and bonds (Supplementary Figure 3), finding very similar distributions.
    We also compared the two packages for other motifs like bulges, loops, and junctions, finding again very similar predictions (graphs are not shown). 
 \smallskip 

\noindent{\bf Exponential growth of the number of strands with strand length $\bf L$}  To illustrate how rapidly the number of sequences grows with length consider the following: There are $L4^L$ nucleotides in the set of all possible sequences of length $L$. The mean mass of a single RNA nucleotide is about  $5 \times 10^{-23}$ kg so that,  for example, the set of all $L=55$ strands weighs about $3.8 \times 10^{10}$ kg, the set of all $L=79$ strands weighs about $1.5 \times 10^{25}$ kg, or about $2.6$ times the mass of our Earth, while  the set of all $L = 126$ RNA strands would have an almost unimaginably large mass of about $5 \times 10^{53}$ kg, more than the mass of the observable universe which is estimated to be about $10^{53}$ kg.    
  \smallskip

\noindent {\bf Generating distributions $\bf P_G(\Omega)$ and $\bf P_P(\Omega)$ by sampling}  We used a standard (Python) random number generator to create sets of random sequences.  For each sequence we used the Vienna package to find the lowest free energy SS.   To determine $\Omega$ for each SS we used the neutral set size estimator (NSSE) described in  ref.~\cite{jorg2008neutral} which employs sampling techniques together with  the inverse fold algorithm from the Vienna package. 
We  used  default settings except for the total number of measurements (set with the -m option) which we set to 1 instead of the default 10, for the sake of speed. We checked that this has a negligible effect (typically $< 1\%$) on the accuracy of our distributions. We also checked the NSSE against the full enumeration for $L=20$, finding an agreement of  $R^2=0.97$ for structures with $\Omega$ larger than the average; it performs slightly less well for rare structures.      For longer lengths the number of samples were: $10^5$ for $L=30$, $3 \times 10^5$ for $L=40$, $20,000$ for $L = 35 - 80$,  $5000$ for $L = 85 -100$ and $1000$ for $L=126$..     Sequences that generate the trivial structure are discarded.  A small fraction of sequences (which increases with increasing length)  were  also discarded due to the inverse folding package failing to converge (See Supplementary Information).

To generate the $P_G(\Omega)$ distribution, we partition the support of the distribution into bins which are uniform on an $S=\log_{10}(\Omega)$ scale. We then determine the probability mass $P_G(\Omega)$ in each bin.  Error bars are simply statistical: 
There is a tradeoff between making smaller bins to give a greater resolution  and minimising statistical errors that increase when there are fewer measurements per bin.
   Bins with too few sampled points were typically not included in the graphs to avoid large error bars.   $P_P(\Omega)$ is generated from the sampled data by dividing though by $\Omega$ (measured at the midpoint of the bin).  $P_P(\Omega)$ is normalised with the $N_P$ calculated from the analytic approximation to $P_P(\Omega)$ (see below). 
   
   \smallskip 

\noindent {\bf  Analytical fit to NS size distributions} 
For analytic fits we make a simple log-binomial ansatz: 
\begin{equation}\label{eq:PP}
P_P(\Omega) =  {N \choose q} (p_P)^q (1-p_P)^{N-q}   
\end{equation}
where $q=\log(\Omega) N/U$,  
$10^U$ is the largest non-trivial $\Omega$,
 and $N$ and $p_P$ are parameters that are fit to measured distributions.  In other words the probability that a P-sampled SS is found with $S=\log(\Omega)$ is distributed binomially.  
 By definition  $P_G(\Omega) \propto \Omega P_P(\Omega)$. Taking normalisation into account is enough to show that $P_G(\Omega)$  has the same  binomial form as Eq.~(\ref{eq:PP}), but with parameter $p_P$ replaced by $p_G  = (p_P 10^{U/N})/(1 - p_P + p_P 10^{U/N})$.  Fixing  the parameters $N$, $U$ and either $p_G$ or $p_P$ thus fixes both distributions.  
   
 For $L = 20$, Eq.~(\ref{eq:PP}) with $U=10$, $N = 8.0$ and $p_P = 0.55$ describes the exact $P_P(\Omega)$  from full enumeration very well, as can be seen in Fig.~1b. The related approximation for $P_G(\Omega)$ performs slightly less well for G-sampled data for $L=20$, but it still captures the main qualitative features.
 
  For larger $L$ we determine the parameters as follows: 
  First, we estimate $U$ from the largest non-trivial NS size found.    This method inevitably provides a lower bound  $U'$ on the true maximum $U$.  However, given the rather sharp upper bound generated by the binomial fit, we expect that the relative errors in $U$ are quite small.  For example, for $L=60$ we used 20,000 samples to determine $U'=30.56$. From the binomial form we estimate that  $U'$ is within an error $\delta U = 0.45$ of the true $U$ with  $90\%$ probability. Next we calculate $\bar{S}_G$, the G-sampled average of $S=\log(\Omega)$, as well as the standard deviation $\bar{\sigma}_G$ in $S$ from G-sampled data. We can then determine the parameter $p_G =\bar{S}_G/U$, derived by taking the mean of $q$ through~Eq.~(\ref{eq:PP}).  The parameter $N$ can subsequently be extracted from the measured G-sampled standard deviation: $\bar{\sigma}_G = \sqrt{N p_G(1-p_G)}U/N$. The P-sampled standard deviation is given by  $\bar{\sigma}_P = \sqrt{N p_P(1-p_P)}U/N$. Since $p_P = \bar{S}_P/U <  p_G$, so also $\bar{\sigma}_P < \bar{\sigma}_G$.
 In this way we obtained the binomial fits shown in Fig.\ref{fig2} and   Supplementary Figs 1,2. The close agreement between the sampled data and our fits for lengths up to $L=126$ suggests that this procedure is fairly robust.  
 
 It is harder to find structures with small $\Omega$, so that the $P_P(\Omega)$ can only be partially sampled, especially for larger $L$. However, given how well our simple ansatz works for predicting $P_G(\Omega)$, and given that for L = 20 RNA the binomial form  works so well for the full range of structures, we expect the full $P_P (\Omega)$ to be at least similar if not very close to Eq.~(\ref{eq:PP}). Further evidence for this form can also be extracted from combinatorial arguments for the distributions of stacks~\cite{hofacker1998combinatorics}, which are correlated with $\log(\Omega)$ (see below).

  \smallskip 

\noindent {\bf Rank plots}  Analytic rank-plots functions are the cumulative density function of $P_P(\Omega)$.  For $L=20$ all structures are known, so a rank plot can be directly made.   For $L > 20$  the measured $\Omega$ of natural SS were used to align points to a rank function calculated from the analytic binomial fit to $P_P(\Omega)$.
   \smallskip

 \noindent {\bf Scaling forms as a function of $\bf L$} We further used the lengths $L=30-100$ to extract scaling forms for several properties as a function of $L$. Linear fits in $L$ are shown for $U$, $T$, $S_G$ and $N$  in  Supplementary Fig. 5.  $T$ is close to $U$ so that the relative difference between $T$ and $U$ decreases as $L$ increases (Supplementary Fig.~5). 
A summary of the scaling forms for different properties in the large $L$ limit can be found in  Table~\ref{table:X1}.

\begin{table}\label{MainTable}
\begin{tabular}{ll}
\multicolumn{2}{c}

 \,
\\
\cline{1-2}
\emph{Quantity} & \emph{Large $L$ scaling form}\\ 
\hline
Total number of genotypes & $N_G     = 4^L$ \\
Total number of SS phenotypes  &$N_P   \approx 0.13\times 1.76^L$\\
Mean  $\Omega$ & $4^L$/$N_P \approx 7.7 \times 2.27^L$\\
Largest non-trivial $\Omega$ & $10^U \approx 0.7 \times 3.27^L$ \\ 
$\Omega$ for the trivial structure & $10^T \approx 21.4\times3.29^L$\\
Probability to sample trivial structure & $P_T \approx 21.4\times 0.82^L$\\
$\Omega$  near peak for phenotype sampling & $10^{\bar{S}_P} \approx 0.01 \times 2.1^L$\\
$\Omega$ near peak for genotype sampling & $10^{\bar{S}_G} \approx 30.2 \times 2.5^L$\\
Shannon entropy of distribution  & $H  \approx 0.675L - 4.92$ \\
`Effective number' of  SS phenotypes  & $\beta N_P   \approx 0.032 \times 1.60^L$ \\
Bias parameter &  $\beta =\frac{2^H}{N_P} \approx 0.25 \times 0.91^L$\\
G-sampled mean number of stacks & $\bar{K}_G = 0.074 L - 0.377$ \\
P-sampled mean number of stacks & $\bar{K}_P =  0.177 L - 0.443 $\\
\hline
\end{tabular} 
 \caption{{\bf  Large $L$ scaling of some key  quantities for RNA SS}.   For the number of sequences $\Omega$ for the NS of the largest non-trivial structure, defined as $\Omega = 10^U$, we find   $U= (0.514 \pm 0.009)L - (0.20 \pm 0.5)$, while for the trivial structure, with $\Omega = 10^T$, we find  $T= (0.5166 \pm 0.0009)L+(1.33 \pm 0.06)$ so that $U$ becomes relatively more close to $T$ as $L$ increases.  The G-sampled mean of $S=\log(\Omega)$ scales as $\bar{S}_G = (0.399 \pm 0.0014)L + (1.48 \pm  0.09)$.
 For large $L$,   $\bar{S}_G \approx 0.78\, U$ while for P-sampling, $\bar{S}_P \approx 0.63\, U$ so that $\bar{S}_P/\bar{S}_G \approx 0.8$.  Similarly, the standard deviations of $\log(\Omega)$  can be directly calculated, and in the large $L$ limit tend to $\bar{\sigma}_P/U \approx 0.37/\sqrt{L}$ and $\bar{\sigma}_G/U \approx 0.31/\sqrt{L}$. This explains analytically what can be observed qualitatively in Fig.~3 and Supplementary  Figs.~1,2: The $P_G(\Omega)$ distribution is slightly narrower than the $P_P(\Omega)$ distribution. As $L$ increases both distributions become more sharply peaked  relative to the total range $[0,U]$ and  $P_G(\Omega)$ highlights SS phenotypes that are deeper into the tails the $P_P(\Omega)$ distribution (and vice versa).
   \label{table:X1}}
\end{table}

  \smallskip 

 \noindent {\bf Probability  $\bf P_T$ to find the trivial structure}
  The probability of finding the trivial structure, $P_T$, decreases exponentially with increasing $L$:
\begin{equation}
P_T=10^T/4^L\approx21.4\times 0.82^L\label{eq:PTriv} 
\end{equation}
For example, for $L=20$ we can directly measure $P_T\approx 33\% $ whereas for $L=55$ it has already dropped down to a mere $0.04\%$.  This rapid decrease in $P_T$ justifies our decision to ignore the trivial structure in our fitting to $P_G(S)$ and $P_P(S)$.

  \smallskip 

  \noindent {\bf The number of SS, $\bf N_P$, as a function of $\bf L$}
For $L=11 - 20$ we used full enumerations~\cite{schaper2014arrival} to calculate $N_P$.  For longer $L$ we used our analytic fit as follows: 
The mean (including trivial structure) of $\Omega$ is $4^L/N_P$ so that $(1-P_T)4^L/N_P = \sum_{S=0}^U  P_P(\Omega)\Omega$. Given the binomial form of $P_P(\Omega)$,  the sum can be carried out analytically, from which it follows that
$
N_P = (1-P_T)4^L/(1-p_P +p_P 10^{U/N})^N.
$
  At each $L$ we evaluated $N, P_T, p_P$ and $U$, and used this to evaluate $N_P$.  A simple linear form
\begin{equation}\label{eq:NP}
N_P \approx (0.13\pm 0.04)\times (1.760\pm 0.007)^L 
\end{equation}
provides a good fit to the data, ( Supplementary Fig. 5).

   \smallskip

\noindent {\bf Scaling forms for the Shannon entropy $\bf H$ and bias parameter $\bf \beta$ as a function of $\bf L$}
From the expression for the entropy derived above it  directly follows  that $H = 2L - \log_2(10)\bar{S}_G$. Using the fit derived above for $S_G$, the entropy grows with $L$ as
$H\approx 0.675 L-4.92$ and the effective number of  states scales as $2^H = 4^L/10^{\bar{S}_G} \approx 0.033 \times 1.60^L$ (This ignores the trivial structure, but the effect is very small for larger $L$). Note that one only has to find $\bar{S}_G$, which is easily obtainable, to determine this important quantity.   By combining with Eq.~(\ref{eq:NP}) we find that the bias parameter scales as $\beta = 2^H/N_P \approx  0.26 \times 0.91^L$.

  \smallskip 

\noindent {\bf Sampling natural RNA from the fRNA database}  To generate the distributions of functional ncRNA we took all available sequences for each length studied  (ranging from $L=40$ to $L=126$) from the non-coding functional RNA database (fRNAdb~\cite{kin2007frnadb}).  For $L = 20$  we took data from {\em Drosophila melanogaster} only, but this made up $77\%$  of all $L=20$ SS in the database.  For each sequence we  found the SS and used the NSSE to estimate its $\Omega$.    A small fraction of the natural RNA sequences contained non-standard nucleotide letters, e.g.\ ÔNÕ or ÔRÕ; such sequences were ignored, because the standard packages cannot treat them. Similarly, a small fraction of sequences were also discarded due to the NSSE failing to converge (see SI).  We checked that there were no repeated sequences in the database of natural RNA.  
 Finally, in the Supplementary Information we provide a breakdown of the identity of the structures for $L=20,55,70$ and $126$. and also take curated subsets of the data to emphasise structures where the SS is known to be important. In Fig.~3 and Supplementary Figs 1,2, we show for $L=55$ and $L=126$ that the close correlation with $P_G(\Omega)$ remains when data is curated.  
  \smallskip 

\noindent{\bf Checking for codon bias}  
Genetic mutations are random in the sense that they don't arise to benefit an organism. Nevertheless, it is well known that in other ways mutations are not uniformly random~\cite{li1999fundamentals}.  One example (among several) is that  transitions (pyramidines $\leftrightarrow$ pyramidines or purines  $\leftrightarrow$ purines) are more frequent that transversions (purines $\leftrightarrow$ pyramidines)~\cite{li1999fundamentals}. For many of these biases, mutations can still effectively sample the whole space  uniformly without preferring certain genotypes over others.   Nevertheless, there are biases that lead, for example, to an excess of $GC$ over $AT$ base pairs, or vice versa in DNA~\cite{li1999fundamentals}.  To test for the effect of strong bias of this type we generated $\Omega$ distributions with $30\% $ GC (AU bias)  and $70 \% $ GC (GC bias) content for different lengths. The overall effect  becomes less pronounced for longer strands (Supplementary Fig 4).  Since natural DNA (and by extension RNA) can show biases for both larger and smaller GC content, we argue that this can to first order be ignored when comparing to natural data sets across many species, although effects may be observable  on datasets with large content bias. 
\smallskip

 \noindent {\bf Calculating P-distribution of stacks} 
   We used a linear relationship between $\log(\Omega)$ and $K$ (Supplementary Fig.~6) to transform $P_P(\Omega)$ into an estimated P-sampled distribution of stacks, as shown in Fig.4 a, and used this to obtain the P-sampled average $\bar{K}_P$.   We also adapted analytic results from ref~\cite{hofacker1998combinatorics} based on combinatorics (see SI)  to calculate estimates for the P-sampled distribution of $K$ and for $\bar{K}_P$.  The close agreement shown  in Fig 4 a,b,   between the two independent methods for estimating the number of stacks from P-sampling increases our confidence in our  binomial fits to the full P-sampled distribution.
    
  \smallskip

\noindent {\bf Calculating the robustness distribution}   A strong positive correlation between mutational robustness and $\Omega$ is well established in the literature~\cite{wagner2005robustness}.
For Fig~6,  the robustness for G-sampled data (taken from $10^5$ random sequences) and natural data (taken from the the $504$ natural sequences for $L=55$), was calculated by folding all $3L$ strands within one point mutation and calculating the fraction $r$ that generate the same SS.   Error bars come from our binning procedure.  
For the P-sampled distribution such a sequence robustness can't be defined, so instead the robustness per sequence was first averaged over a set of sequences for each secondary structure to estimate a mean robustness (phenotype robustness) per SS.  We  next generated a cubic fit to the mean $r$ per SS v.s.\ $\log\Omega$. The fit was constrained to have $r=0$ at $\Omega=1$ and had $R^2 = 0.93$.   This was then combined with $P_P(\Omega)$ to generate an estimate of the $r$ distribution for P-sampled data.   Since we don't have many data points at small $\Omega$, the fit is partially an extrapolation. Further error may arise from the partial sampling of the phenotype robustness. Nevertheless,  we don't expect this procedure to lead to a significant difference in the qualitative comparison between P- and G-sampled data for robustness, given that the two are so different.

\medskip

\noindent {\bf Acknowledgements}  We acknowledge discussions with S.E. Ahnert, F.Q. Camargo, J.P.K. Doye, J.L. England,  T.M.A. Fink, W. Haerty,  I.G. Johnston, D.C. Lahti, A.C. Love, O. Martin, T.C.B. McLeish, T.E. Ouldridge,  C.P. Ponting, F. Randisi, and P. Sulc. 

\noindent {\bf Funding}  KD acknowledges funding from the EPSRC/EP/G03706X/1


\bibliographystyle{rspublicnatwithsort}
\bibliography{RNADistnRefs} 

\begin{thebibliography}{58}
\providecommand{\natexlab}[1]{#1}
\expandafter\ifx\csname urlstyle\endcsname\relax
  \providecommand{\doi}[1]{doi:\discretionary{}{}{}#1}\else
  \providecommand{\doi}{doi:\discretionary{}{}{}\begingroup
  \urlstyle{rm}\Url}\fi

\bibitem[{Gould(1989)}]{gould1989wonderful}
Gould, S.~J. 1989 \emph{{Wonderful life: the Burgess Shale and the nature of
  history}}.
\newblock W. W. Norton and Co.

\bibitem[{Morris(2003)}]{morris2003life}
Morris, S. 2003 \emph{Life's solution: inevitable humans in a lonely universe}.
\newblock Cambridge University Press.

\bibitem[{McGhee(2006)}]{mcghee2006geometry}
McGhee, G.~R. 2006 \emph{The geometry of evolution: adaptive landscapes and
  theoretical morphospaces}.
\newblock Cambridge University Press.

\bibitem[{Salisbury(1969)}]{salisbury1969natural}
Salisbury, F.~B. 1969 Natural selection and the complexity of the gene.
\newblock \emph{Nature}, \textbf{224}, 342--343.

\bibitem[{Maynard~Smith(1970)}]{smith1970natural}
Maynard~Smith, J. 1970 Natural selection and the concept of a protein space.
\newblock \emph{Nature}, \textbf{225}, 563--564.

\bibitem[{Wagner(2005)}]{wagner2005robustness}
Wagner, A. 2005 \emph{{Robustness and evolvability in living systems}}.
\newblock Princeton University Press Princeton, NJ:.

\bibitem[{Mattick \& Makunin(2006)}]{mattick2006non}
Mattick, J.~S. \& Makunin, I.~V. 2006 Non-coding rna.
\newblock \emph{Human molecular genetics}, \textbf{15}(suppl 1), R17--R29.

\bibitem[{Carothers \emph{et~al.}(2004)Carothers, Oestreich, Davis \&
  Szostak}]{carothers2004informational}
Carothers, J., Oestreich, S., Davis, J. \& Szostak, J. 2004 {Informational
  complexity and functional activity of {RNA} structures}.
\newblock \emph{Journal of the American Chemical Society}, \textbf{126}(16),
  5130--5137.

\bibitem[{Bailor \emph{et~al.}(2011)Bailor, Mustoe, III \&
  Al-Hashimi}]{bailor2011topological}
Bailor, M.~H., Mustoe, A.~M., III, C. L.~B. \& Al-Hashimi, H.~M. 2011
  Topological constraints: using {RNA} secondary structure to model 3d
  conformation, folding pathways, and dynamic adaptation.
\newblock \emph{Current Opinion in Structural Biology}, \textbf{21}(3), 296 --
  305.

\bibitem[{Hofacker \emph{et~al.}(1994)Hofacker, Fontana, Stadler, Bonhoeffer,
  Tacker \& Schuster}]{hofacker1994fast}
Hofacker, I., Fontana, W., Stadler, P., Bonhoeffer, L., Tacker, M. \& Schuster,
  P. 1994 Fast folding and comparison of {RNA} secondary structures.
\newblock \emph{Monatshefte f{\"u}r Chemie/Chemical Monthly}, \textbf{125}(2),
  167--188.

\bibitem[{Zuker \emph{et~al.}(1999)Zuker, Mathews \&
  Turner}]{zuker1999algorithms}
Zuker, M., Mathews, D. \& Turner, D. 1999 Algorithms and thermodynamics for
  {RNA} secondary structure prediction: a practical guide.
\newblock \emph{RNA biochemistry and biotechnology}, \textbf{70}, 11--44.

\bibitem[{Schuster \emph{et~al.}(1994)Schuster, Fontana, Stadler \&
  Hofacker}]{schuster1994sequences}
Schuster, P., Fontana, W., Stadler, P. \& Hofacker, I. 1994 {From sequences to
  shapes and back: A case study in {RNA} secondary structures}.
\newblock \emph{Proceedings: Biological Sciences}, \textbf{255}(1344),
  279--284.

\bibitem[{Fontana \emph{et~al.}(1993)Fontana, Konings, Stadler \&
  Schuster}]{fontana1993statistics}
Fontana, W., Konings, D.~A., Stadler, P.~F. \& Schuster, P. 1993 Statistics of
  {RNA} secondary structures.
\newblock \emph{Biopolymers}, \textbf{33}(9), 1389--1404.

\bibitem[{Schultes \emph{et~al.}(2005)Schultes, Spasic, Mohanty \&
  Bartel}]{schultes2005compact}
Schultes, E.~A., Spasic, A., Mohanty, U. \& Bartel, D.~P. 2005 Compact and
  ordered collapse of randomly generated {RNA} sequences.
\newblock \emph{Nature structural \& molecular biology}, \textbf{12}(12),
  1130--1136.

\bibitem[{Schuster(2006)}]{schuster2006prediction}
Schuster, P. 2006 Prediction of rna secondary structures: from theory to models
  and real molecules.
\newblock \emph{Reports on Progress in Physics}, \textbf{69}(5), 1419.

\bibitem[{Smit \emph{et~al.}(2006)Smit, Yarus \& Knight}]{smit2006natural}
Smit, S., Yarus, M. \& Knight, R. 2006 Natural selection is not required to
  explain universal compositional patterns in r{RNA} secondary structure
  categories.
\newblock \emph{RNA}, \textbf{12}(1), 1--14.

\bibitem[{Cowperthwaite \emph{et~al.}(2008)Cowperthwaite, Economo, Harcombe,
  Miller \& Meyers}]{cowperthwaite2008ascent}
Cowperthwaite, M., Economo, E., Harcombe, W., Miller, E. \& Meyers, L. 2008 The
  ascent of the abundant: how mutational networks constrain evolution.
\newblock \emph{PLoS computational biology}, \textbf{4}(7), e1000\,110.

\bibitem[{Jorg \emph{et~al.}(2008)Jorg, Martin \& Wagner}]{jorg2008neutral}
Jorg, T., Martin, O. \& Wagner, A. 2008 {Neutral network sizes of biological
  {RNA} molecules can be computed and are not atypically small}.
\newblock \emph{BMC bioinformatics}, \textbf{9}(1), 464.

\bibitem[{Aguirre \emph{et~al.}(2011)Aguirre, Buld{\'u}, Stich \&
  Manrubia}]{aguirre2011topological}
Aguirre, J., Buld{\'u}, J.~M., Stich, M. \& Manrubia, S.~C. 2011 Topological
  structure of the space of phenotypes: the case of {RNA} neutral networks.
\newblock \emph{PloS one}, \textbf{6}(10), e26\,324.

\bibitem[{Schaper \& Louis(2014)}]{schaper2014arrival}
Schaper, S. \& Louis, A.~A. 2014 The arrival of the frequent: How bias in
  genotype-phenotype maps can steer populations to local optima.
\newblock \emph{PloS one}, \textbf{9}(2), e86\,635.

\bibitem[{Kin \emph{et~al.}(2007)Kin, Yamada, Terai, Okida, Yoshinari, Ono,
  Kojima, Kimura, Komori \emph{et~al.}}]{kin2007frnadb}
Kin, T., Yamada, K., Terai, G., Okida, H., Yoshinari, Y., Ono, Y., Kojima, A.,
  Kimura, Y., Komori, T. \emph{et~al.} 2007 frnadb: a platform for
  mining/annotating functional {RNA} candidates from non-coding {RNA}
  sequences.
\newblock \emph{Nucleic acids research}, \textbf{35}(suppl 1), D145--D148.

\bibitem[{de~Vries(1904)}]{devries1904mutation}
de~Vries, H. 1904 \emph{Species and Varieties, Their Origin by Mutation}.
\newblock {The Open Court Publishing Company}.

\bibitem[{Wagner(2014)}]{wagner2014arrival}
Wagner, A. 2014 \emph{Arrival of the Fittest: Solving Evolution's Greatest
  Puzzle}.
\newblock Penguin.

\bibitem[{Hofacker \emph{et~al.}(1998)Hofacker, Schuster \&
  Stadler}]{hofacker1998combinatorics}
Hofacker, I.~L., Schuster, P. \& Stadler, P.~F. 1998 Combinatorics of {RNA}
  secondary structures.
\newblock \emph{Discrete Applied Mathematics}, \textbf{88}(1), 207--237.

\bibitem[{Andronescu \emph{et~al.}(2008)Andronescu, Bereg, Hoos \&
  Condon}]{andronescu2008rna}
Andronescu, M., Bereg, V., Hoos, H.~H. \& Condon, A. 2008 {RNA STRAND}: the
  {RNA} secondary structure and statistical analysis database.
\newblock \emph{BMC bioinformatics}, \textbf{9}(1), 340.

\bibitem[{Greenbury \emph{et~al.}(2015)Greenbury, Ahnert \&
  Louis}]{greenbury2015genetic}
Greenbury, S.~F., Ahnert, S.~E. \& Louis, A.~A. 2015 Genetic correlations
  greatly increase mutational robustness and can both reduce and enhance
  evolvability.
\newblock \emph{http://arxiv.org/abs/1505.07821}.

\bibitem[{Mezard \& Montanari(2009)}]{mezard2009information}
Mezard, M. \& Montanari, A. 2009 \emph{Information, physics, and computation}.
\newblock Oxford University Press, USA.

\bibitem[{Salehi-Ashtiani \& Szostak(2001)}]{salehi2001vitro}
Salehi-Ashtiani, K. \& Szostak, J. 2001 In vitro evolution suggests multiple
  origins for the hammerhead ribozyme.
\newblock \emph{Nature}, \textbf{414}(6859), 82--83.

\bibitem[{Hammann \emph{et~al.}(2012)Hammann, Luptak, Perreault \& de~la
  Pe{\~n}a}]{hammann2012ubiquitous}
Hammann, C., Luptak, A., Perreault, J. \& de~la Pe{\~n}a, M. 2012 The
  ubiquitous hammerhead ribozyme.
\newblock \emph{RNA}, \textbf{18}(5), 871--885.

\bibitem[{Gan \emph{et~al.}(2003)Gan, Pasquali \& Schlick}]{Gan03exploringthe}
Gan, H.~H., Pasquali, S. \& Schlick, T. 2003 Exploring the repertoire of rna
  secondary motifs using graph theory; implications for rna design.
\newblock \emph{Nucleic Acids Res}, \textbf{31}, 2926--2943.

\bibitem[{Athavale \emph{et~al.}(2014)Athavale, Spicer \&
  Chen}]{athavale2014experimental}
Athavale, S.~S., Spicer, B. \& Chen, I.~A. 2014 Experimental fitness landscapes
  to understand the molecular evolution of rna-based life.
\newblock \emph{Current opinion in chemical biology}, \textbf{22}, 35--39.

\bibitem[{Yampolsky \& Stoltzfus(2001)}]{yampolsky2001bias}
Yampolsky, L. \& Stoltzfus, A. 2001 {Bias in the introduction of variation as
  an orienting factor in evolution}.
\newblock \emph{Evolution \& Development}, \textbf{3}(2), 73--83.

\bibitem[{Amundson(2005)}]{amundson2005changing}
Amundson, R. 2005 \emph{The changing role of the embryo in evolutionary
  thought: roots of evo-devo}.
\newblock Cambridge University Press.

\bibitem[{Laland \emph{et~al.}(2014)Laland, Uller, Feldman, Sterelny,
  M{\"u}ller, Moczek, Jablonka \& Odling-Smee}]{laland2014does}
Laland, K., Uller, T., Feldman, M., Sterelny, K., M{\"u}ller, G.~B., Moczek,
  A., Jablonka, E. \& Odling-Smee, J. 2014 Does evolutionary theory need a
  rethink? yes, urgently.
\newblock \emph{Nature}, \textbf{514}(7521), 161--164.

\bibitem[{Wray \emph{et~al.}(2014)Wray, Hoekstra, Futuyma, Lenski, Mackay,
  Schluter \& Strassman}]{wray2014does}
Wray, G.~A., Hoekstra, H.~E., Futuyma, D., Lenski, R., Mackay, T., Schluter, D.
  \& Strassman, J. 2014 Does evolutionary theory need a rethink? no, all is
  well.
\newblock \emph{Nature}, \textbf{514}(7521), 161--164.

\bibitem[{Losos(2011)}]{losos2011convergence}
Losos, J.~B. 2011 Convergence, adaptation, and constraint.
\newblock \emph{Evolution}, \textbf{65}(7), 1827--1840.

\bibitem[{Scotland(2011)}]{scotland2011parallelism}
Scotland, R.~W. 2011 What is parallelism?
\newblock \emph{Evolution \& development}, \textbf{13}(2), 214--227.

\bibitem[{Pearce(2011)}]{pearce2011convergence}
Pearce, T. 2011 Convergence and parallelism in evolution: a neo-{G}ouldian
  account.
\newblock \emph{The British Journal for the Philosophy of Science}, p. axr046.

\bibitem[{Powell(2012)}]{powell2012convergent}
Powell, R. 2012 Convergent evolution and the limits of natural selection.
\newblock \emph{European Journal for Philosophy of Science}, \textbf{2}(3),
  355--373.

\bibitem[{Weismann(1893)}]{weissmann1893all}
Weismann, A. 1893 The all-sufficiency of natural selection: A reply to
  {H}erbert {S}pencer.
\newblock \emph{Contemporary Review}, \textbf{64}, 309--338.

\bibitem[{Morris(2008)}]{morris2008deep}
Morris, S.~C. 2008 \emph{The deep structure of biology: is convergence
  sufficiently ubiquitous to give a directional signal}.
\newblock 45. Templeton Foundation Press.

\bibitem[{Mayr(1961)}]{mayr1961cause}
Mayr, E. 1961 Cause and effect in biology.
\newblock \emph{Science (New York, NY)}, \textbf{134}(3489), 1501--1506.

\bibitem[{Laland \emph{et~al.}(2011)Laland, Sterelny, Odling-Smee, Hoppitt \&
  Uller}]{laland2011cause}
Laland, K.~N., Sterelny, K., Odling-Smee, J., Hoppitt, W. \& Uller, T. 2011
  Cause and effect in biology revisited: is mayr'€™s proximate-ultimate
  dichotomy still useful?
\newblock \emph{Science}, \textbf{334}(6062), 1512--1516.

\bibitem[{Scholl \& Pigliucci(2014)}]{schollproximate}
Scholl, R. \& Pigliucci, M. 2014 The proximate--ultimate distinction and
  evolutionary developmental biology: causal irrelevance versus explanatory
  abstraction.
\newblock \emph{Biology \& Philosophy}, pp. 1--18.

\bibitem[{Schwenk \& Wagner(2004)}]{schwenk2004relativism}
Schwenk, K. \& Wagner, G.~P. 2004 The relativism of constraints on phenotypic
  evolution.
\newblock \emph{in M. Pigliucci and K. Preston, eds. Phenotypic integration:
  studying the ecology and evolution of complex phenotypes. Oxford Univ. Press,
  Oxford, UK.}, pp. 390--408.

\bibitem[{Gould \& Lewontin(1979)}]{gould1979spandrels}
Gould, S. \& Lewontin, R. 1979 {The spandrels of San Marco and the Panglossian
  paradigm: a critique of the adaptationist programme}.
\newblock \emph{Proceedings of the Royal Society of London. Series B,
  Biological Sciences}, \textbf{205}(1161), 581--598.

\bibitem[{Gilbert(2000)}]{gilbert2000developmental}
Gilbert, S.~F. 2000 \emph{Developmental Biology. Sunderland, MA}.
\newblock Sinauer Associates, Inc.

\bibitem[{Li \emph{et~al.}(1996)Li, Helling, Tang \&
  Wingreen}]{li1996emergence}
Li, H., Helling, R., Tang, C. \& Wingreen, N. 1996 {Emergence of preferred
  structures in a simple model of protein folding}.
\newblock \emph{Science}, \textbf{273}(5275), 666--669.

\bibitem[{England \& Shakhnovich(2003)}]{england2003structural}
England, J. \& Shakhnovich, E. 2003 Structural determinant of protein
  designability.
\newblock \emph{Physical Review Letters}, \textbf{90}(21), 218\,101.

\bibitem[{Greenbury \emph{et~al.}(2014)Greenbury, Johnston, Louis \&
  Ahnert}]{greenbury2014tractable}
Greenbury, S.~F., Johnston, I.~G., Louis, A.~A. \& Ahnert, S.~E. 2014 A
  tractable genotype--phenotype map modelling the self-assembly of protein
  quaternary structure.
\newblock \emph{Journal of The Royal Society Interface}, \textbf{11}(95),
  20140\,249.

\bibitem[{Nochomovitz \& Li(2006)}]{nochomovitz2006highly}
Nochomovitz, Y. \& Li, H. 2006 Highly designable phenotypes and mutational
  buffers emerge from a systematic mapping between network topology and dynamic
  output.
\newblock \emph{Proceedings of the National Academy of Sciences of the United
  States of America}, \textbf{103}(11), 4180.

\bibitem[{Raman \& Wagner(2011)}]{raman2011evolvabilitysig}
Raman, K. \& Wagner, A. 2011 {Evolvability and robustness in a complex
  signalling circuit}.
\newblock \emph{Mol. BioSyst.}

\bibitem[{Borenstein \& Krakauer(2008)}]{borenstein2008end}
Borenstein, E. \& Krakauer, D. 2008 {An end to endless forms: epistasis,
  phenotype distribution bias, and nonuniform evolution}.
\newblock \emph{PLoS Comput Biol}, \textbf{4}(10), e1000\,202.

\bibitem[{Gerber(2014)}]{gerber2014not}
Gerber, S. 2014 Not all roads can be taken: development induces anisotropic
  accessibility in morphospace.
\newblock \emph{Evolution \& development}, \textbf{16}(6), 373--381.

\bibitem[{Mathews \emph{et~al.}(2004)Mathews, Disney, Childs, Schroeder, Zuker
  \& Turner}]{mathews2004incorporating}
Mathews, D.~H., Disney, M.~D., Childs, J.~L., Schroeder, S.~J., Zuker, M. \&
  Turner, D.~H. 2004 Incorporating chemical modification constraints into a
  dynamic programming algorithm for prediction of {RNA} secondary structure.
\newblock \emph{Proceedings of the National Academy of Sciences of the United
  States of America}, \textbf{101}(19), 7287--7292.

\bibitem[{Reuter \& Mathews(2010)}]{reuter2010rnastructure}
Reuter, J.~S. \& Mathews, D.~H. 2010 Rnastructure: software for {RNA} secondary
  structure prediction and analysis.
\newblock \emph{BMC bioinformatics}, \textbf{11}(1), 129.

\bibitem[{Gardner \& Giegerich(2004)}]{gardner2004comprehensive}
Gardner, P.~P. \& Giegerich, R. 2004 A comprehensive comparison of comparative
  {RNA} structure prediction approaches.
\newblock \emph{BMC bioinformatics}, \textbf{5}(1), 140.

\bibitem[{Li \& Graur(1999)}]{li1999fundamentals}
Li, W. \& Graur, D. 1999 \emph{Fundamentals of molecular evolution}.
\newblock Sunderland: Sinauer.

\end{thebibliography}


\pagebreak
\widetext
\begin{center}
\textbf{\large Supplemental Materials: The structure of the genotype-phenotype map strongly constrains the evolution of non-coding RNA }
\end{center}
\setcounter{equation}{0}
\setcounter{figure}{0}
\setcounter{table}{0}
\setcounter{page}{1}
\makeatletter
\renewcommand{\theequation}{S\arabic{equation}}
\renewcommand{\thefigure}{S\arabic{figure}}
\renewcommand{\bibnumfmt}[1]{[S#1]}

\section{Supplementary Methods}

\noindent{\bf Correlation between number of stacks and $\bf \Omega$} 
 In Supplementary  Fig.~S6  we plot the number of stacks (i.e.\ contiguous base pairs), $K$, in sampled SS, which is linearly correlated with  $\log(\Omega)$.  Only two lengths are shown, but the correlation is very similar for all lengths.

\medskip





  \noindent   {\bf P-distribution of stacks from combinatorial analyses}\label{p}
Another approach to calculating the $P$ sampled number of stacks can be found in an important paper by Hofacker \emph{et al}~\cite{hofacker1998combinatorics}, who analyse all possible SS under various constraints.  Their arguments essentially count all ways of connecting bonds together, given physical constraints such as a minimum loop size $m$.      This set is not  the same as the set of all  structures that sequences fold to, as there may be structures that are not the lowest free-energy structure for any sequence at the temperature investigated.     Nevertheless, It is instructive to compare the distributions they derive to our P-sampling results.

More  specifically, Eqs.\ 6 and 7 of  Hofacker \emph{et al}~\cite{hofacker1998combinatorics}  give a recursion formula for the number of SS of length $L$ with $K$ stacks, denoted $N_{L}(K):$
\begin{equation}
N_{L+1}(K)=N_{L}(K)+\sum_{k=m}^{L-1}\sum_{l=0}^{K}Z_{k+2}(l)N_{L-k-1}(K-l)
\end{equation}
for $K>0$ and $L\geq m+1$ (where $m$ is the minimum loop size), and
\begin{equation}
N_L(0)=1, N_{L}(K)=0, K>0, L\leq m+1
\end{equation}
Here $Z_{L}(K)$ is an auxillary variable defined as
\begin{equation}
Z_L(K)=Z_{L-2}(K)+N_{L-2}(K-1)-Z_{L-2}(K-1)
\end{equation} 
 with $Z_0(K)=Z_1(K)=0$. 
  
These recursion relationships can be solved to find the estimated stack distribution as a function of length. We note that in order to implement their recursion formulas we had to make two simple minor adjustments to the original formulae, both of which were confirmed by Peter Stadler, one of the original authors of ref.~\cite{hofacker1998combinatorics}, in a private communication.  The adjustments were to set $Z_{L}(0)=0$, and to set $Z_L(b)=0$ if $L<m+2$. With this in hand,  we calculated these distributions using the constraint that the minimum loop size is $m=3$, which is also a constraint in the Vienna package. We find that the distributions are unimodal and narrowly peaked.  The peak  is roughly mid-way between the minimum value of $K$ (which is 0) and maximum value of $K$, which we found to be $\approx L/3$.  More precisely, we calculated the mean $\bar{K}_P$ for lengths $20 \ldots 200$, which can be fit to the following linear relationship:  $\bar{K}_P=(0.19152 \pm0.0001) L   - (0.476 \pm 0.001)$.
 The standard deviation can be fit quite accurately with $\sigma_{KP} = 0.17\sqrt{L}$ so that the width of the stack distribution, divided by the mean number of stacks, goes down as $1/\sqrt{L}$ for large $L$:  the peaks become relatively more sharp. This  scaling of the distribution width  with $L$ is the same as that found for $P_P(\sigma)$ and for the P-sampled stack distribution described above an shown in Fig 3 of the main text. 




Hofacker \emph{et al}~\cite{hofacker1998combinatorics}  also gave analytic asymptotic large $L$ estimates for the P-mean number of stacks. These estimates varied slightly depending on assumptions such as the value of $m$, and the type of bonding the RNA sequence was assumed to make. The most relevant of these estimates for our work is a P-mean which assumes the minimum loop size is $m=3$, and that permits G-U bonds. The corresponding large $L$  estimate is $\bar{K}_P = 0.1717L$.  

These two analytic approximations for $\bar{K}_P$ are fairly close to our estimate from combining the linear relationship between $K$ and $S=\log(\Omega)$ with $P_P(\Omega)$, namely $\bar{K}_P = 0.177 L  -0.443 $.   It is gratifying to find that these independent estimates, one using our empirical relationship between $K$ and $\log (\Omega)$ together with a fit to $P_P(\Omega)$,  and the others from combinatorics, are so similar. This gives us confidence that the P-sampled mean numbers of stacks is considerably larger than the G-sampled $\bar{K}_G$, which can be approximated as $\bar{K}_G = 0.074L - 0.377$, and which is close to the mean number of stacks found in the RNA STRAND database for solved structures.   All these different strands of evidence strengthen the case for our simple binomial fit approximation for $P_P(\Omega)$, which we could only partially sample.





 \,

\medskip
\section{ fRNAdb datasets}\label{datasets}

In this section, we describe in more detail  the contents the natural datasets for $L=20,55, 70\,  \& \, 126$, taken from fRNAdb~\cite{kin2007frnadb} database.

\subsection{$\bf L=20$}

The full fRNAdb database~\cite{kin2007frnadb} had $14350$ RNA sequences of length $L=20$ when we accessed it in  August 2014.  They were categorised as follows in the fRNAdb file: \\

\noindent
{\footnotesize
17 \% Putative conserved noncoding region (EvoFold)  \\
4.3 \% Piwi-interacting RNA (piRNA) \\
1.2 \% Mature microRNA \\
77 \% Fly small RNA \\
total number of sequences = 14350
}
\\

The vast majority of these sequences came from the fruit fly {\em Drosophila melanogaster}.  The sequences labeled as `putative conserved noncoding region'  are derived by bioinformatic methods and are expected to be functional ncRNA, but the function still needs to be confirmed.   We therefore removed these sequences from our $L=20$ data set.  We also removed  the piRNA for which it is not yet clear that the SS is important. 
We  only used ``Fly small RNA'' for $L=20$ from \emph{D. melanogaster}, which reduced the total number of sequences to $11,050$.  Of these, about 34$\%$ map to the unbound trivial structure, 
which is close to what is expected from random G-sampling since the trivial structure takes up about $P_T=33\%$ of sequences in the mapping  for $L=20$ (this percentage drops rapidly for increasing $L$).  It remains somewhat unclear what fraction of the  $7327$ non-trivial structures for  \emph{D. melanogaster} have SS that are important for function.  Nevertheless, we used this whole set to compare to the sampled structures.


\subsection{ $\bf L=55$}
 
 The contents of the fRNAdb file for $L=55$, when we accessed it in August 2014, were:\\ 

\noindent
{\footnotesize
0.4 \%   self-splicing ribozyme RNA \\
1.2 \%   non-protein coding (noncoding) transcript\\ 
0.4 \%   HIV gag stem loop 3 (GSL3) \\
0.2 \%   small nucleolar RNA (snoRNA) SNORD82 / SNORD83A / SNORD83B / U82 / U83A / U83B / Z25 \\
0.2 \%   Japanese encephalitis virus (JEV) hairpin structure \\
0.2 \%   precursor micro RNA (miRNA) mir-BART2 \\
0.8 \%   guide RNA (gRNA) \\
16.9 \%    Unagi (eel) L2 (UnaL2) LINE 3' element \\
3.7 \%   small nucleolar RNA (snoRNA) \\
 1.0 \%    Simian virus 40 late polyadenylation signal (SVLPA) \\
  0.2 \%    small nuclear RNA (snRNA) \\
 0.6 \%    Pyrococcus C/D box guide small nucleolar RNA (snoRNA) \\
 40.9 \%  Putative conserved noncoding region (EvoFold) \\
0.2 \%   C/D box guide small nucleolar RNA (snoRNA) HBII-240 / SNORD72 \\
16.3 \%    Putative conserved noncoding region (RNAz) \\
 0.2 \%    C/D box guide small nucleolar RNA (snoRNA) HBII-210 / SNORD69 \\
 0.2 \%    C/D box small nucleolar RNA (snoRNA) HBII-429 / SNORD100 \\
 2.4 \%    Trans-activation response element (TAR) \\
 0.2 \%  No description given   \\
1.0 \%   small non-messenger RNA (snmRNA) \\
2.2 \%   Hammerhead ribozyme (type III) \\
 0.2 \%    C/D box guide small nucleolar RNA (snoRNA) Z266 \\
0.2 \%   small nucleolar RNA (snoRNA) snoR185 \\
 0.2 \%    predicted precursor micro RNA (miRNA) HP-67 [false negative] \\
 0.2 \%    Rous sarcoma virus (RSV) primer binding site (PBS) \\
 0.4 \%    Ribosomal protein L13 leader \\
 6.5 \%    HIV Ribosomal frameshift signal \\
 0.8 \%    C/D box small nucleolar RNA (snoRNA) \\
 0.2 \%    putative archaeal H/ACA box small nucleolar RNA (snoRNA) u-46 \\
 1.2 \%    Group II intron \\
 0.2 \%    C/D box guide small nucleolar RNA (snoRNA) HBII-239 / SNORD71 \\
0.6 \%   Bovine leukaemia virus RNA packaging signal \\
0.2 \%   sok RNA \\
Total number of sequences = 509 
}
\\

We first note that over half (41\%, 16\% and 0.2\%) are labelled putative.   Of the remaining non-putative RNA described above, 17\% are eel LINE elements, which are known to have a stem-loop structure that is important for function. The next most abundant (7\%) are HIV ribsomal frameshift signal RNA, whose structure is important for interacting with the ribosome. Next is snoRNA (4\%) which has conserved and functionally relevant secondary structures.  
  The Hammerhead ribozyme (2.2\%) and Group 2 intron (1.2\%) are both  catalytic ribozymes and so have SS important for their functions.   Smaller abundance contributions include Bovine leukaemia virus signal, precursor RNA, JEV hairpin structure, TAR and RSV, all of which are known to have functionally relevant SS. In sum, for this dataset, a significant fraction of the non-putative RNA have known functionally relevant SS. 
  
     In the main text Fig 2, we plot the NS size distribution for 504 structures.   There were 5 structures (about $1 \%$) which either had non-standard nucleotides or else the NSSE estimator did not converge.
We also investigated the $L=55$ dataset with all putative RNA discarded (leaving 213 sequences remaining). As argued above, most of the rest of the sequences have SS that are thought to be important for function.    The distributions for this curated set of natural RNA structures are  not notably different from the distributions which included putative ncRNA and therefore remain very close to the G-sampled distribution (Extended Data Figs. 1,2.)

%
%
%
%
%
%
%
%
%


\subsection{ $\bf L=70$}
The contents of the fRNAdb file for $L=70$ we accessed August 2014 are:\\ 

\noindent
{\footnotesize
 0.04 \%  precursor micro RNA (miRNA) mir-99b  \\
0.04 \%  small nucleolar RNA (snoRNA) Esmeraldo SLA1  \\
0.91 \%  transfer RNA (tRNA), GCC (Gly/G) Glycine  \\
0.3 \%  non-protein coding (noncoding) transcript  \\
0.04 \%  precursor micro RNA (miRNA) mir-N367  \\
0.69 \%  transfer RNA (tRNA), CAT (Met/M) Methionine  \\
0.04 \%  precursor micro RNA (miRNA) mir-140  \\
0.09 \%  C/D box guide small nucleolar RNA (snoRNA) SNORD38 / U38  \\
0.04 \%  predicted precursor micro RNA (miRNA) RP-88 [false negative]  \\
0.04 \%  precursor micro RNA (miRNA) mir-16c  \\
0.04 \%  precursor micro RNA (miRNA) mir-24-1  \\
0.04 \%  precursor micro RNA (miRNA) mir-303  \\
0.04 \%  C/D box guide small nucleolar RNA (snoRNA) SNORD18A / U18A  \\
0.04 \%  C/D box guide small nucleolar RNA (snoRNA) Z37  \\
1.09 \%  transfer RNA (tRNA), TTC (Glu/E) Glutamic acid  \\
5.12 \%  Putative conserved noncoding region (EvoFold)  \\
0.04 \%  precursor micro RNA (miRNA) mir-K12-10b  \\
0.04 \%  precursor micro RNA (miRNA) mir-200b  \\
0.04 \%  transfer RNA (tRNA), GCG (Arg/R) Arginine  \\
0.04 \%  C/D box small nucleolar RNA (snoRNA) HBII-429 / SNORD100  \\
0.04 \%  precursor micro RNA (miRNA) mir-8  \\
0.04 \%  precursor micro RNA (miRNA) mir-K12-3  \\
0.04 \%  precursor micro RNA (miRNA) mir-199a-1  \\
0.61 \%    \\
0.13 \%  small non-messenger RNA (snmRNA)  \\
0.17 \%  C/D box guide small nucleolar RNA (snoRNA) SNORD113 / SNORD114  \\
0.04 \%  precursor micro RNA (miRNA) mir-196 / mir-196-1 / mir-196a-1  \\
0.04 \%  predicted precursor micro RNA (miRNA) HP-27 [false negative]  \\
0.13 \%  C/D box guide small nucleolar RNA (snoRNA) SNORD82 / U82 / Z25  \\
0.04 \%  precursor micro RNA (miRNA) mir-412  \\
0.04 \%  Plasmid R1162 RNA  \\
2.78 \%  transfer RNA (tRNA), TCG (Arg/R) Arginine  \\
0.04 \%  precursor micro RNA (miRNA) mir-23a  \\
0.04 \%  precursor micro RNA (miRNA) mir-K12-8  \\
0.04 \%  C/D box guide small nucleolar RNA (snoRNA) SNORD63 / U63  \\
0.04 \%  precursor micro RNA (miRNA) mir-30 / mir-30d  \\
0.04 \%  precursor micro RNA (miRNA) mir-US25-1  \\
0.3 \%  precursor micro RNA (miRNA) lin-4  \\
0.04 \%  predicted precursor micro RNA (miRNA) MP-64 [false negative]  \\
0.04 \%  C/D box guide small nucleolar RNA (snoRNA) Z40  \\
0.13 \%  transfer RNA (tRNA), GGT (Thr/T) Threonine  \\
0.04 \%  precursor micro RNA (miRNA) mir-10b-2  \\
14.41 \%  transfer RNA (tRNA), TGG (Pro/P) Proline  \\
0.04 \%  C/D box guide small nucleolar RNA (snoRNA) SNORD2 / snR39B  \\
0.04 \%  predicted precursor micro RNA (miRNA) HP-61 [all confirmed]  \\
0.04 \%  precursor micro RNA (miRNA) mir-128a  \\
1.13 \%  transfer RNA (tRNA), TGC (Ala/A) Alanine  \\
0.09 \%  Enterovirus 5' cloverleaf cis-acting replication element  \\
0.04 \%  S-element  \\
0.04 \%  C/D box guide small nucleolar RNA (snoRNA) R38  \\
0.09 \%  transfer RNA (tRNA), CAA (Leu/L) Leucine  \\
0.04 \%  small nucleolar RNA (snoRNA)  \\
1.17 \%  transfer RNA (tRNA), TGA (Ser/S) Serine  \\
0.04 \%  C/D box guide small nucleolar RNA (snoRNA) SNORD18B / U18B  \\
0.04 \%  C/D box small nucleolar RNA (snoRNA) 14q(I-1) / SNORD113-1  \\
0.04 \%  C/D box guide small nucleolar RNA (snoRNA) SNORD41 / U41  \\
0.04 \%  precursor micro RNA (miRNA) mir-US33  \\
3.69 \%  Putative conserved noncoding region (RNAz)  \\
0.13 \%  Selenocysteine insertion sequence  \\
0.04 \%  Tombusvirus internal replication element (IRE)  \\
0.04 \%  precursor micro RNA (miRNA) mir-10b-1  \\
0.04 \%  C/D box guide small nucleolar RNA (snoRNA) SNORD34 / U34  \\
0.04 \%  C/D box guide small nucleolar RNA (snoRNA) SNORD51 / U51  \\
0.3 \%  HIV primer binding site (PBS)  \\
0.04 \%  suhB  \\
0.52 \%  HIV gag stem loop 3 (GSL3)  \\
0.04 \%  precursor micro RNA (miRNA) mir-219  \\
1.17 \%  transfer RNA (tRNA), GAA (Phe/F) Phenylalanine  \\
0.22 \%  C/D box small nucleolar RNA (snoRNA)  \\
0.09 \%  C/D box guide small nucleolar RNA (snoRNA) SNORD77 / U77  \\
0.04 \%  sar RNA  \\
3.6 \%  Group II intron  \\
0.04 \%  C/D box guide small nucleolar RNA (snoRNA) Me28S-Cm788  \\
0.13 \%  transfer RNA (tRNA), TCT (Arg/R) Arginine  \\
0.09 \%  Vimentin 3' untraslated region (UTR) protein-binding region  \\
1.3 \%  transfer RNA (tRNA), GAT (Ile/I) Isoleucine  \\
0.04 \%  transfer RNA (tRNA), AGA (Ser/S) Serine  \\
0.09 \%  precursor micro RNA (miRNA) mir-383  \\
0.04 \%  precursor micro RNA (miRNA) mir-M30  \\
0.13 \%  C/D box guide small nucleolar RNA (snoRNA) SNORD30 / U30  \\
0.04 \%  small nucleolar RNA (snoRNA) Sylvio X10 SLA1  \\
0.04 \%  sroB RNA  \\
0.04 \%  small nuclear RNA (snRNA)  \\
0.04 \%  transfer RNA (tRNA), CAC (Val/V) Valine  \\
0.22 \%  C/D box guide small nucleolar RNA (snoRNA) SNORD18 / U18  \\
6.77 \%  Selenocysteine transfer RNA (tRNA), TCA  \\
0.13 \%  precursor micro RNA (miRNA) mir-30  \\
0.04 \%  precursor micro RNA (miRNA) mir-K12-10a  \\
0.17 \%  C/D box guide small nucleolar RNA (snoRNA) U2-30  \\
0.09 \%  precursor micro RNA (miRNA) mir-BART1  \\
0.04 \%  predicted precursor micro RNA (miRNA) RP-104 [false negative]  \\
0.35 \%  RNA-OUT  \\
0.04 \%  transfer RNA (tRNA), GGC (Ala/A) Alanine  \\
0.04 \%  transfer RNA (tRNA), AGC (Ala/A) Alanine  \\
10.76 \%  transfer RNA (tRNA), GTA (Tyr/Y) Tyrosine  \\
0.17 \%  Rous sarcoma virus (RSV) primer binding site (PBS)  \\
0.04 \%  C/D box guide small nucleolar RNA (snoRNA) SNORD58 / U58  \\
4.9 \%  transfer RNA (tRNA), TAC (Val/V) Valine  \\
0.04 \%  precursor micro RNA (miRNA) mir-694  \\
0.09 \%  C/D box guide small nucleolar RNA (snoRNA) SNORD24 / U24  \\
0.09 \%  precursor micro RNA (miRNA) mir-145  \\
0.04 \%  C/D box guide small nucleolar RNA (snoRNA) SNORD56 / U56  \\
0.87 \%  Tombus virus defective interfering (DI) RNA region 3  \\
0.04 \%  precursor micro RNA (miRNA) mir-30d  \\
0.43 \%  Flavivirus DB element  \\
0.09 \%  precursor micro RNA (miRNA) mir-361  \\
10.11 \%  transfer RNA (tRNA), TTG (Gln/Q) Glutamine  \\
0.04 \%  transfer RNA (tRNA), GGG (Pro/P) Proline  \\
0.35 \%  transfer RNA (tRNA), GTT (Asn/N) Asparagine  \\
0.04 \%  transfer RNA (tRNA), GCT (Ser/S) Serine  \\
0.04 \%  precursor micro RNA (miRNA) mir-K12-4  \\
0.69 \%  transfer RNA (tRNA), GTC (Asp/D) Aspartic acid  \\
0.04 \%  transfer RNA (tRNA), AGG (Pro/P) Proline  \\
3.26 \%  transfer RNA (tRNA), TAG (Leu/L) Leucine  \\
0.04 \%  small nucleolar RNA (snoRNA) Trypanosoma brucei RNA 9 (TBR9)  \\
0.04 \%  precursor micro RNA (miRNA) mir-125b  \\
0.04 \%  transfer RNA (tRNA), ACC (Gly/G) Glycine  \\
0.04 \%  precursor micro RNA (miRNA) mir-137  \\
0.09 \%  transfer RNA (tRNA), CCC (Gly/G) Glycine  \\
0.04 \%  C/D box guide small nucleolar RNA (snoRNA) SNORD73 / U73  \\
0.04 \%  precursor micro RNA (miRNA) mir-127  \\
0.04 \%  transfer RNA (tRNA), ACA (Cys/C) Cysteine  \\
0.04 \%  precursor micro RNA (miRNA) mir-M3  \\
0.04 \%  msr RNA  \\
0.04 \%  guide RNA (gRNA)  \\
0.35 \%  transfer RNA (tRNA), TAA (Leu/L) Leucine  \\
0.04 \%  predicted precursor micro RNA (miRNA) HN-3 [false negative]  \\
0.04 \%  Glycine riboswitch  \\
0.04 \%  Y RNA  \\
0.13 \%  precursor micro RNA (miRNA) mir-183  \\
3.26 \%  transfer RNA (tRNA), TGT (Thr/T) Threonine  \\
0.04 \%  precursor micro RNA (miRNA) mir-206  \\
0.04 \%  small nucleolar RNA (snoRNA) SNORD21 / U21  \\
0.26 \%  transfer RNA (tRNA) with undetermined isotype  \\
0.13 \%  C/D box guide small nucleolar RNA (snoRNA) Z12  \\
0.04 \%  precursor micro RNA (miRNA) mir-13b-2  \\
0.04 \%  precursor micro RNA (miRNA) mir-K12-5  \\
0.52 \%  transfer RNA (tRNA), CTT (Lys/K) Lysine  \\
0.17 \%  precursor micro RNA (miRNA) mir-32  \\
0.04 \%  small nucleolar RNA (snoRNA) CL Brenner SLA1  \\
0.04 \%  precursor micro RNA (miRNA) mir-147-1 / mir-147-2  \\
0.04 \%  precursor micro RNA (miRNA) mir-790  \\
0.04 \%  precursor micro RNA (miRNA) mir-302a  \\
0.09 \%  C/D box guide small nucleolar RNA (snoRNA) SNORD50 / SNORD50A / U50  \\
0.13 \%  transfer RNA (tRNA), GCA (Cys/C) Cysteine  \\
0.04 \%  precursor micro RNA (miRNA) mir-369  \\
5.69 \%  transfer RNA (tRNA), GTG (His/H) Histidine  \\
0.09 \%  transfer RNA (tRNA), CCA (Trp/W) Tryptophan  \\
0.04 \%  small nucleolar RNA (snoRNA) SNORD50A / U50  \\
0.26 \%  Retroviral Psi packaging element  \\
0.95 \%  transfer RNA (tRNA), TTT (Lys/K) Lysine  \\
0.26 \%  precursor micro RNA (miRNA) mir-10  \\
3.43 \%  transfer RNA (tRNA), TCC (Gly/G) Glycine  \\
0.04 \%  C/D box guide small nucleolar RNA (snoRNA) R160  \\
Total number of sequences = 2304
}
\\

The descriptions above show that the largest category  comes from tRNA, which clearly has a functionally relevant structure. Additionally there are many precursor microRNA, Group 2 introns, and snoRNA,  which are thought have functionally relevant structures. Finally, this dataset contains a relatively small proportion ($\sim$9\%) of `putative' RNA. In sum, for $L=70$, the vast majority of RNA in the dataset have SS that are thought to be important for function.    In the main text Fig 2, we plot the NS size distribution for 2263 structures.   There were 41  structures ($1.8 \%$) either with non-standard nucleotides or where the NSSE estimator did not converge.  Given that the majority of structures have SS thought to be important for function, we did not separately plot curated data as we did for $L=55$ and $L=126$.

\subsection{ $\bf L=126$}
The contents of the fRNAdb file for $L=126$ we accessed August 2014 are:\\ 

\noindent
{\footnotesize
1.57 \%  5S ribosomal RNA  \\
0.31 \%  H/ACA box small nucleolar RNA (snoRNA)  \\
0.31 \%  predicted precursor micro RNA (miRNA) HP-58 [false negative]  \\
0.31 \%  BC200 RNA  \\
1.57 \%  Thiamin pyrophosphate (TPP) riboswitch (THI element)  \\
0.31 \%  H/ACA box guide small nucleolar RNA (snoRNA) SNORA68 / U68  \\
0.63 \%  H/ACA box guide small nucleolar RNA (snoRNA) ACA27 / SNORA27  \\
5.33 \%    \\
2.82 \%  non-protein coding (noncoding) transcript  \\
0.31 \%  precursor micro RNA (miRNA) mir-29a-2  \\
0.31 \%  precursor micro RNA (miRNA) mir-393  \\
0.31 \%  H/ACA box guide small nucleolar RNA (snoRNA) ACA28 / SNORA28  \\
0.31 \%  Pseudomonas small noncoding RNA (sRNA) P9  \\
15.36 \%  5.8S ribosomal RNA (rRNA)  \\
0.31 \%  precursor micro RNA (miRNA) mir-394a  \\
0.31 \%  Picornavirus internal ribosome entry site (IRES)  \\
0.31 \%  precursor micro RNA (miRNA) mir-1223  \\
0.31 \%  H/ACA box guide small nucleolar RNA (snoRNA) SNORA70 / U70  \\
1.25 \%  Ribosomal protein L20 leader  \\
0.31 \%  H/ACA box guide small nucleolar RNA (snoRNA) ACA46 / SNORA46  \\
0.31 \%  precursor micro RNA (miRNA) mir-29a-1  \\
0.31 \%  precursor micro RNA (miRNA) mir-399b / mir-399c  \\
0.31 \%  U4 spliceosomal RNA  \\
0.31 \%  precursor micro RNA (miRNA) mir-859  \\
0.31 \%  precursor micro RNA (miRNA) mir-171e  \\
0.31 \%  HgcE RNA  \\
7.52 \%  small nuclear RNA (snRNA)  \\
0.31 \%  Enteroviral 3' untraslated region (UTR) element  \\
8.15 \%  Putative conserved noncoding region (EvoFold)  \\
0.31 \%  Threonine operon leader  \\
0.31 \%  precursor micro RNA (miRNA) mir-860  \\
0.31 \%  HgcC family RNA  \\
33.86 \%  Putative conserved noncoding region (RNAz)  \\
0.63 \%  ydaO / yuaA leader  \\
0.31 \%  precursor micro RNA (miRNA) mir-169p  \\
0.31 \%  precursor micro RNA (miRNA) mir-396b  \\
0.31 \%  precursor micro RNA (miRNA) mir-169g  \\
0.31 \%  precursor micro RNA (miRNA) mir-866  \\
0.63 \%  Hepatitis C alternative reading frame stem-loop  \\
0.31 \%  precursor micro RNA (miRNA) mir-169c  \\
0.63 \%  small non-messenger RNA (snmRNA)  \\
0.31 \%  FMN riboswitch (RFN element)  \\
0.31 \%  H/ACA box guide small nucleolar RNA (snoRNA) ACA42 / SNORA42  \\
0.31 \%  ylbH leader  \\
0.94 \%  H/ACA box guide small nucleolar RNA (snoRNA) ACA25 / SNORA25  \\
0.63 \%  Ribosomal protein L10 leader  \\
0.94 \%  C/D box guide small nucleolar RNA (snoRNA) SNORD14 / U14  \\
0.31 \%  H/ACA box guide small nucleolar RNA (snoRNA) ACA44 / SNORA44  \\
0.31 \%  precursor micro RNA (miRNA) mir-164  \\
0.31 \%  repair RNA  \\
0.31 \%  H/ACA box guide small nucleolar RNA (snoRNA) ACA58 / SNORA58  \\
0.31 \%  precursor micro RNA (miRNA) mir-190  \\
0.31 \%  GcvB RNA  \\
0.31 \%  precursor micro RNA (miRNA) mir-172b  \\
0.31 \%  C/D box guide small nucleolar RNA (snoRNA) SNORD118 / U8  \\
0.31 \%  precursor micro RNA (miRNA) mir-166g  \\
0.31 \%  H/ACA box small nucleolar RNA (snoRNA) ACA63 / SNORA77  \\
0.31 \%  precursor micro RNA (miRNA) mir-156e  \\
0.31 \%  precursor micro RNA (miRNA) mir-399e  \\
0.63 \%  C/D box guide small nucleolar RNA (snoRNA) R9  \\
0.63 \%  Group II intron  \\
0.63 \%  yybP-ykoY leader  \\
0.31 \%  H/ACA box guide small nucleolar RNA (snoRNA) ACA61 / SNORA61  \\
0.63 \%  C/D box guide small nucleolar RNA (snoRNA) SNORD22 / U22  \\
0.94 \%  H/ACA box guide small nucleolar RNA (snoRNA) E3 / SNORA63  \\
0.31 \%  H/ACA box guide small nucleolar RNA (snoRNA) ACA40 / SNORA40  \\
Total number of sequences = 319
}
\\

 In the main text, Fig, 2, we  plotted 291 of the 319  structures.  For 28 structures ($9.6\%$) either there were non-standard nucleotides or the NSSE estimator did not converge.  We note that the NSSE has more difficulty converging for longer $L$, which is one reason this length was the longest we study here.  About $40\%$ of the structures are labelled putative, that is they are expected to be functional ncRNA, but the function still needs to be confirmed.  When these were left out of the data, we ended up with $172$ sequences for which the NS size of the structures could be found. From their descriptions above, most have SS that are thought to be important for function.   As can be seen in Extended Data Figs 1,2, The NS size distribution for this curated data set is very close to the G-sampled estimate and the the full data set.  
 
   One concern could be that some sequences in the database are from closely related organisms, which could bias the data.  To double check, we took one sequence for each of the $66$ categories above from the  $L=126$ fRNAdb file. Of these, 1 sequence contained non-standard nucleotides, and 4 failed in the NNSE. This leaves $61$ sequences. The mean $S=\log\Omega$ for these $61$ sequences is $ \bar{S} = 51.92$, and the standard deviation is $1.76$.  This is very close to the results obtained when using all $291$  sequences that we could fold where we find that the mean is $\bar{S} = 51.94$, and the standard deviation is $1.94$.  Both these results are close to our estimates from $1000$ randomly sampled $L=126$ sequences where we found $\bar{S}_G =  51.5$ with standard deviation $\bar{\sigma}_G = 2.1$.

\section{Supplementary Figures}

\renewcommand{\tablename}{{}}

\setcounter{table}{0}
\renewcommand{\thetable}{{\bf\hspace*{-0.2cm}  Table} S\arabic{table}}
\
\renewcommand{\figurename}{{}}

\renewcommand\thefigure{{\bf\hspace*{-0.2cm}  Figure} S\arabic{figure}}    
\setcounter{figure}{0}    
\

 \begin{figure*}[h]
\includegraphics[height=3.4cm,width=4.65cm]{L20-simple.pdf}
\includegraphics[height=3.4cm,width=4.65cm]{L40.pdf}
\includegraphics[height=3.4cm,width=4.65cm]{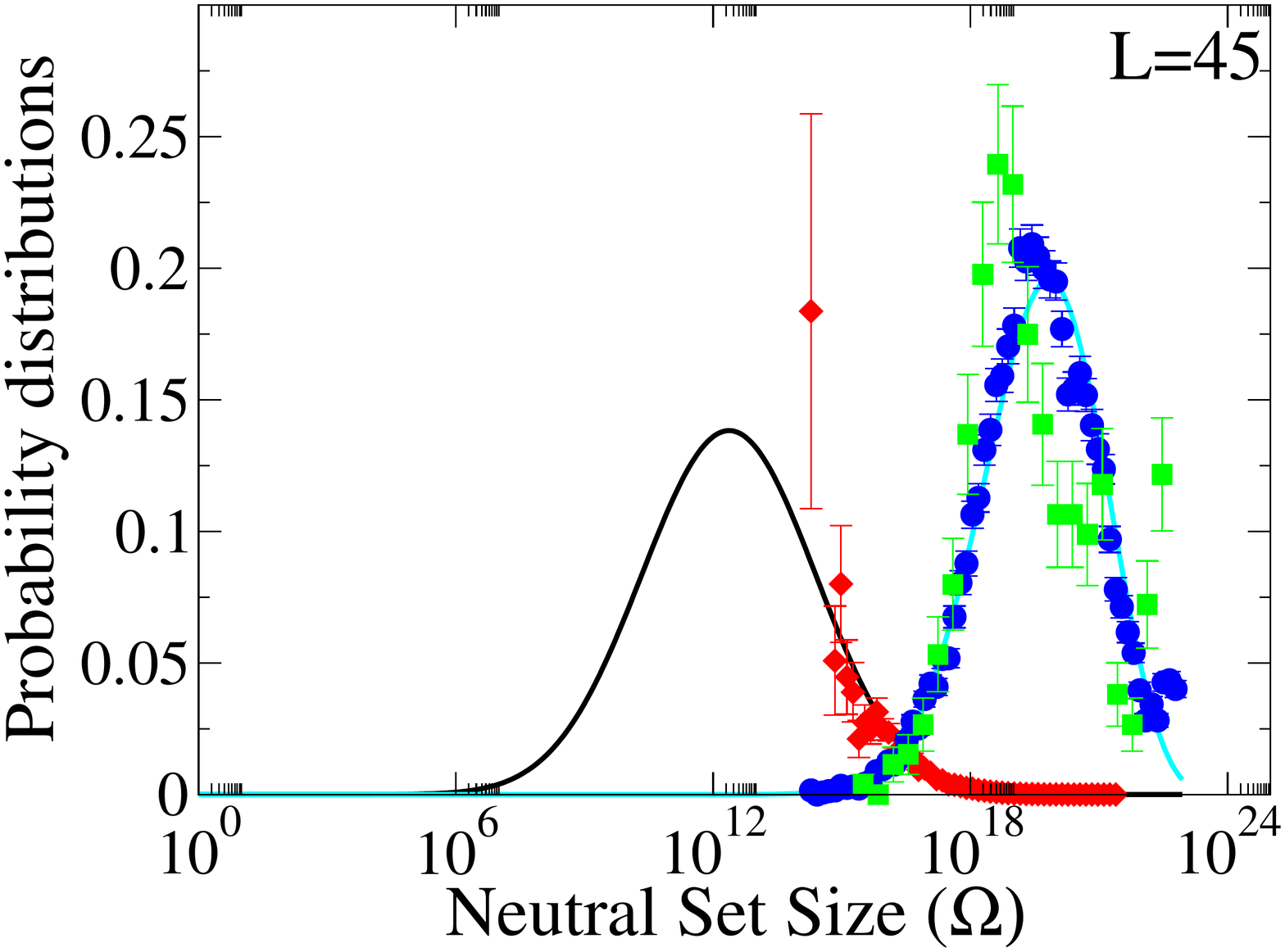}
\includegraphics[height=3.4cm,width=4.65cm]{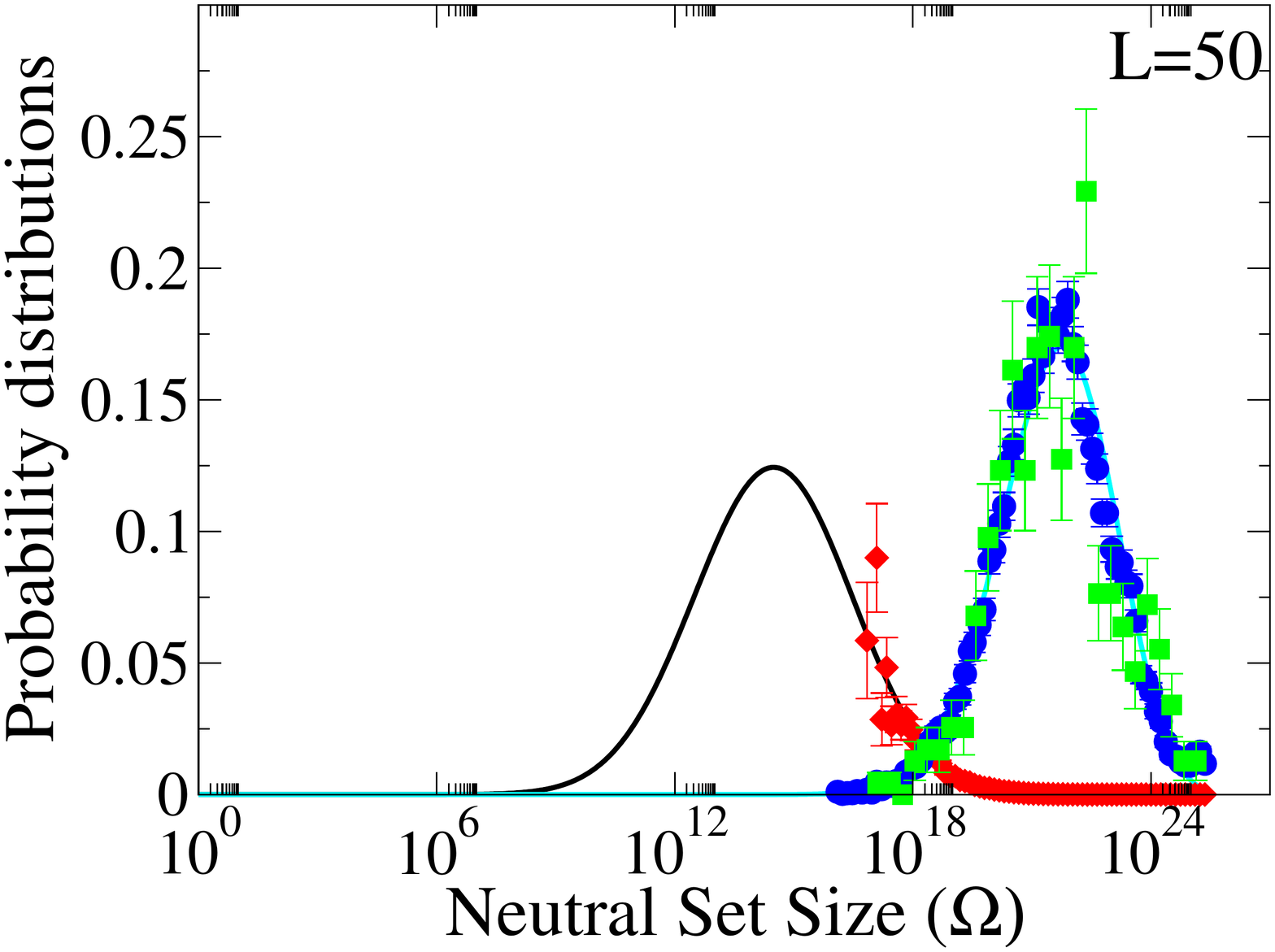}
\includegraphics[height=3.4cm,width=4.65cm]{L55-simple.pdf}
\includegraphics[height=3.4cm,width=4.65cm]{L60.pdf}
\includegraphics[height=3.4cm,width=4.65cm]{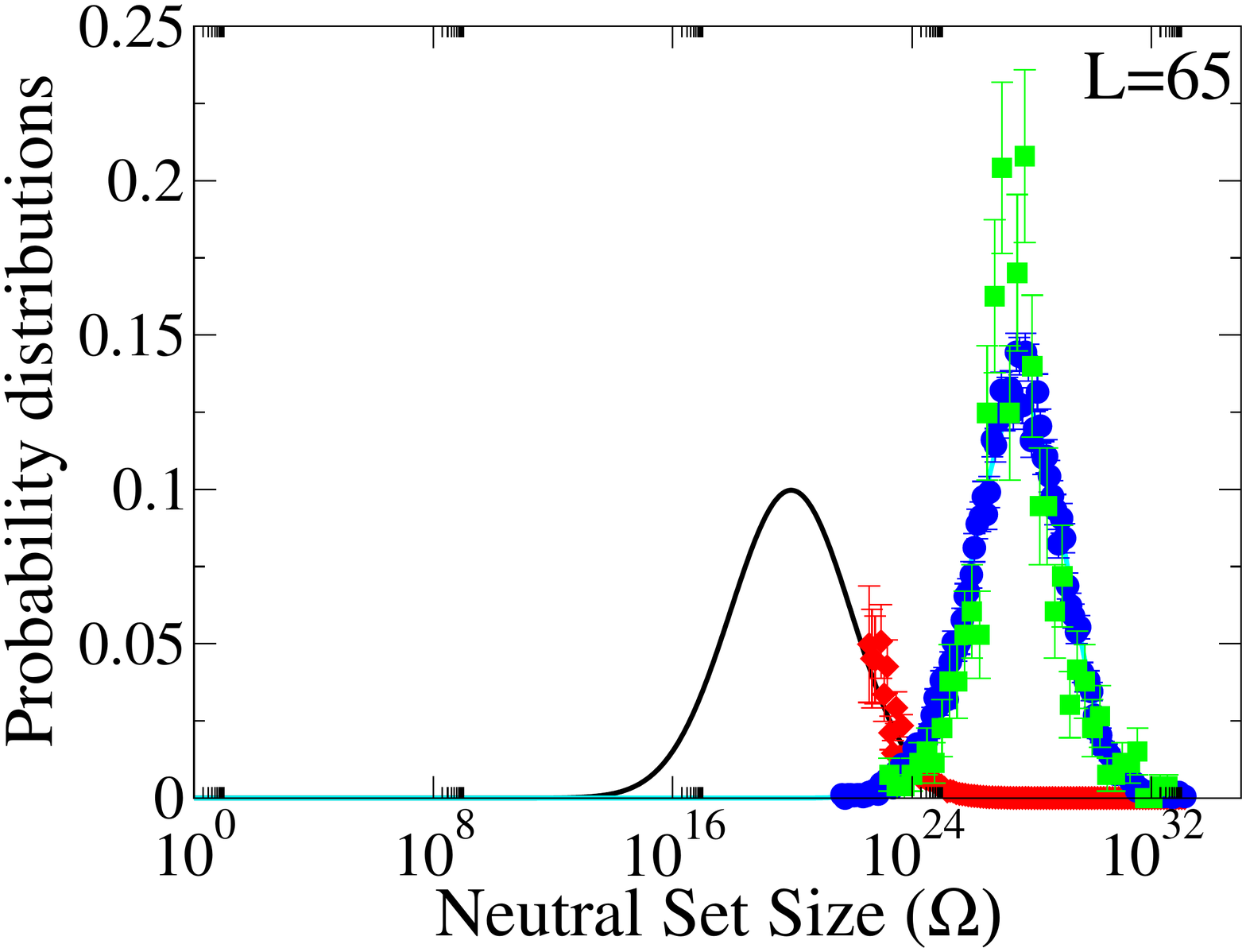}
\includegraphics[height=3.4cm,width=4.65cm]{L70-simple.pdf}
\includegraphics[height=3.4cm,width=4.65cm]{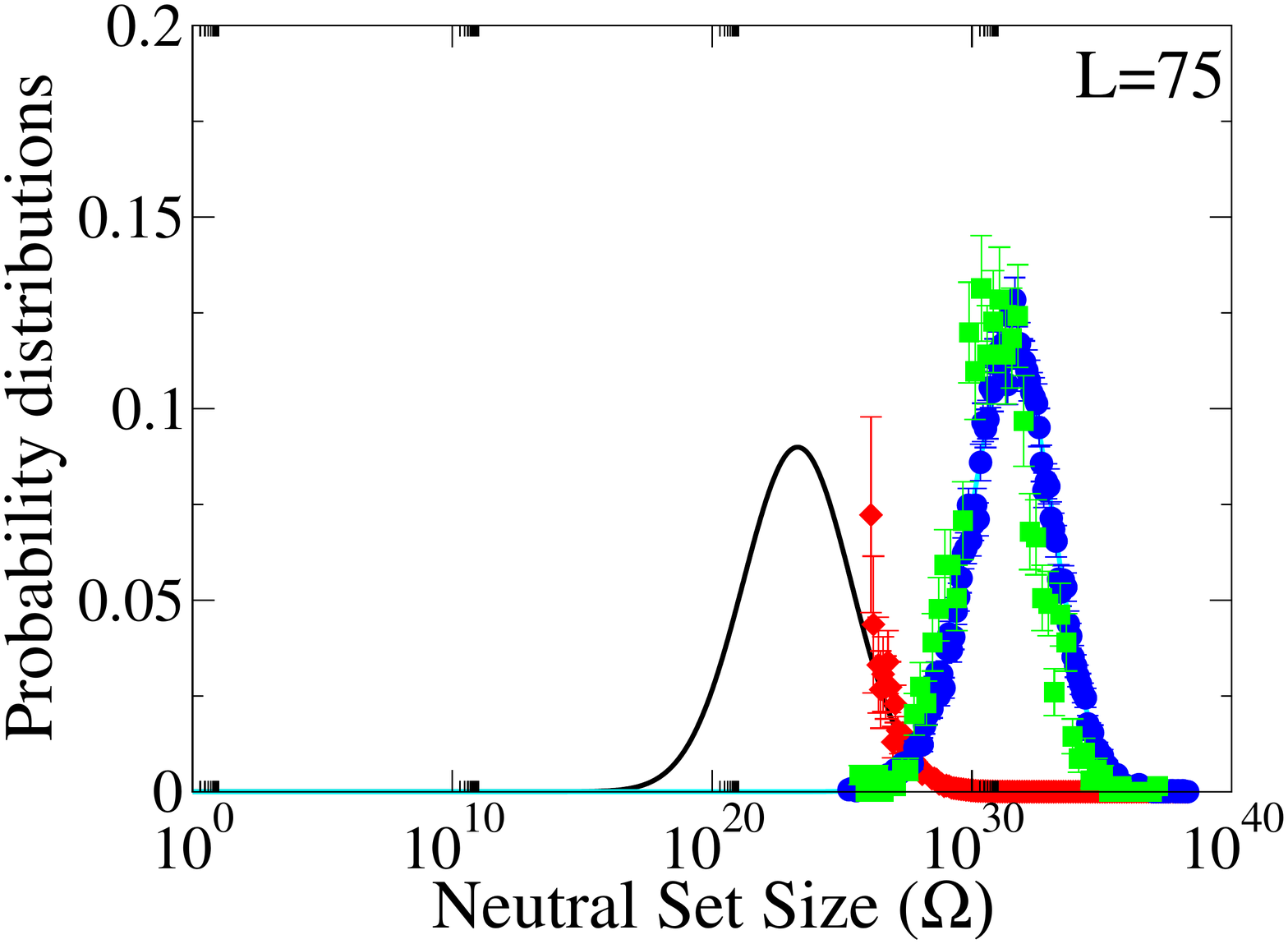}
\includegraphics[height=3.4cm,width=4.65cm]{L80.pdf}
\includegraphics[height=3.4cm,width=4.65cm]{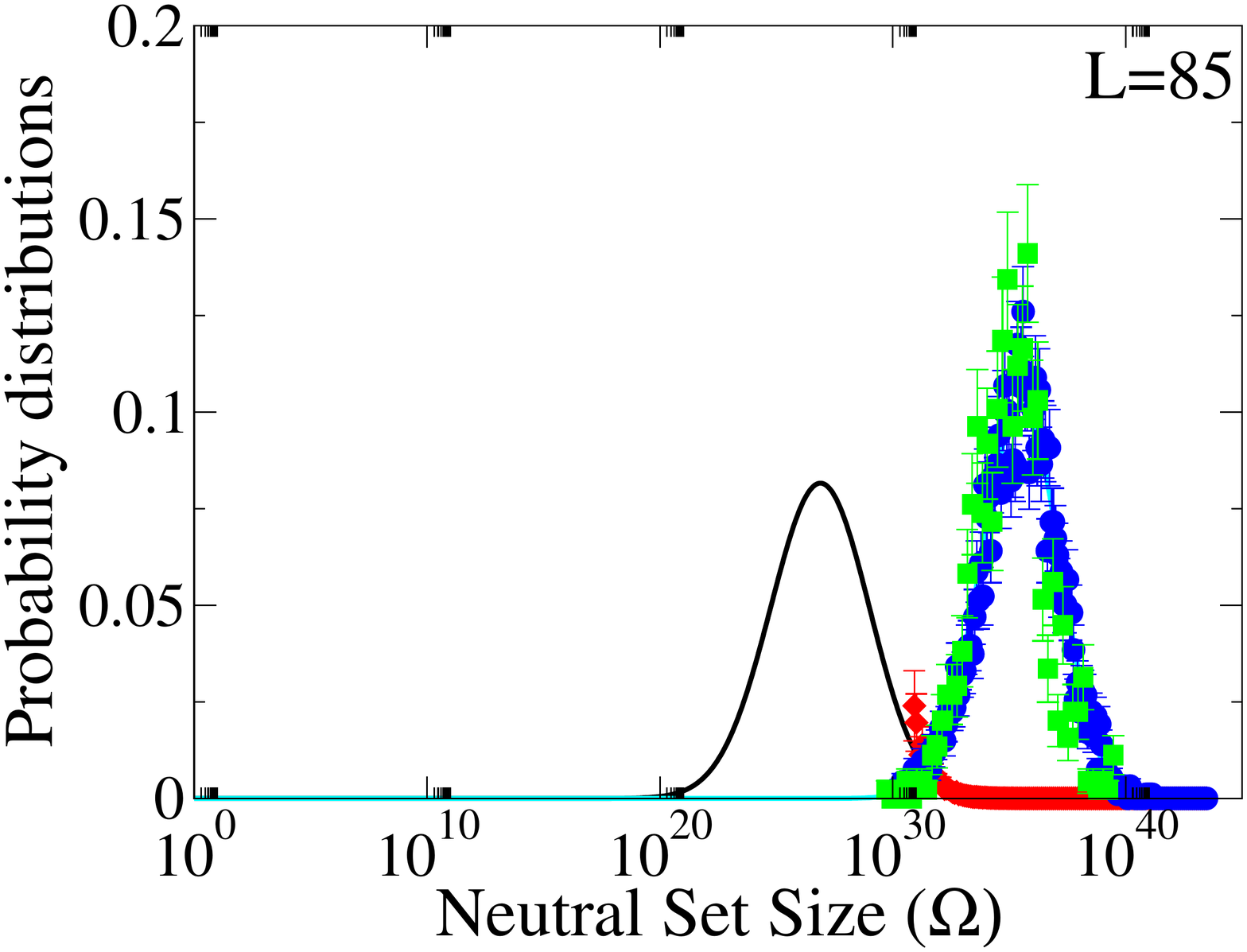}
\includegraphics[height=3.4cm,width=4.65cm]{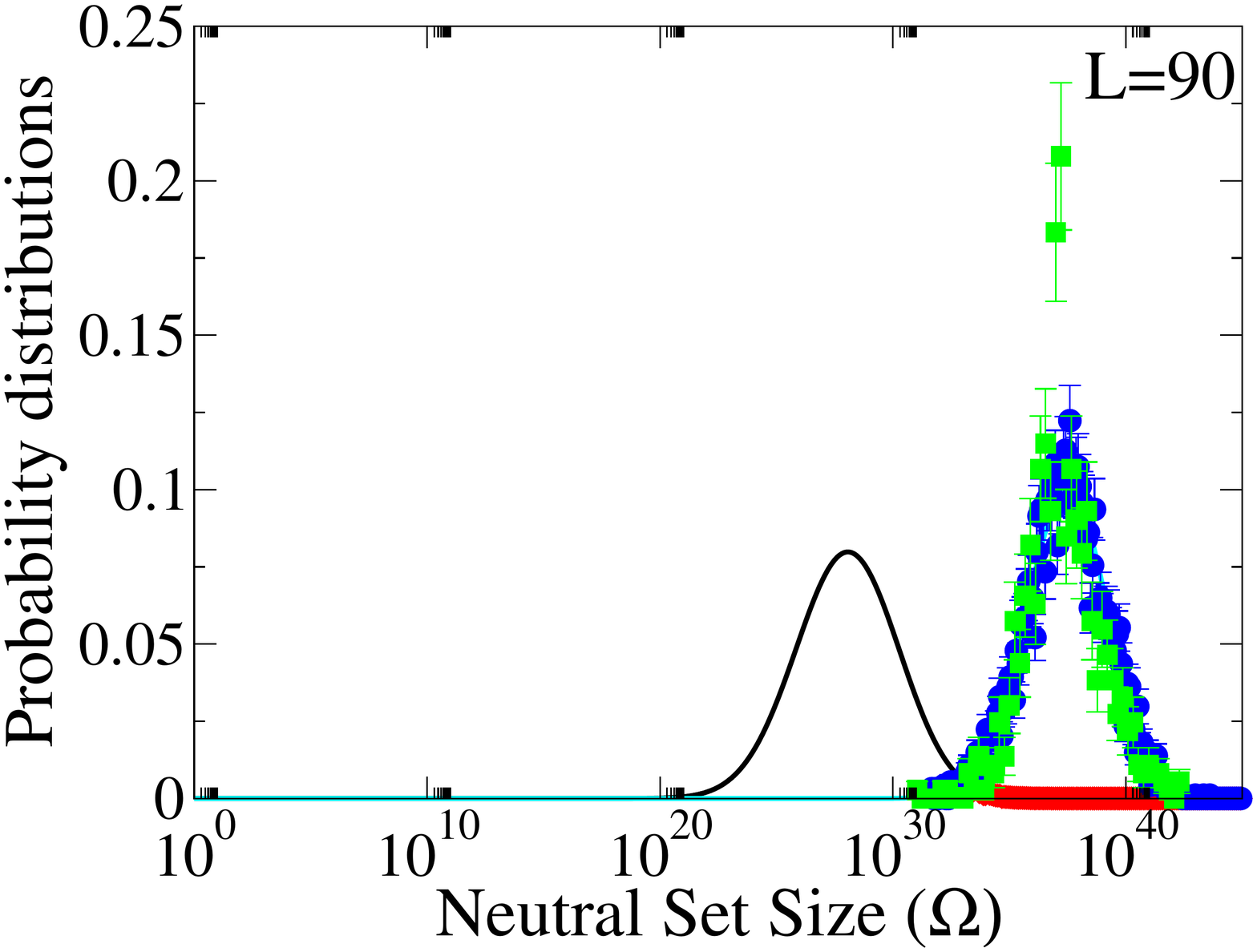}
\includegraphics[height=3.4cm,width=4.65cm]{L100.pdf}
\includegraphics[height=3.4cm,width=4.65cm]{L126-simple.pdf}
\caption{ {\bf Neutral set size distributions for RNA secondary structures}.  Data as in the main text, Fig.~3, but with a larger set of lengths. 
 Randomly sampled sequences are used to estimate  $P_P(\Omega)$ (red diamonds) and $P_G(\Omega)$ (blue circles). Black and cyan lines are theoretical approximations to  $P_P(\Omega)$ and $P_G(\Omega)$ respectively.  Green squares denote the probability distribution for natural ncRNA secondary structures taken from the fRNAdb \cite{kin2007frnadb} database (green squares), which are close to the   G-sampled $P_G(\Omega)$, but quite far from the P-sampled $P_P(\Omega)$. The number of natural structures plotted are:  $7327$ for $L=20$;\,\, $658$ for $L=40$;\,\, $533$ for $L=45$;\,\, $472$ for $L=50$;\,\, $504$ for $L=55$;\,\, $350$ for $ L=60$;\,\, $529$ for $L=65$;\,\, $2263$ for $L=70$;\,\, $1385$ for $L=75$;\,\, $553$ for $L=80$;\,\, $893$ for $L=85$;\,\, $731$ for $L=90$;\,\, $558$ for $L=95$;\,\, $891$ for $L=100$;\,\, and $291$ for $L=126$. 
  In each case all structures in the  fRNAdb \cite{kin2007frnadb} database are used, except for $L=20$ where only structures for {\it Drosophila melanogaster} are used, and for $L=55$ and $L=126$ where we also plot a curated data set  of 213 and 172 structures respectively  (magenta triangles) where the SS is known to be important (see subsection~\ref{datasets}).  Smaller numbers of data lead to larger binning error bars, but  the curated sets are clearly very similar to the full set of fRNAdb structures.     }
\label{fig:X1}
\end{figure*}

\begin{figure*}[h]
\includegraphics[height=3.4cm,width=4.65cm]{L20-simple-log.pdf}
\includegraphics[height=3.4cm,width=4.65cm]{L40-log.pdf}
\includegraphics[height=3.4cm,width=4.65cm]{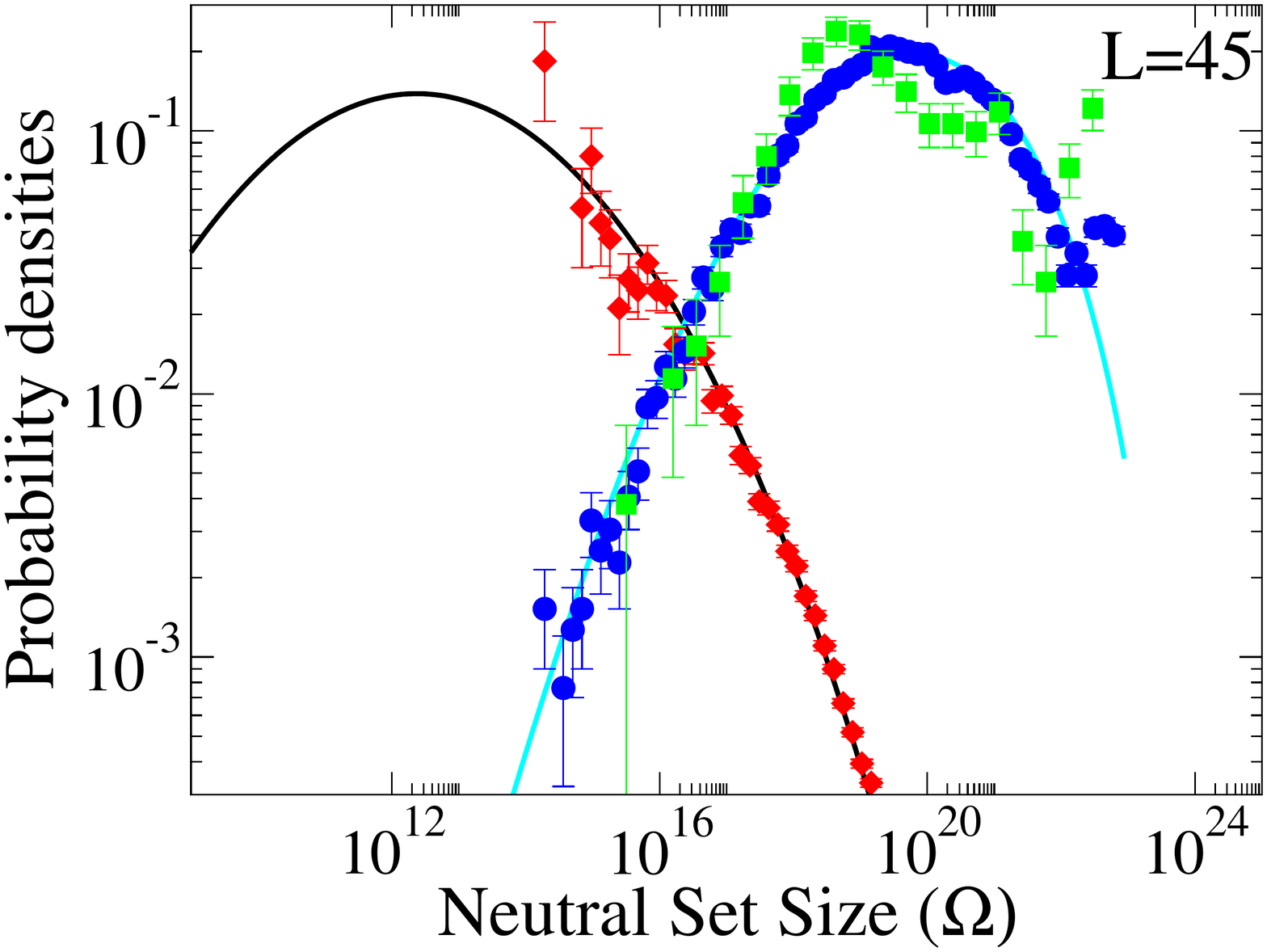}
\includegraphics[height=3.4cm,width=4.65cm]{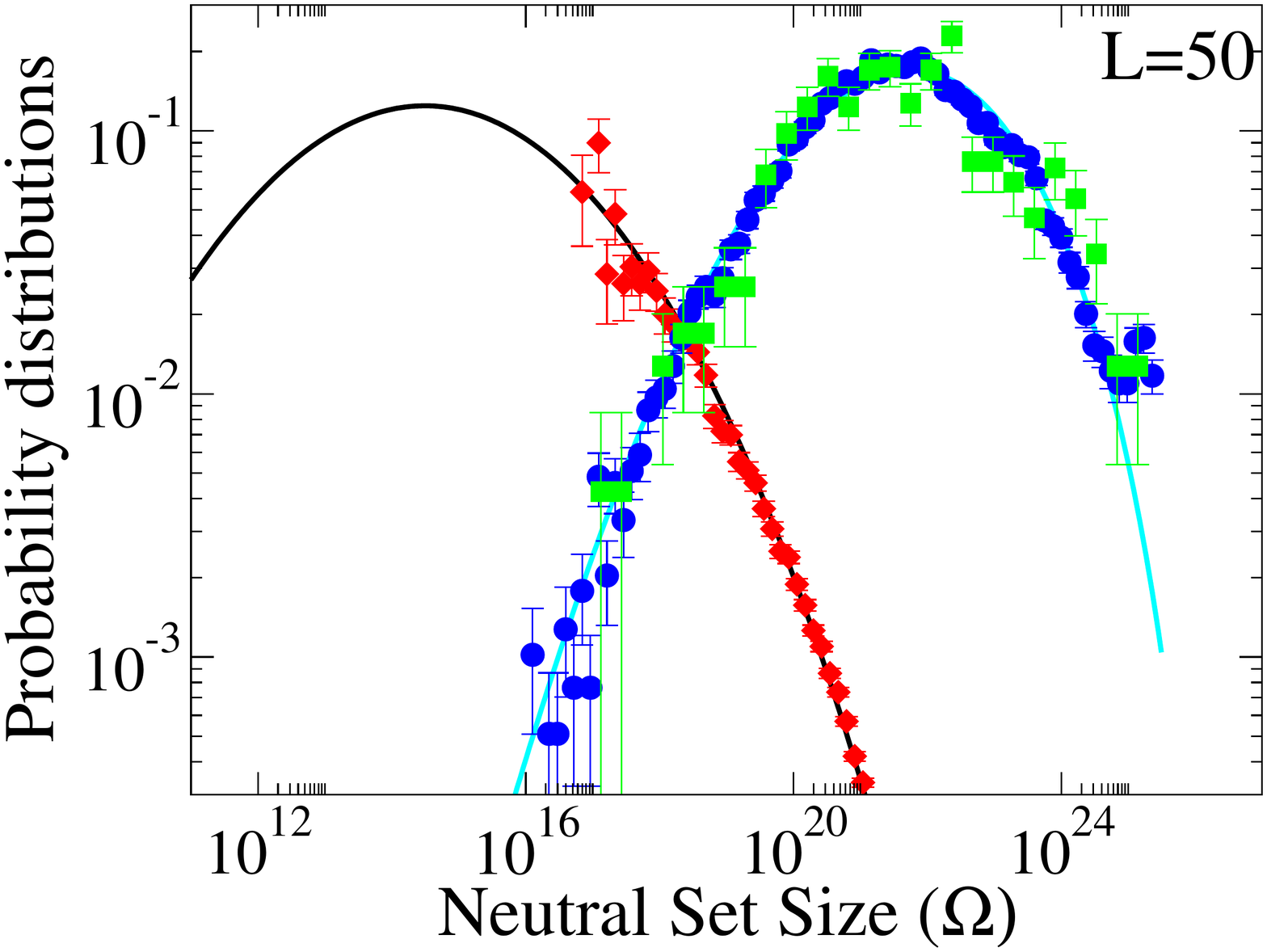}
\includegraphics[height=3.4cm,width=4.65cm]{L55-simple-log.pdf}
\includegraphics[height=3.4cm,width=4.65cm]{L60-log.pdf}
\includegraphics[height=3.4cm,width=4.65cm]{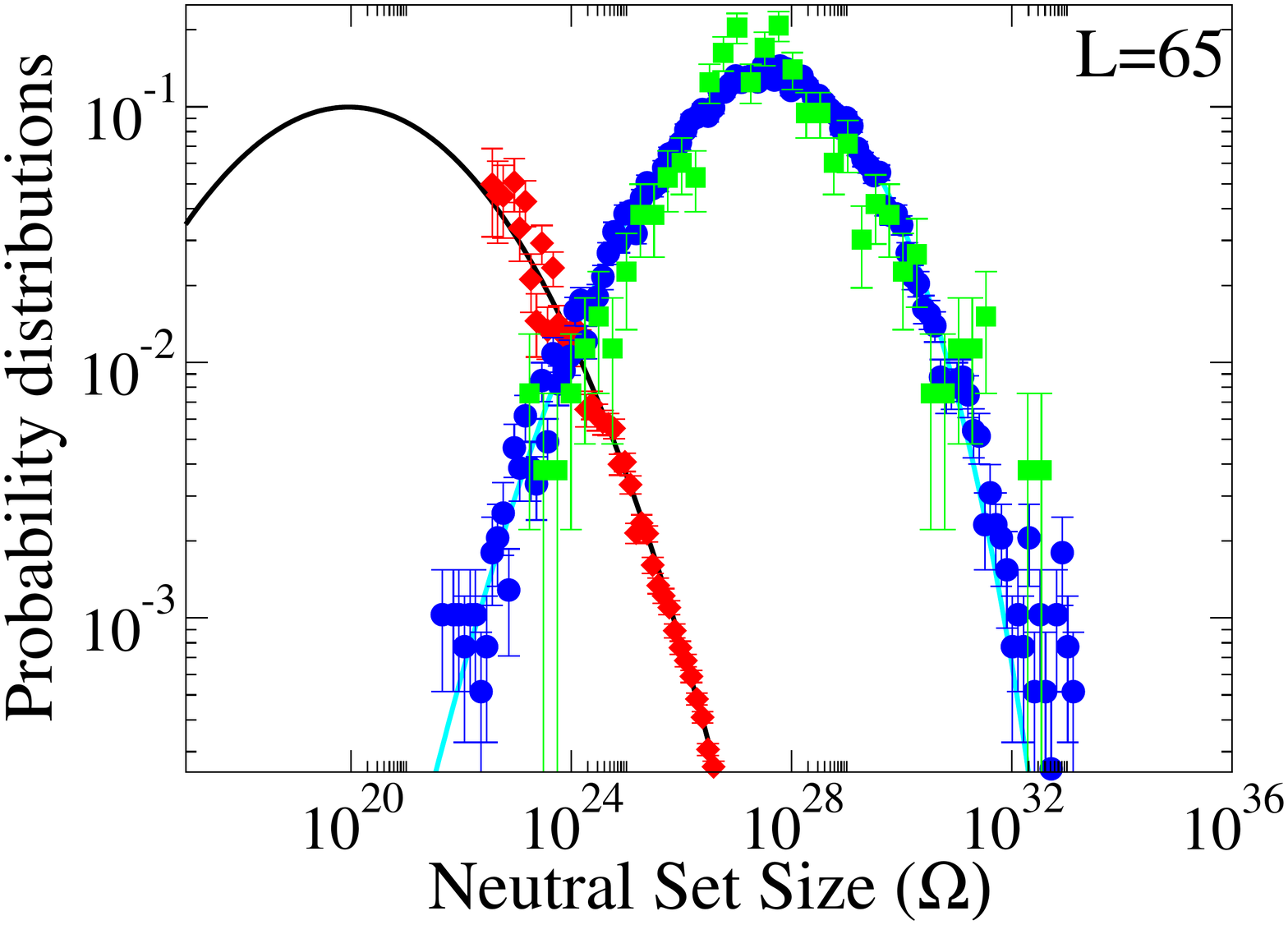}
\includegraphics[height=3.4cm,width=4.65cm]{L70-simple-log.pdf}
\includegraphics[height=3.4cm,width=4.65cm]{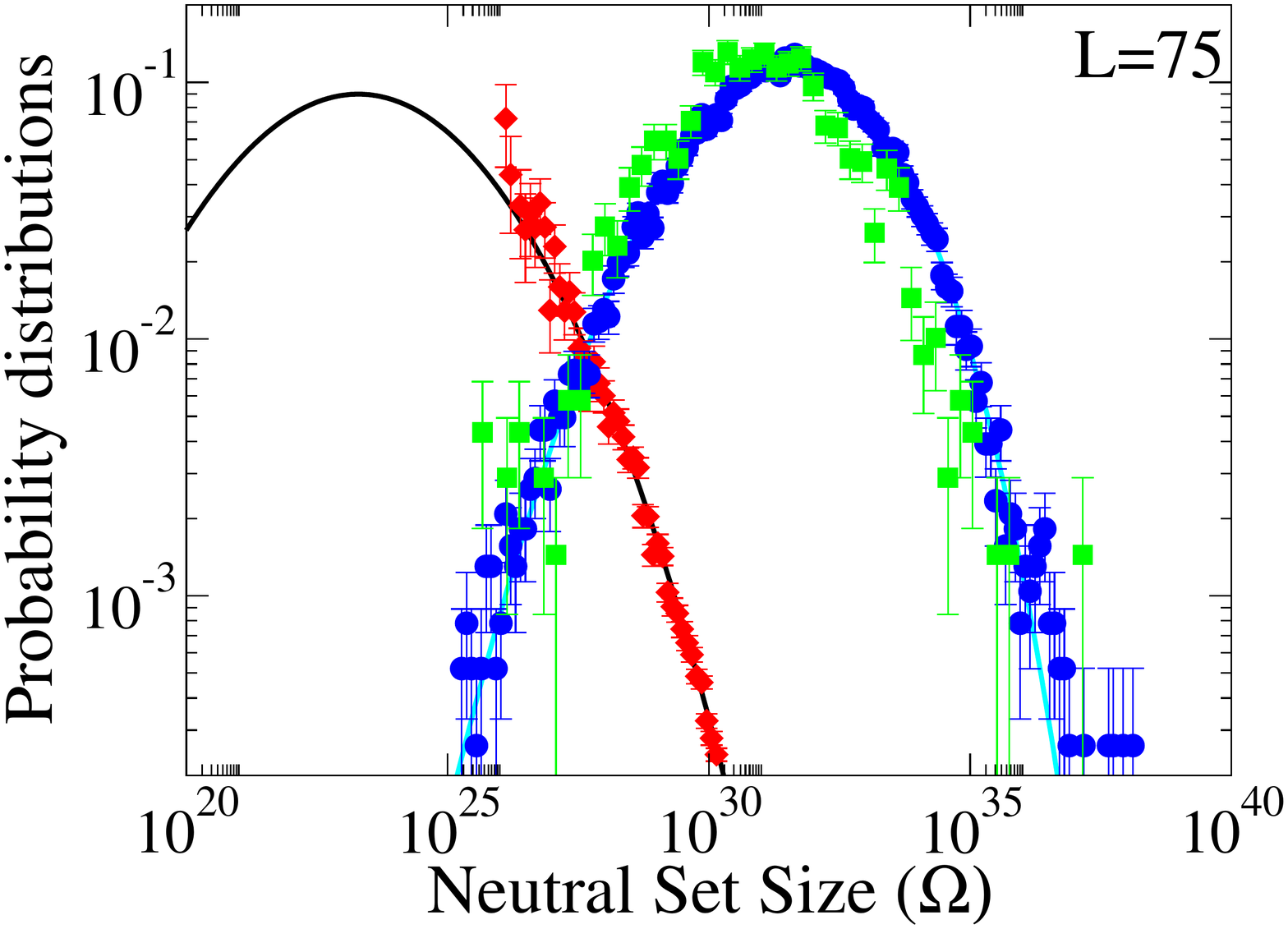}
\includegraphics[height=3.4cm,width=4.65cm]{L80-log.pdf}
\includegraphics[height=3.4cm,width=4.65cm]{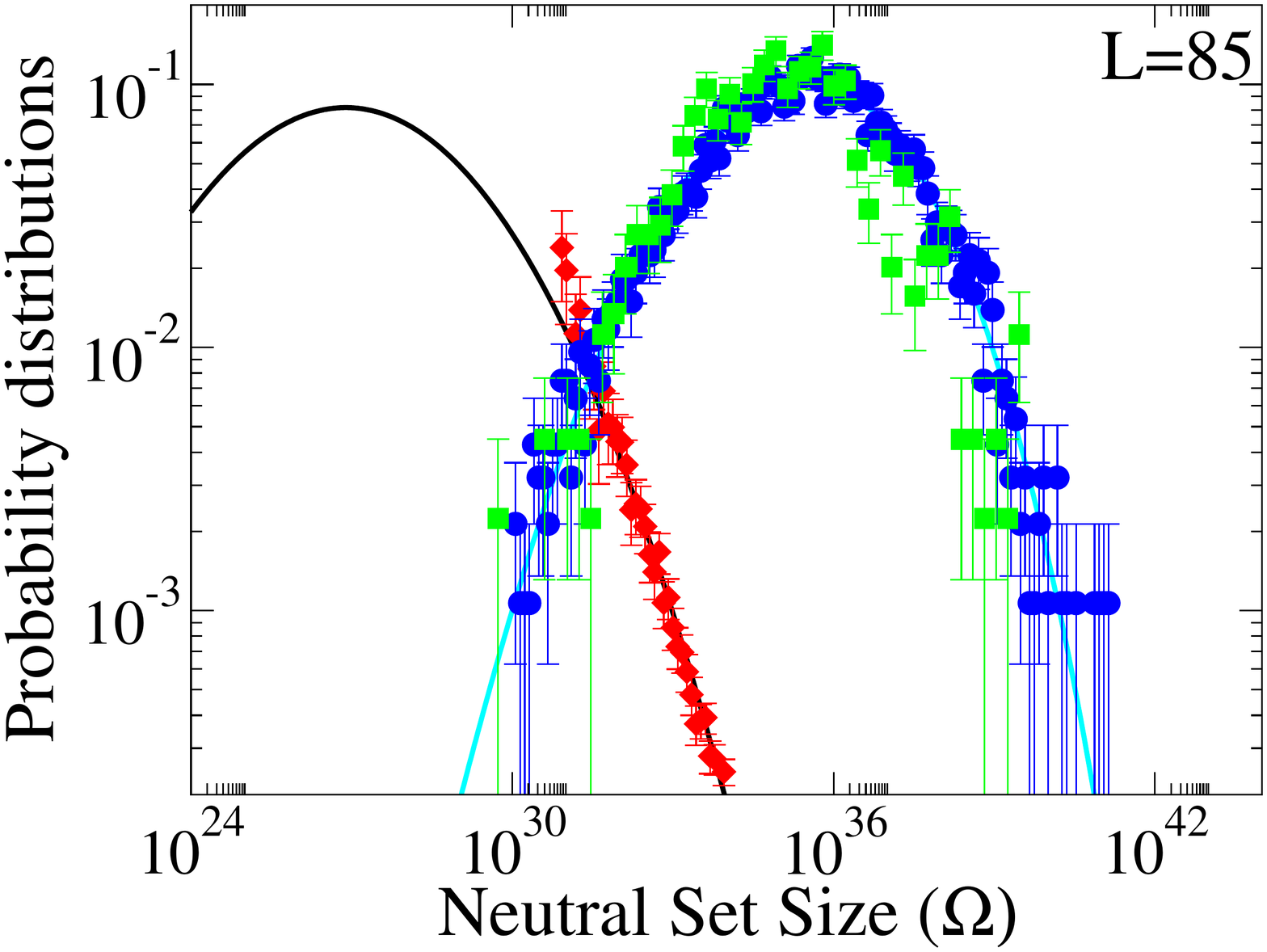}
\includegraphics[height=3.4cm,width=4.65cm]{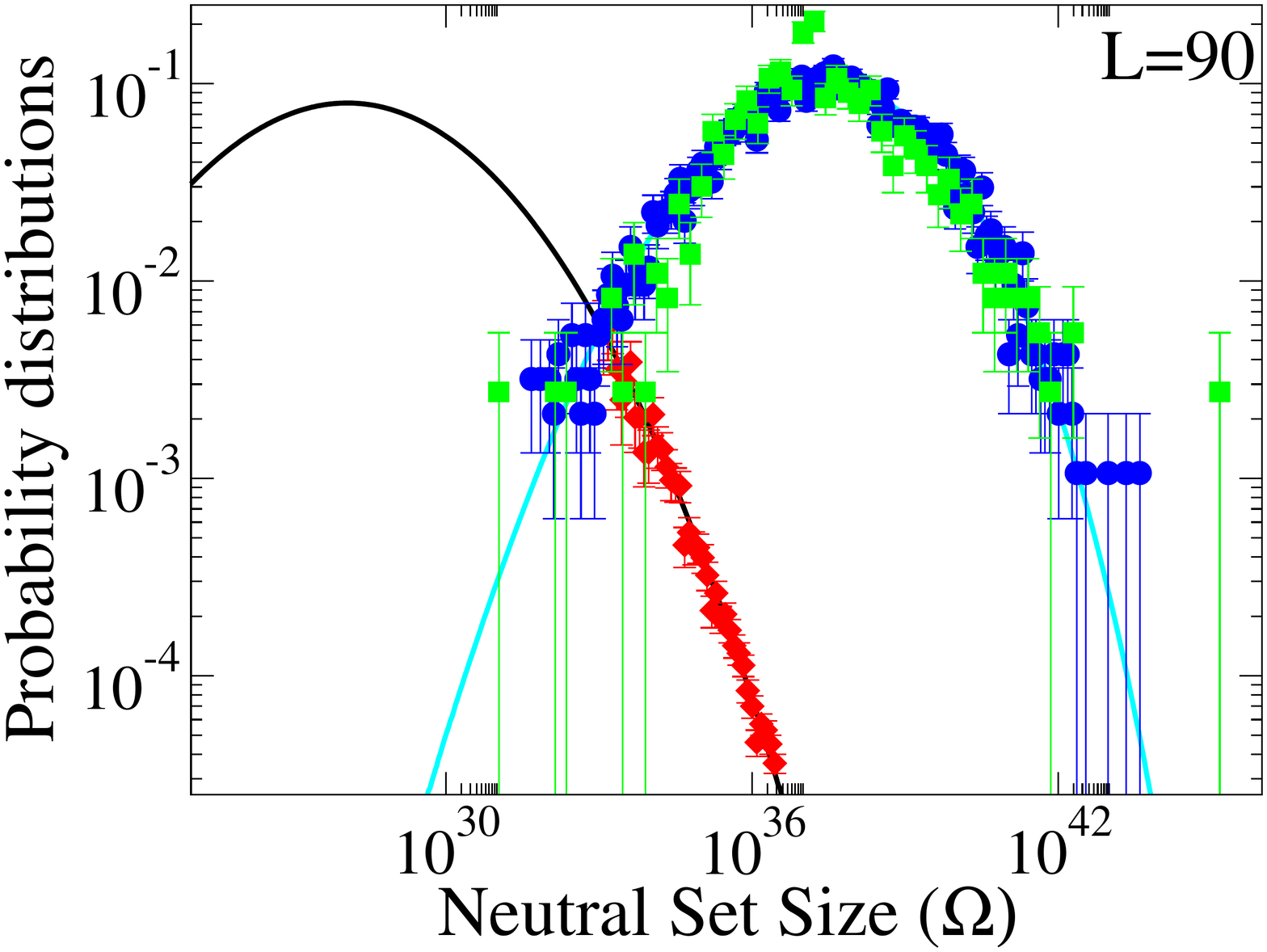}
\includegraphics[height=3.4cm,width=4.65cm]{L100-log.pdf}
\includegraphics[height=3.4cm,width=4.65cm]{L126-simple-log.pdf}
\caption{  {\bf Neutral set size distributions for RNA secondary structures } Same data as Supplementary  Fig S1, but on a log-log scale.}
\label{fig:X1b}
\end{figure*}

\begin{figure*}[htp] \centerline{
\includegraphics[height=5cm,width=6cm]{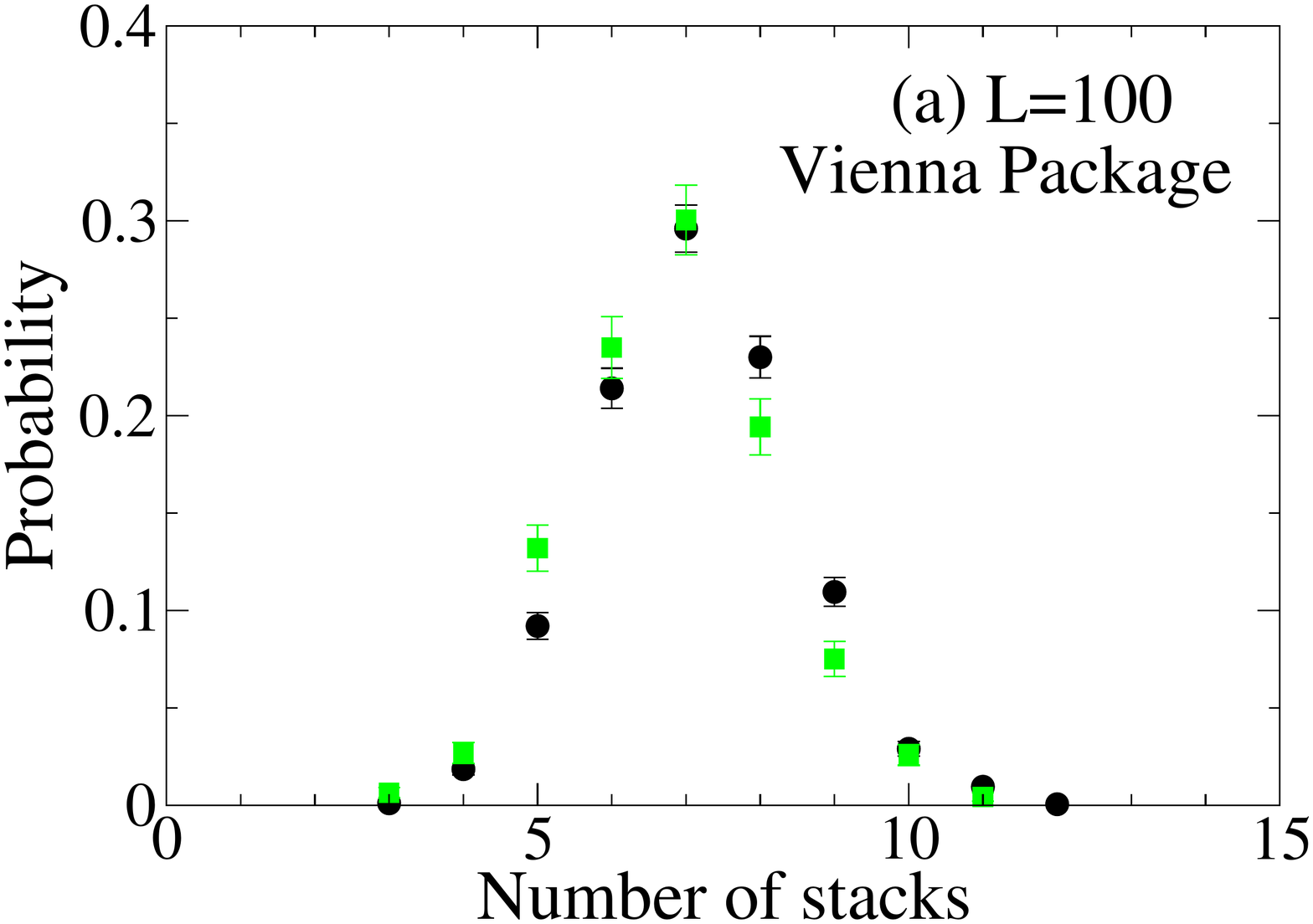}
\includegraphics[height=5cm,width=6cm]{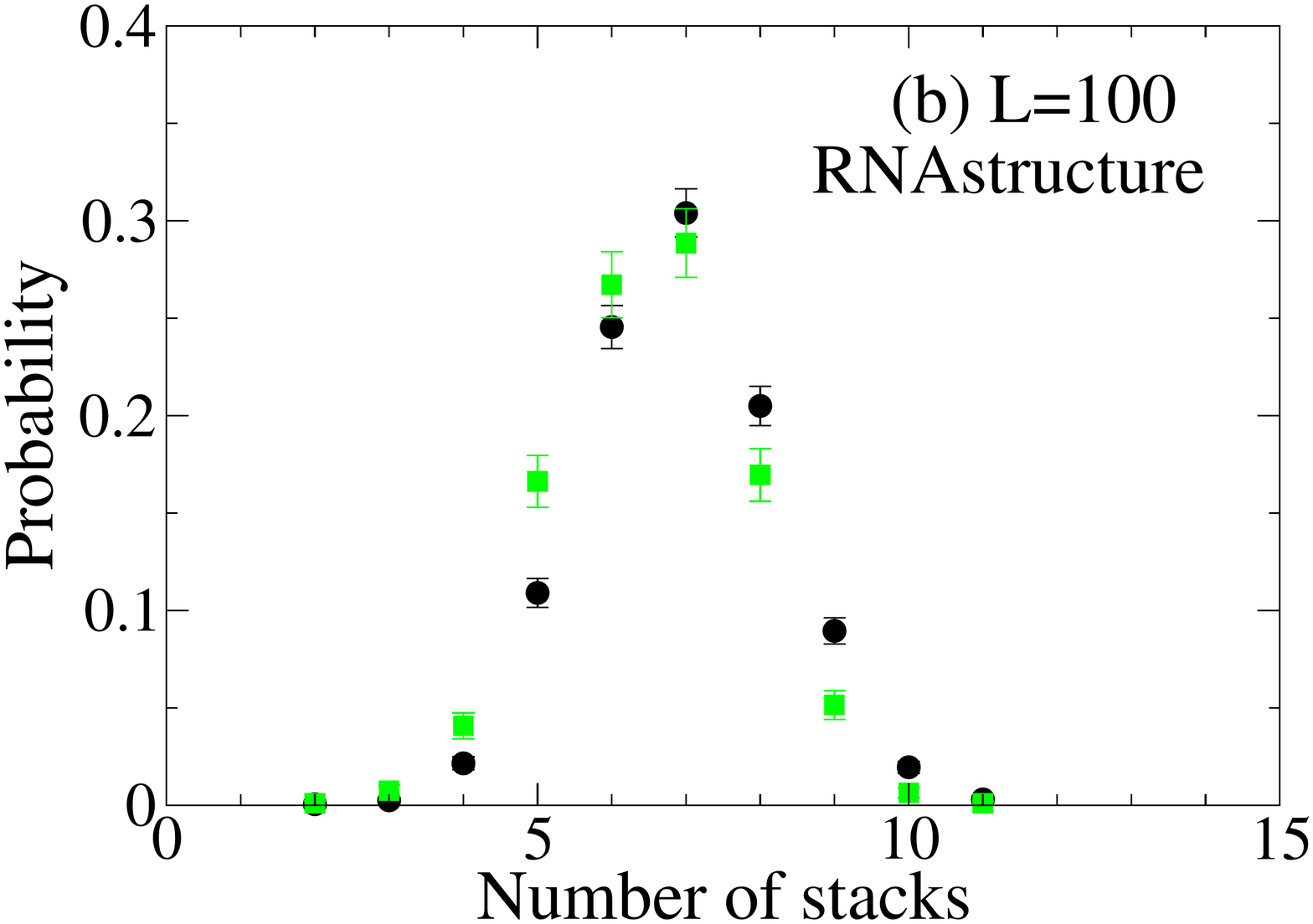}} 
\centerline{
\includegraphics[height=5cm,width=6cm]{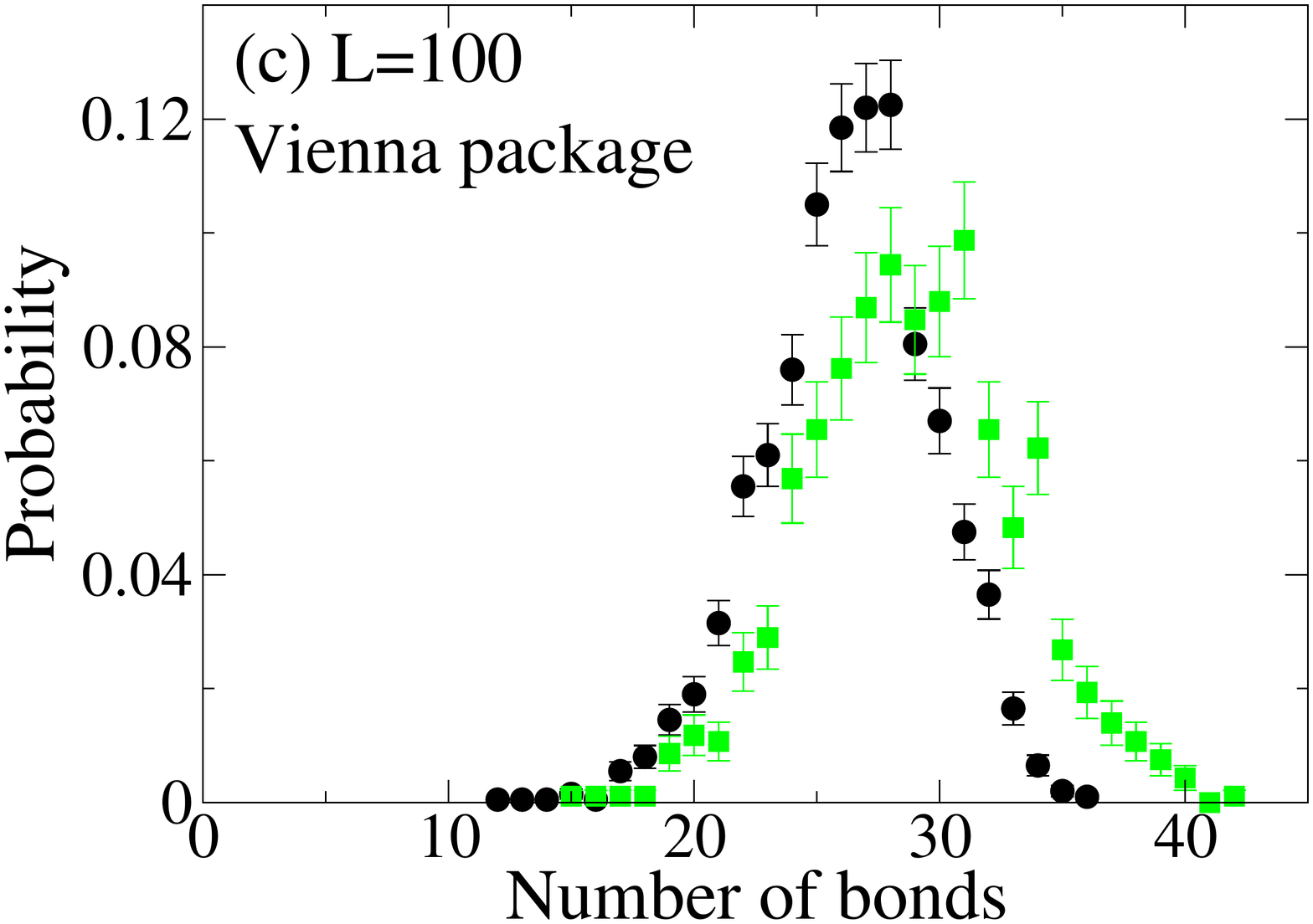}
\includegraphics[height=5cm,width=6cm]{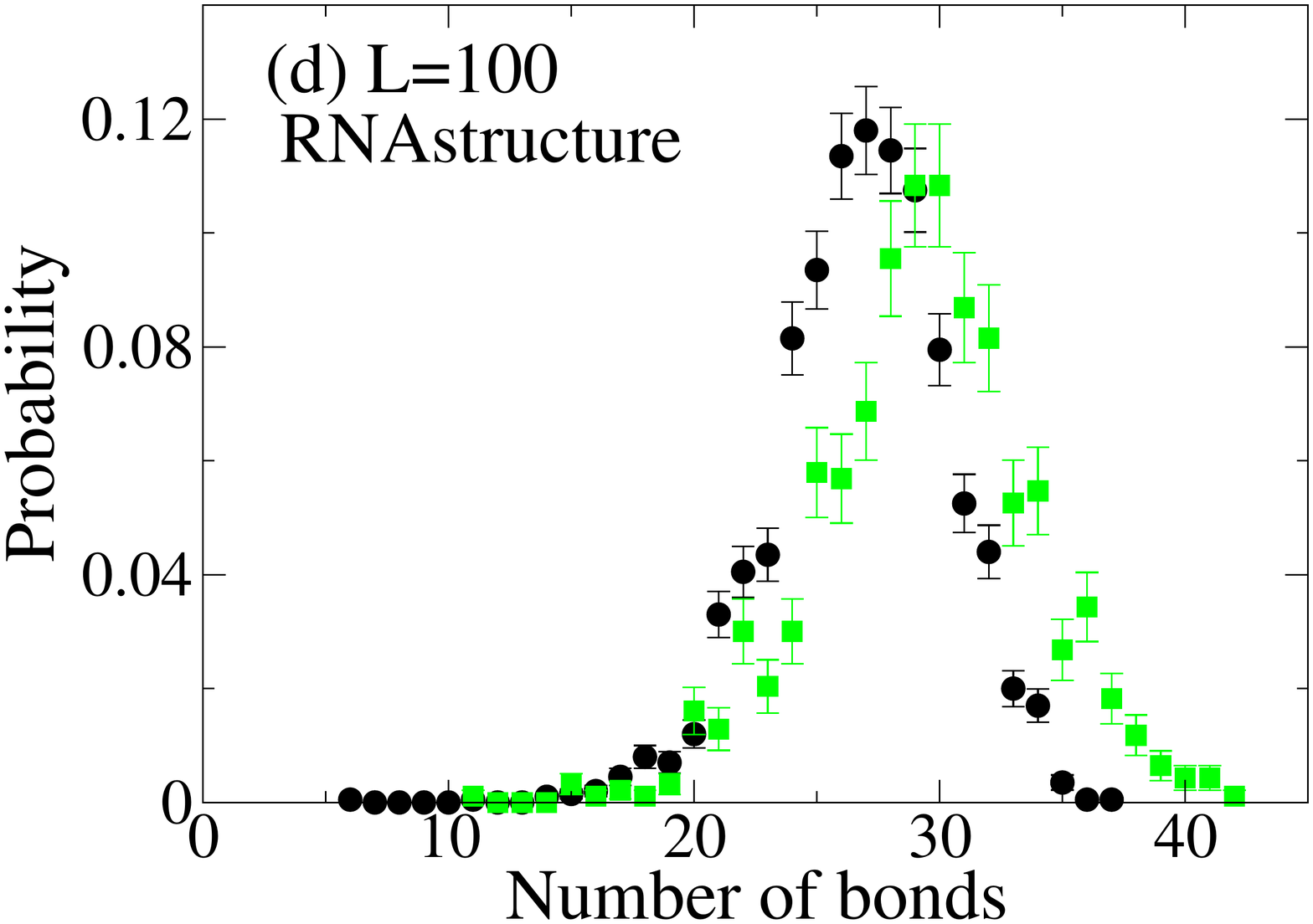}}
\caption{ {\bf Comparison of Vienna package and RNAstructure package predictions.}  In each case green symbols are for natural sequences, with  their SS  and number of stacks calculated with the respective method and the black symbols are for G-sampled sequences. {\bf (a)} Stacks for $L=100$ (Vienna) {\bf (b)} Stacks for $L=100$ (RNAstructure) {\bf (c)} Bonds for $L=100$ (Vienna) {\bf (d)} Bonds for $L=100$ RNAstructure.   Note the overall good agreement between the two methods.  While the natural and G-sampled stack distributions are very close, the natural RNA have slightly more bonds (about 2.5 more) on average than the G-sampled distributions do (28.9 v.s.\ 26.3 respectively for the Vienna package and 29.3 v.s.\  26.8 respectively for RNAstruture), in agreement with the expectation that they are slightly more stable than G-sampled structures\protect~\cite{schultes2005compact}.  Similar agreement between the two methods  is found  for other motifs such as junctions, bulges and loops, where again natural RNA distributions are very close to G-sampled distributions. }
\label{fig:X4}
\end{figure*}

\begin{figure*}[htp] \centerline{
\includegraphics[height=5cm,width=6cm]{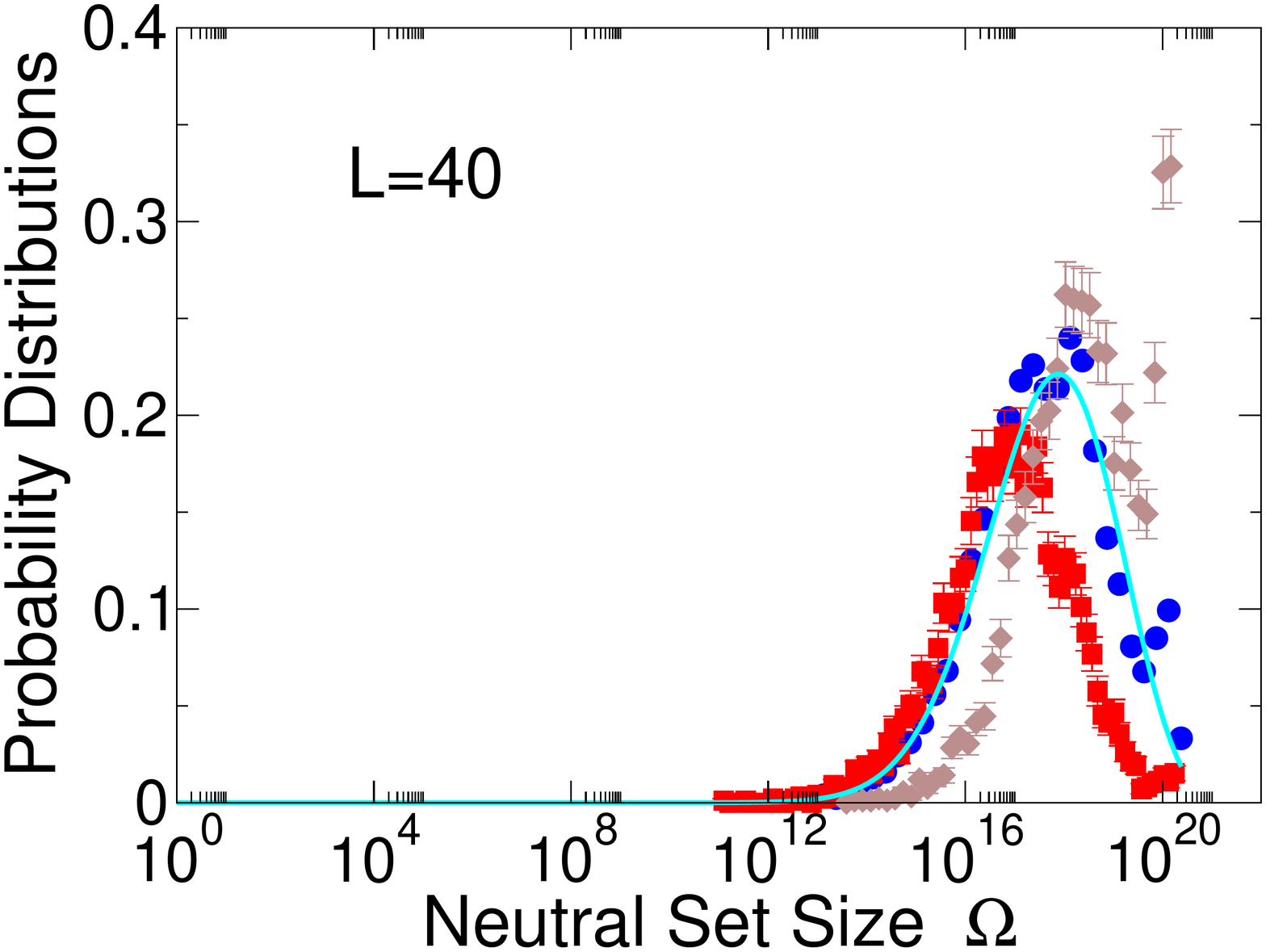}
\includegraphics[height=5cm,width=6cm]{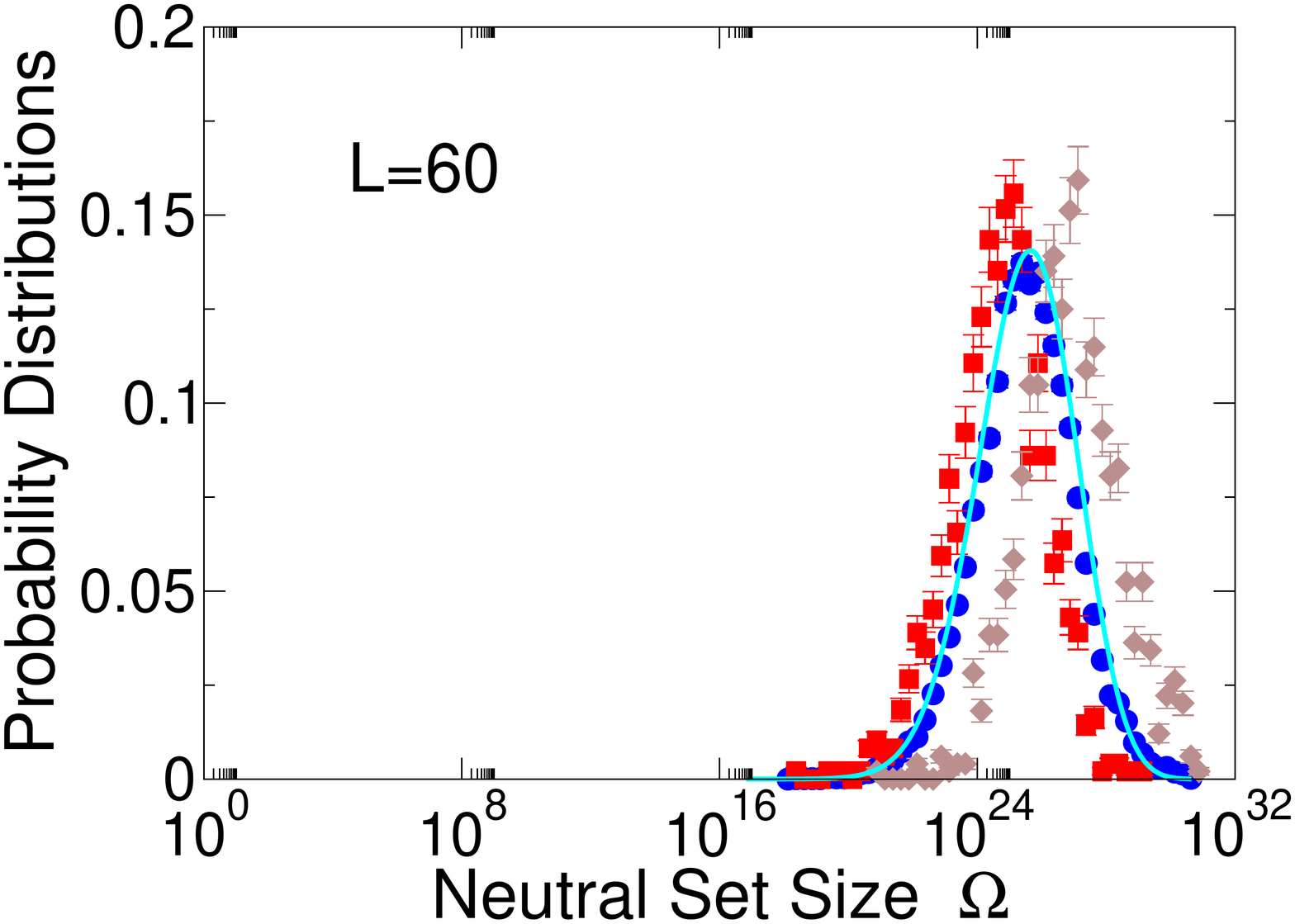}}
\caption{\hspace*{-0.15cm} {\bf  $\bf P_G(\Omega)$  with $\bf 30\%$ GC (AU bias)  or $\bf 70\%$ GC (GC bias). }  The G-sampled distributions were  taken from 5000  samples for each bias and length. These are compared to the unbiased results obtained from $3 \times 10^5$ samples for $L=40$ and $2 \times 10^4$ samples for $L=60$ (blue circles) and our approximation for $P_G(\Omega)$ (cyan line).   Strong AU bias (brown diamonds) favours  structures with slightly  larger NS size, while strong GC bias (red squares) favours structures with slightly smaller NS size.  Although the overall effect is rather small, it would be interesting to compare RNA structures for genomes that contain a large bias to see if such shifts can be observed. }
\label{fig:X3}
\end{figure*}

\begin{figure*}[htp] \centerline{
\includegraphics[height=5.0cm,width=6.5cm]{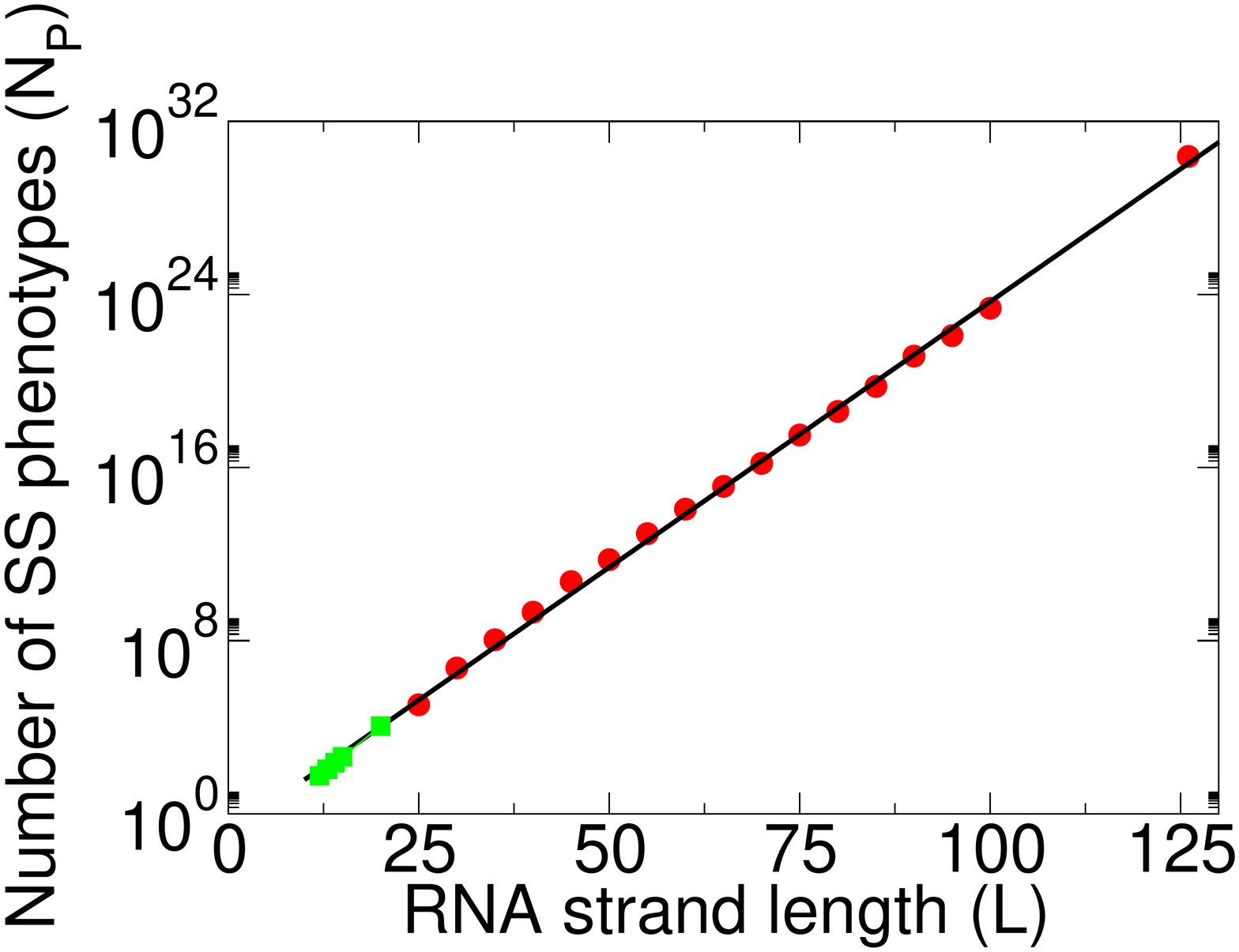}
\includegraphics[height=5.0cm,width=6.5cm]{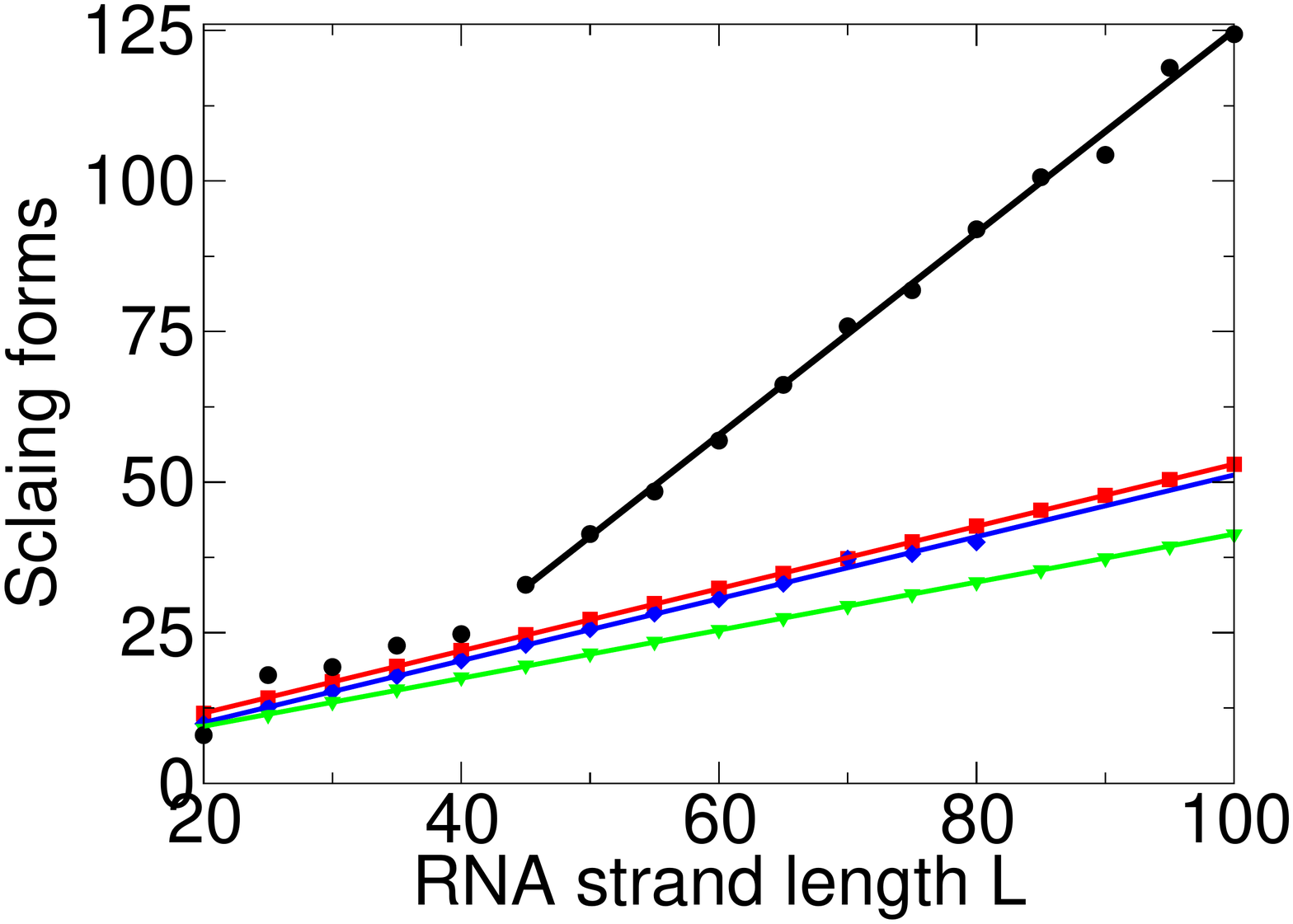} }
\caption{\hspace*{-0.15cm} {\bf  Measured or extracted properties compared to linear fits as a function of RNA strand length $\bf L$}. Left figure: Green diamonds are the number of SS, $N_P$, from full enumerations (up to $L=20$).  Red circles are $N_P$ extracted from sampling and fitting to a binomial form for longer $L$ (Methods).  $N_P = 0.13 \times1.76^L$ (straight black line) provides a good fit ($R^2 = 0.9986$) to the data.
  Right figure:  The   logarithm of the largest non-trivial NS size, $\log(\Omega) = U$, (blue diamonds) can be fit with  $U= (0.514 \pm 0.009)L - (0.20 \pm 0.5)$ (blue line). The logarithm of the NS size of the trivial structure, $\log(\Omega) =T$, (red squares) can be fit with $T= (0.5166 \pm 0.0009)L+(1.33 \pm 0.06)$ (red line), and is close to $U$.  The mean $\bar{S}_G$ of the G-sampled distribution of $S=\log(\Omega)$ (green triangles)  can be fit with $\bar{S}_G = (0.399 \pm 0.0014)L + (1.48 \pm 0.09)$ (green line). Measured data for the binomial fit parameter $N$ (black circles) can be fit with $N = 1.68 L - 43$ (black line) for  $N > 40$.  For lower $N$ a linear fit does not capture the data as well. For these shorter length there tends to be more structure in the measured G-distribution than in the smoother binomial fit.
    }
\label{fig:X2}
\end{figure*}

\begin{figure*}[htp]\centerline{
\includegraphics[height=5cm,width=6cm]{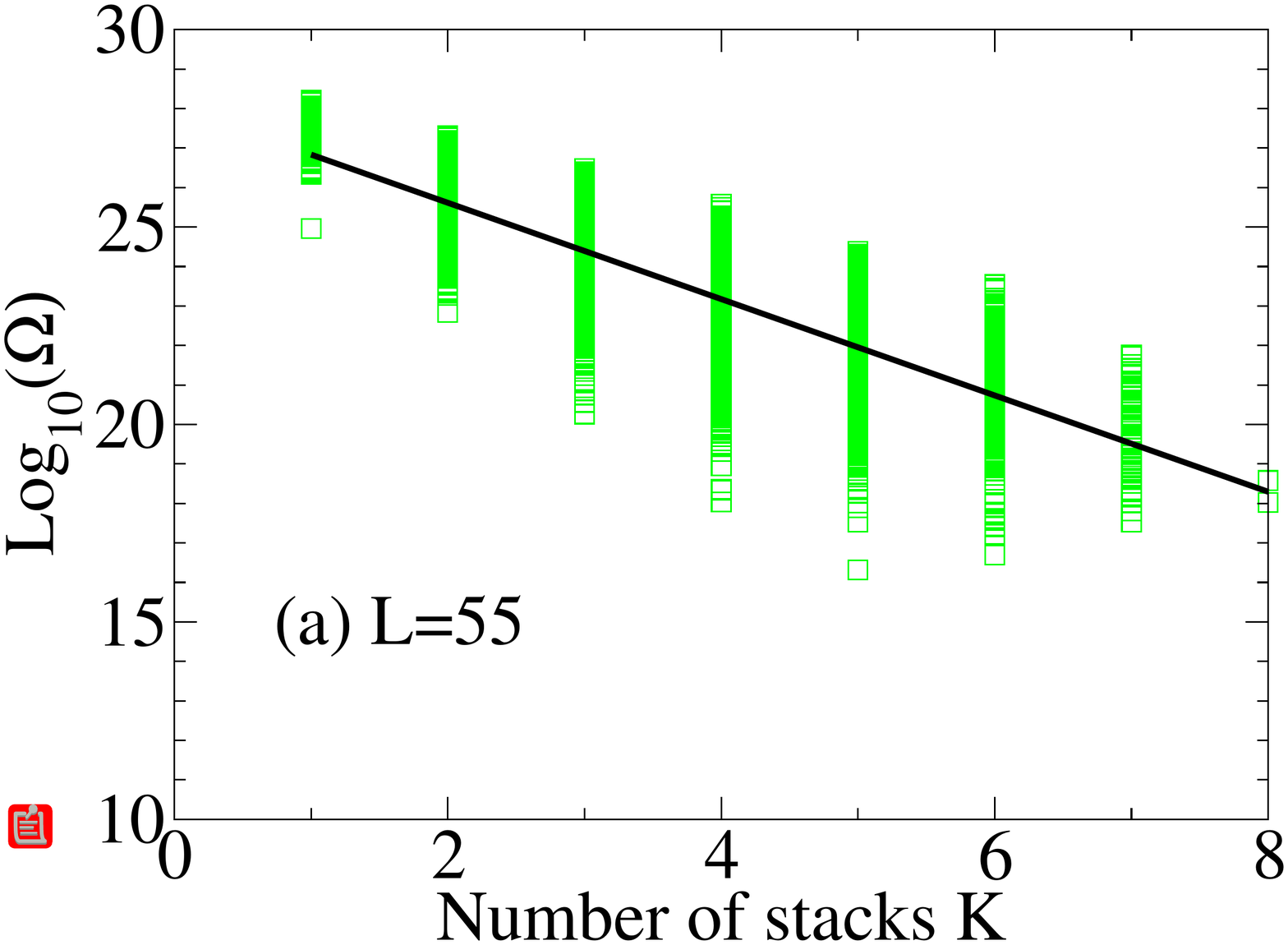}
\includegraphics[height=5cm,width=6cm]{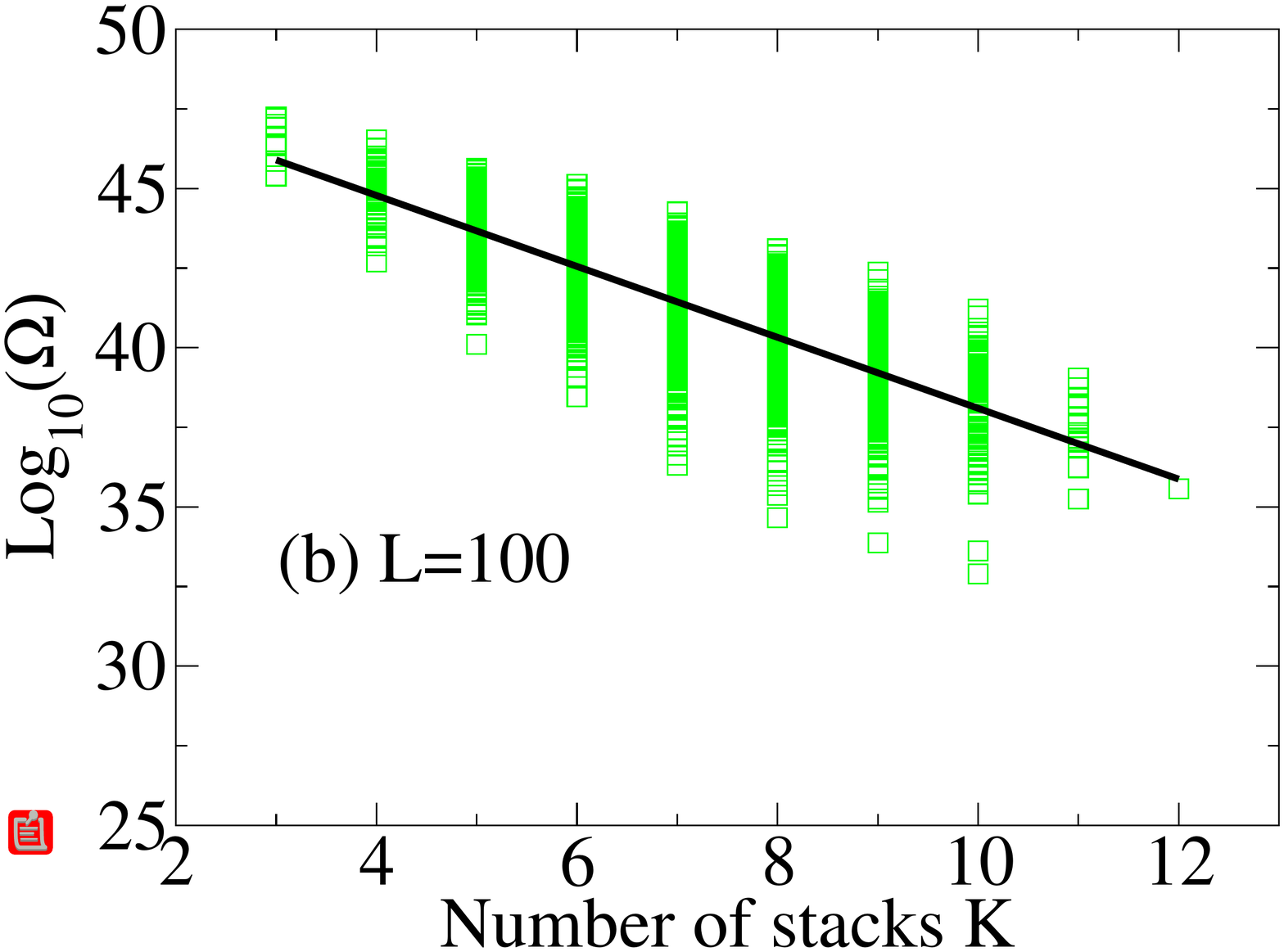}}
\caption{{\bf The number of stacks (continuous sets of base-pairs) $\bf K$ }.  {\bf (a)}  For $L=55$,  the relationship between NS size and number of stacks (green squares), calculated by G-sampling, 
 can be fit with  $ \log(\Omega) = 28.053 - 1.22K$ ($R^2 = -0.84$) (black line).   {\bf (b)}  For $L=100$ we find $\log(\Omega) = 49.238 - 1.114  K$ ($R^2=0.80$).  Similar linear correlations are found for other lengths.     }
\label{fig:X5}
\end{figure*}

\end{document}